\documentclass[12pt]{article}
\usepackage{amssymb,lscape}
\usepackage{amsmath,multirow}
\usepackage{amsthm}
\usepackage{pdflscape}
\usepackage{diagbox}
\usepackage{subfig}
\usepackage[ruled,vlined]{algorithm2e}
\usepackage[normalem]{ulem}

\usepackage{hyperref}

\usepackage{graphicx}
\usepackage[curve,matrix,arrow,color]{xy}

\usepackage[usenames,dvipsnames]{color}

\numberwithin{equation}{section}
\def \bes{\begin{eqnarray}}
\def \ees{\end{eqnarray}}
\def \bns{\begin{eqnarray*}}
\def \ens{\end{eqnarray*}}

\newcommand{\ffrac}[2]{\mbox{\footnotesize$\displaystyle\frac{#1}{#2}$}}

\newcommand{\BTnew}[1]{\mathbb{B}(#1)}
\newcommand{\BTmunew}[1]{\mathbb{B}_{\mu}(#1)}

\newcommand{\vecT}[2]{|\!| #1;#2\rangle\!\rangle}
\newcommand{\vecTmu}[2]{|\!| #1;#2\rangle\!\rangle_{\mu}}

\newcommand{\vecG}[2]{|\!|\!| #2\rangle\!\rangle\!\rangle^{(#1)}}
\newcommand{\vecGomega}[2]{|\!|\!| #2\rangle\!\rangle\!\rangle^{(#1)}_{\omega}}

\newcommand{\be}{\begin{eqnarray}}
\newcommand{\ee}{\end{eqnarray}}
\newcommand{\non}{\nonumber}
\newcommand{\id}{\mathbb{I}}
\newcommand{\tr}{\mathop{\rm tr}\nolimits}
\newcommand{\diag}{\mathop{\rm diag}\nolimits}
\newcommand{\sh}{\mathop{\rm sh}\nolimits}
\newcommand{\ch}{\mathop{\rm ch}\nolimits}

\newcommand{\qg}{\mathop{U_{q} sl(2)}\nolimits}
\newcommand{\lam}{\mathop{v}\nolimits}

\newcommand{\Irr}[1]{\langle#1\rangle}
\newcommand{\ef}{\mathsf{f}}

\newcommand{\PP}{\mathsf{P}}

\newcommand{\q}{q}
\newcommand{\toppr}{\mathsf{t}}
\newcommand{\botpr}{\mathsf{b}}
\newcommand{\leftpr}{\mathsf{l}}
\newcommand{\rightpr}{\mathsf{r}}
\newcommand{\stprp}{\mathsf{a}}
\newcommand{\half}{%
  \mathchoice{\ffrac{1}{2}}{\frac{1}{2}}{\frac{1}{2}}{\frac{1}{2}}}

\newcommand{\vac}{|\Omega\rangle}

\newtheorem{Prop}[subsection]{Proposition}

\newtheorem{Cor}[subsection]{Corollary}

\theoremstyle{definition}

\newtheorem{Rem}[subsection]{Remark}

\begin{document}

\begin{titlepage}
\strut\hfill UMTG--283
\vspace{.5in}
\begin{center}

\LARGE 
Algebraic Bethe ansatz \\
for the quantum group invariant open XXZ chain\\ 
at roots of unity\\
\vspace{1in}
\large Azat M. Gainutdinov \footnote{
DESY, Theory Group, Notkestra\ss e 85, Bldg. 2a, 22603 Hamburg,
Germany;
Laboratoire de Math\'ematiques et Physique Th\'eorique CNRS/UMR 7350,
 F\'ed\'eration Denis Poisson FR2964,
Universit\'e de Tours,
Parc de Grammont, 37200 Tours, 
France,
azat.gainutdinov@lmpt.univ-tours.fr}
and Rafael I. Nepomechie \footnote{
Physics Department,
P.O. Box 248046, University of Miami, Coral Gables, FL 33124 USA,
nepomechie@physics.miami.edu}\\[0.8in]
\end{center}

\vspace{.5in}

\begin{abstract}
    For generic values of $q$, all the eigenvectors of 
    the transfer matrix of the $\qg$-invariant open spin-1/2 XXZ chain
    with finite length $N$
    can be constructed using the algebraic Bethe ansatz (ABA) 
    formalism of Sklyanin. However, when $q$ is a  root of unity
    ($q=e^{i \pi/p}$ with integer $p\ge 2$),
    the Bethe equations
    acquire continuous solutions, and the transfer matrix develops 
    Jordan cells. Hence, there appear eigenvectors of two new types:
    eigenvectors corresponding to continuous solutions
    (exact complete $p$-strings), and generalized eigenvectors. We 
  propose general ABA constructions for these two new types of 
    eigenvectors.
    We present many explicit examples, and we construct complete sets 
    of (generalized) eigenvectors for various values of $p$ and $N$.
    \end{abstract}

\end{titlepage}

\setcounter{footnote}{0}

\section{Introduction}\label{sec:intro}

In the pantheon of anisotropic integrable quantum spin chains, one
model stands out for its high degree of symmetry: the $\qg$-invariant
open spin-1/2 XXZ quantum spin chain, whose Hamiltonian is given by
\cite{Pasquier:1989kd}
\be
H = \sum_{k=1}^{N-1} \left[ \sigma^x_k \sigma^x_{k+1} +
\sigma^y_k \sigma^y_{k+1} + \tfrac{1}{2}( q + q^{-1}) \sigma^z_k
\sigma^z_{k+1}
\right]  - \tfrac{1}{2}( q - q^{-1})
\left( \sigma^z_1 - \sigma^z_N \right) \,, \label{Hamiltonian}
\ee
where $N$ is the length of the chain, $\vec\sigma$ are the usual Pauli
spin matrices, and $q=e^{\eta}$ is an arbitrary complex parameter.  As
is true for generic quantum integrable models, the Hamiltonian is a
member of a family of commuting operators that can be obtained from a
transfer matrix \cite{Sklyanin:1988yz}; and the eigenvalues of the
transfer matrix can be obtained in terms of admissible solutions
$\{\lambda_{k}\}$ of the corresponding set of Bethe equations
\cite{Alcaraz:1987uk, Sklyanin:1988yz, Pasquier:1989kd}  \footnote{In order to reduce the size of 
formulas, we denote the hyperbolic sine function ($\sinh$) by 
$\sh$.}
\be
\lefteqn{\sh^{2N}\left(\lambda_{k} + \tfrac{\eta}{2}\right)
\prod_{\scriptstyle{j \ne k}\atop \scriptstyle{j=1}}^M
\sh(\lambda_{k} - \lambda_{j} - \eta)\sh(\lambda_{k} + \lambda_{j} - \eta)
} \non \\
&&=\sh^{2N}\left(\lambda_{k} - \tfrac{\eta}{2}\right)
\prod_{\scriptstyle{j \ne k}\atop \scriptstyle{j=1}}^M
\sh(\lambda_{k} - \lambda_{j} + \eta)\sh(\lambda_{k} + \lambda_{j} + \eta) \,, \non \\
&&\qquad k = 1 \,, 2\,, \ldots \,, M \,, \qquad
M = 0\,, 1\,, \ldots\,,  
\lfloor\tfrac{N}{2}\rfloor \,,
\label{BAE}
\ee
where $\lfloor k \rfloor$ denotes the integer not greater than $k$.

When the anisotropy parameter $\eta$ takes the values $\eta=i\pi/p$
with integer $p \ge 2$, and therefore $q=e^{\eta}$ is a root of unity,
several interesting new features appear.  In particular, the symmetry
of the model is enhanced (for example, an $sl(2)$ symmetry arises from
the so-called divided powers of the quantum group generators); the
Hamiltonian has Jordan
cells~\cite{Dubail:2010zz,Vasseur:2011fi,MorinDuchesne:2011kd}; and
the Bethe equations (\ref{BAE}) admit continuous solutions
\cite{Gainutdinov:2015vba}, in addition to the usual discrete
solutions (the latter phenomenon also occurs for the closed XXZ chain
\cite{Baxter:1972wg, Fabricius:2000yx, Fabricius:2001yy, Baxter:2001sx,
Tarasov:2003xz}). 

We have recently found \cite{Gainutdinov:2015vba} significant
numerical evidence that the Bethe equations have precisely the right
number of admissible solutions to describe all the distinct (generalized)
eigenvalues of the model's transfer matrix, even at roots of unity.
 
We focus here on the related problem of constructing, via the algebraic
Bethe ansatz, all $2^{N}$ (generalized) eigenvectors of the transfer
matrix.
For generic $q$, the construction of these eigenvectors is similar to the one 
for the simpler spin-1/2 XXX chain: to each admissible solution of the
Bethe equations, there corresponds a Bethe vector, which is a
highest-weight state of $\qg$ \cite{Pasquier:1989kd, Hou:1991tc,
Mezincescu:1991rb}; and lower-weight states can be obtained by acting
on the Bethe vector with the quantum-group lowering 
operator $F$.

However, at roots of unity $q=e^{i\pi/p}$ with integer $p \ge 2 $, 
we find that there are two additional features:
\renewcommand{\theenumi}{\roman{enumi}}
\begin{enumerate}
    \item Certain eigenvectors must be constructed using the continuous
    solutions noted above.  These solutions contain $p$ equally-spaced
    roots (so-called exact complete $p$-strings), whose centers are
    arbitrary, see Prop.~\ref{main-prop-1} for more details.  This construction is a generalization of the one
    proposed by Tarasov for the closed chain at roots of unity~\cite{Tarasov:2003xz}.
    \item We propose that the generalized eigenvectors can be constructed
    using similar string configurations of length up to $p-1$, except
    the centers tend to infinity.  We refer to  Prop.~\ref{prop:gen-vect} for more details.
\end{enumerate}
We demonstrate explicitly for several values of $p$ and $N$ that the 
complete set of (generalized) eigenvectors can indeed be obtained in this way.

The outline of this paper is as follows. In section 
\ref{sec:preliminary} we briefly review results and notations 
(specifically, the construction of the transfer matrix, the algebraic Bethe ansatz, and 
$\qg$ symmetry) that are used later in the paper. In section 
\ref{sec:pstrings} we work out in detail the construction noted in 
item i above with the result formulated in Prop.~\ref{main-prop-1},  see in particular Eqs. 
(\ref{Tarasovstate}) and (\ref{Tarasovstategen}). In section \ref{sec:generalized} we describe the 
construction noted in item ii above with the final result in Prop.~\ref{prop:gen-vect}, see in particular 
Eq. (\ref{genvec2}).
These two constructions are then used 
in section \ref{sec:p2} to construct all the (generalized) 
eigenvectors for the $p=2$ root of unity case with $N=4,5,6$, 
as well as selected eigenvectors with $N=7, 9$.
We present all the (generalized) 
eigenvectors for various values of $p>2$ and $N$ in section \ref{sec:pgt2}.
We conclude with a brief discussion in section \ref{sec:discuss}. 
Some ancillary results are collected in four appendices. 
In Appendix \ref{app:proj-mod-base}, we explicitly describe the
action of $\qg$ in tilting modules at roots of unity.
In Appendix \ref{sec:numerics}, we present numerical evidence for the 
string solutions used in section 
\ref{sec:generalized} for constructing generalized 
eigenvectors. In Appendix \ref{sec:special}, we derive a special 
off-shell relation (similar to the one found by Izergin and Korepin 
\cite{Izergin:1982hy} for repeated Bethe roots), which we use in 
Appendix \ref{sec:proof} to derive an off-shell relation
for generalized eigenvectors.

\section{Preliminaries}\label{sec:preliminary}

The transfer matrix and algebraic Bethe ansatz for the model 
(\ref{Hamiltonian}) follow from the work of Sklyanin 
\cite{Sklyanin:1988yz}, which was already reviewed in 
\cite{Gainutdinov:2015vba}. However, we repeat here the main 
results, both for the convenience of the reader and also to explain a 
useful change in notation (see (\ref{shiftedB}) and subsequent formulas).

\subsection{Transfer matrix}\label{sec:transfer}

The basic ingredients of the transfer matrix are the R-matrix (solution 
of the Yang-Baxter equation)
\be
R(u) = \left(\begin{array}{cccc}
\sh(u+\eta) & 0 & 0 & 0\\
0 & \sh(u) & \sh(\eta) & 0 \\
0 & \sh(\eta) & \sh(u) & 0 \\
0 & 0 & 0 & \sh(u+\eta)
\end{array} \right) \,, 
\ee
and the left and right K-matrices (solutions of the boundary 
Yang-Baxter equations) given by the diagonal matrices
\be
K^{+}(u)=\diag(e^{-u-\eta}\,, e^{u+\eta}) \,, \qquad 
K^{-}(u)=\diag(e^{u}\,, e^{-u})\,, 
\ee 
respectively. The R-matrix is used to construct the monodromy matrices
\be
T_{a}(u) = R_{a1}(u) \cdots R_{aN}(u)\,, \qquad
\hat T_{a}(u) = R_{aN}(u) \cdots R_{a1}(u) \,.
\ee
Finally, the transfer matrix $t(u)$ is given by \cite{Sklyanin:1988yz}
\be
t(u) = \tr_{a} K^{+}_{a}(u)\, {\cal U}_{a}(u)
\,,
\label{transfer}
\ee
where
\be
{\cal U}_{a}(u) = T_{a}(u)\, K^{-}_{a}(u)\, \hat T_{a}(u) \,.
\ee 
The transfer matrix commutes for different values of the spectral parameter
\be
\left[ t(u) \,, t(v) \right] = 0 \,,
\label{commutativity}
\ee
and contains the Hamiltonian (\ref{Hamiltonian}) $H \sim t'(0)$  up 
to multiplicative and additive constants. 

\subsection{Algebraic Bethe ansatz}\label{sec:ABA}

The $A$, $B$, $C$, and $D$ operators of the algebraic Bethe ansatz 
are defined by \cite{Sklyanin:1988yz}
\be
{\cal U}_{a}(u) 
= \left( \begin{array}{cc}
A(u) & B(u) \\
C(u) & D(u) + \frac{\sh \eta}{\sh(2u+\eta)} A(u) 
\end{array} \right) \,,
\ee 
where $\left[B(u)\,, B(v) \right]=0$. 
However, in order to avoid a later shift of the Bethe roots (see 
e.g. Eq. (A.24) in \cite{Gainutdinov:2015vba}),
we now introduce a shifted $B$ operator
\be
{\cal B}(u) \equiv B(u-\tfrac{\eta}{2})
\label{shiftedB} \,.
\ee
We define the Bethe states using this shifted $B$ operator 
\be
|\lambda_{1} \ldots \lambda_{M} \rangle = \prod_{k=1}^{M} {\cal 
B}(\lambda_{k})  \vac\,,
\label{Bethestate}
\ee
where $\vac$ is the reference state with all spins up 
\be
\vac = {1\choose 0}^{\otimes N} \,,
\label{reference}
\ee
and $\lambda_{1}\,, \ldots \,, \lambda_{M}$ remain to be specified.  
The Bethe states satisfy the off-shell relation 
\be
t(u) |\lambda_{1} \ldots \lambda_{M} \rangle = \Lambda(u) 
|\lambda_{1} \ldots \lambda_{M} \rangle + 
\sum_{m=1}^{M} \Lambda^{\lambda_{m}}(u)\, B(u) 
\prod_{\scriptstyle{k \ne m}\atop \scriptstyle{k=1}}^M {\cal 
B}(\lambda_{k}) \vac
\,,
\label{offshell}
\ee
where $\Lambda(u)$ is given by the T-Q 
equation
\be
\Lambda(u) = 
\frac{\sh(2u+2\eta)}{\sh(2u+\eta)}\sh^{2N}(u+\eta)\frac{Q(u-\eta)}{Q(u)} +
\frac{\sh(2u)}{\sh(2u+\eta)}\sh^{2N}(u)\frac{Q(u+\eta)}{Q(u)} \,,
\label{Lambda}
\ee
with
\be
Q(u) 
=\prod_{k=1}^{M}\sh\left(u-\lambda_{k}+\tfrac{\eta}{2}\right)\sh\left(u+\lambda_{k}+\tfrac{\eta}{2}\right) = Q(-u-\eta)\,.
\label{Q}
\ee 
Furthermore,
\be
\Lambda^{\lambda_{m}}(u) &=& \ef(u,\lambda_{m}-\tfrac{\eta}{2})
\Bigg[\sh^{2N}(\lambda_{m}+\tfrac{\eta}{2})
\prod_{\scriptstyle{k \ne m}\atop \scriptstyle{k=1}}^M
\frac{\sh(\lambda_{m}-\lambda_{k}-\eta) 
\sh(\lambda_{m}+\lambda_{k}-\eta)}
{\sh(\lambda_{m}-\lambda_{k}) \sh(\lambda_{m}+\lambda_{k})} 
\non \\
&& -\sh^{2N}(\lambda_{m}-\tfrac{\eta}{2})
\prod_{\scriptstyle{k \ne m}\atop \scriptstyle{k=1}}^M
\frac{\sh(\lambda_{m}-\lambda_{k}+\eta) 
\sh(\lambda_{m}+\lambda_{k}+\eta)}
{\sh(\lambda_{m}-\lambda_{k}) 
\sh(\lambda_{m}+\lambda_{k})}\Bigg] \,,
\label{Lambdam}
\ee
where 
\be
\ef(u,v) = \frac{\sh(2u+2\eta)\sh(2v) \sh \eta}
{ \sh(u-v) \sh(u+v+\eta) \sh(2v+\eta)} \,.
\label{fuv}
\ee
It follows from the off-shell equation (\ref{offshell}) that the Bethe state $|\lambda_{1} \ldots 
\lambda_{M} \rangle$ (\ref{Bethestate}) is an eigenstate 
of the transfer matrix $t(u)$ (\ref{transfer}) with eigenvalue $\Lambda(u)$ (\ref{Lambda})
if the coefficients $\Lambda^{\lambda_{m}}$ of all the ``unwanted'' terms vanish; that is, according to (\ref{Lambdam}),
if $\lambda_{1}\,, \ldots \,, \lambda_{M}$ satisfy the Bethe 
equations (\ref{BAE}).    In particular, the 
eigenvalues of the Hamiltonian (\ref{Hamiltonian}) are given by
\be
E = 2 \sh^{2}\eta \sum_{k=1}^{M}
\frac{1}{\sh(\lambda_{k}-\frac{\eta}{2})\,
\sh(\lambda_{k}+\frac{\eta}{2})} + (N-1)\ch\eta \,.
\label{energy}
\ee

We can restrict to solutions that are 
\textit{admissible} \cite{Gainutdinov:2015vba}: all the $\lambda_{k}$'s are finite and pairwise distinct (no
two are equal), and each $\lambda_{k}$ satisfies
either
\be
\Re e(\lambda_{k} ) > 0  \qquad  \mbox{ and  } \qquad
-\frac{\pi}{2} <  \Im m(\lambda_{k} ) \le \frac{\pi}{2}
\label{admissibleXXZa}
\ee
or
\be
\Re e(\lambda_{k} ) = 0  \qquad \mbox{ and  } \qquad 0 <  \Im m(\lambda_{k} ) < \frac{\pi}{2}   \,.
\label{admissibleXXZb}
\ee
 Moreover, for the root of unity case $\eta=i\pi/p$
with integer $p \ge 2$, we exclude 
solutions containing exact complete $p$-strings, see section \ref{sec:pstrings} 
below. All the admissible solutions of the Bethe equations 
(\ref{BAE}) for small values of $p$ and $N$ are given in 
\cite{Gainutdinov:2015vba}.

\subsection{$\qg$ symmetry}\label{sec:qg}

For generic $q$, the quantum group $\qg$ has generators $E\,, F\,, K$ 
that satisfy the relations
\be
K\, E\, K^{-1} = q^{2} E\,, \qquad K\, F\, K^{-1} = q^{-2} F\,,\qquad
\left[ E\,, F \right] = \frac{K - K^{-1}}{q-q^{-1}} \,.
\ee
These generators are represented on the spin chain by (see e.g. \cite{Gainutdinov:2012qy})
\be
E &=& \sum_{k=1}^{N} \id \otimes \cdots \otimes \id \otimes 
\sigma_{k}^{+} \otimes q^{\sigma^{z}_{k+1}}\otimes \cdots \otimes 
q^{\sigma^{z}_{N}} \,, \non \\
F &=& \sum_{k=1}^{N} q^{-\sigma^{z}_{1}} \otimes \cdots \otimes 
q^{-\sigma^{z}_{k-1}}  \otimes 
\sigma_{k}^{-} \otimes \id \otimes \cdots \otimes 
\id \,, \non \\
K &=& q^{\sigma^{z}_{1}} \otimes \cdots \otimes q^{\sigma^{z}_{N}} \,.
\ee
The transfer matrix has $\qg$ symmetry  \cite{Kulish:1991np}
\be
\left[ t(u) \,, E \right] = \left[ t(u) \,, F \right] = \left[ t(u) 
\,, K \right]  = 0 \,.
\label{qgsymm}
\ee
Moreover, the transfer matrix commutes with $S^{z}$
\be
\left[ t(u) \,, S^{z} \right] = 0 \,, \qquad S^{z}=\tfrac{1}{2}\sum_{k=1}^N \sigma^z_k 
\,,
\ee
and the Bethe states satisfy
\be
S^{z} |v_{1} \ldots v_{M} \rangle = (\tfrac{N}{2}-M)  |v_{1} \ldots 
v_{M} \rangle \,.
\label{Szeig}
\ee 
As reviewed in \cite{Gainutdinov:2015vba}, the Bethe states are 
$\qg$ highest-weight states of spin-$j$ representations $V_{j}$ 
with 
\be
j=\ffrac{N}{2}-M\,,
\label{spinj}
\ee
and dimension
\be
\dim V_{j} = 2j+1=N-2M+1 \,.
\label{dimVj}
\ee

For the root of unity case $q=e^{i \pi/p}$, the generators satisfy 
the additional relations
\be
E^{p}= F^{p} = 0\,, \qquad K^{2p}=1 \,.
\ee
The Lusztig's ``divided powers''~\cite{Chari:1994pz} are defined by (see e.g. \cite{Bushlanov:2009cv})
\be
e=\frac{1}{[p]_{q}!}K^{p}\, E^{p}\,, \qquad 
f=\frac{(-1)^{p}}{[p]_{q}!} F^{p}\,, \qquad 
h=\tfrac{1}{2}\left[e\,, f\right] \,,
\ee
where 
\be
[n]_{q} = \frac{q^{n}-q^{-n}}{q-q^{-1}}\,, \qquad [n]_{q}! = 
\prod_{k=1}^{n}[k]_{q}\,.
\ee
The generators $e, f, h$ obey the usual $sl(2)$ relations
\be
\left[h \,, e \right] = e\,, \qquad \left[h \,, f \right] = - f \,.
\ee
The transfer matrix also has this $sl(2)$ symmetry at roots of unity.

The space of states of the spin chain is given by the $N$-fold tensor
product of spin-1/2 representations $V_{1/2}$.  As already reviewed in
\cite{Gainutdinov:2015vba}, for $q=e^{i \pi/p}$, this vector space
decomposes into a direct sum of tilting $\qg$-modules $T_j$
characterized by spin $j$,
\be
\bigl(V_{\frac{1}{2}}\bigr)^{\otimes N} = \bigoplus_{j=0(1/2)}^{N/2} d^0_j T_j,
\label{decomposition}
\ee
where the sum starts from $j=0$ for even $N$ and $j=1/2$ for odd $N$.
The multiplicities $d^0_j$ of these $T_j$ modules are given
by~\cite{Gainutdinov:2012mr}
\be
d_{j}^{0} = \sum_{n\ge 0} d_{j+ n p}
- \sum_{n \ge t(j)+1} d_{j+ n p -1 -2(j
\, {\rm  mod } \, p)} \,, \qquad (j\ {\rm mod} \ p) \ne p-\frac{1}{2}
\,, \frac{p-1}{2} \,,
\label{dj0}
\ee
where $d_{j}$ is given by 
\be
d_{j} = {N\choose \frac{N}{2}-j} - {N\choose \frac{N}{2}-j-1} \,,
\qquad \qquad d_{j} = 0 \quad {\rm for  }\quad j > \frac{N}{2} \,,
\label{dj}
\ee
and
\be
t(j) = \left\{ \begin{array}{ll}
1 & \ {\rm for  }\  (j\ {\rm mod} \ p) > \frac{p-1}{2} \,, \\
0 & \ {\rm for  }\  (j\ {\rm mod} \ p) < \frac{p-1}{2} \,.
\end{array} \right.
\label{tj}
\ee
If $(j\ {\rm mod} \ p) = p-\frac{1}{2} \,, \frac{p-1}{2}$, then
$d_{j}^{0} =d_{j}$.

The dimensions of the tilting modules are given by \cite{Gainutdinov:2015vba}
\be
\dim T_j =
\begin{cases}
2j+1,&\qquad 2j+1\leq p \quad\text{or}\quad s(j)=0,\\
2(2j+1-s(j)),&\qquad \text{otherwise}\,,
\end{cases}
\label{dimTj}
\ee
where we set~\footnote{$(j\ {\rm mod} \ p)$ is the remainder on division of
$j$ by $p$.}
\be\label{sj}
s(j)=(2j+1)\;{\rm mod}\;p \ .
\ee

\subsection{General structure of the tilting modules}\label{sec:Tj-str}
For our analysis, we need an explicit structure and the $\qg$ action on the tilting modules $T_j$ that appear in  the decomposition (\ref{decomposition}).
 The structure of the tilting
$\qg$-modules was studied in
many works~\cite{Pasquier:1989kd,Martin:1991pk,Chari:1994pz,Read:2007qq,
Gainutdinov:2012mr}.
The tilting $U_q sl(2)$-modules $T_j$
in~\eqref{decomposition}  for $2j+1$ less than $p$ or  divisible by $p$ are
irreducible and isomorphic to the spin-$j$ modules (or $V_j$ in our 
notations).\footnote{The tilting modules $T_j$ with $2j+1<p$ are 
the type-II representations in \cite{Pasquier:1989kd}, while all 
others  are of type~I.}
Otherwise, each  $T_j$ is indecomposable but reducible and contains
$V_j$ as a submodule while  the quotient $T_j/V_j$ is isomorphic to 
$V_{j-s(j)}$, where $s(j)$ is defined in~\eqref{sj}. Both the components $V_{j}$ and  $V_{j-s(j)}$ are further
reducible but indecomposable: $V_j$ has the unique submodule isomorphic
to the head (or irreducible quotient) of the $V_{j-s(j)}$ module, and $V_{j-s(j)}$ has the unique submodule isomorphic
to the head of the $V_{j-p}$ module. We denote
the head of  $V_j$  by $\Irr{j}$.  Then, the sub-quotient structure of $T_j$
in terms of the irreducible modules $\Irr{j}$ can be depicted as
\begin{equation}
\xymatrix@R=22pt@C=1pt{
\mbox{}&\\
&T_j\quad:\
\mbox{}&\\
}
\xymatrix@R=22pt@C=10pt@W=4pt@M=6pt{
&\Irr{j-s(j)}\ar[dl]\ar[dr]&\\
\Irr{j-p}\ar[dr]
&&\mbox{}\;\Irr{j}\;\;\;\;\;\ar[dl]\\
&\Irr{j-s(j)}&
} 
\label{Tj-diag}
\end{equation}
where arrows correspond to
irreversible action of $U_q sl(2)$ generators and we set 
$\Irr{j}=0$ for $j<0$.

To compute dimensions $\dim \Irr{j}$ of the irreducible subquotients in~\eqref{Tj-diag}, we note the relation $\dim \Irr{j} = 2j+1 - \dim \Irr{j-s(j)}$
that follows from the discussion above~\eqref{Tj-diag}.
It is then easy to check the following formula 
for dimensions\footnote{If $s(j)=0$, 
then $2j+1$ is divisible by $p$, so the tilting module is 
irreducible (of dimension $2j+1$ as noted above), and therefore the sub-quotient 
structure is trivial.} by induction in $r\geq0$:
 \be
 \dim \Irr{j} = s(j)(r+1),\qquad \text{where}\quad 2j+1 \equiv rp + s(j).
 \label{dimj}
 \ee
 Note that the 
highest-weight vector in the irreducible module $\Irr{j}$ has $S^{z}=j$.

We shall refer to the four irreducible subquotients in 
(\ref{Tj-diag}), starting from 
the top $\Irr{j-s(j)}$ and going around clockwise, as 
the ``top'' ${\bf T}_{j}$, ``right'' ${\bf R}_{j}$, ``bottom'' ${\bf 
B}_{j}$, and ``left'' ${\bf L}_{j}$ nodes, respectively.
We refer the interested reader to Appendix~\ref{app:proj-mod-base} for the description of the basis and $\qg$-action in $T_j$.

\section{Bethe states for exact complete $p$-strings}\label{sec:pstrings}

For $\eta=i\pi/p$ with integer $p \ge 2 $ (so that $q=e^{\eta}$ is a
root of unity), the Bethe equations (\ref{BAE}) admit exact solutions
consisting of $p$ $\lambda$'s differing by $\eta$, e.g. 
\be
\{ \lam  \,, \lam + \eta\,,  \lam + 2\eta\,, \ldots \,, 
\lam + (p-1)\eta \}
\label{exactcompleterstring}
\ee
where $\lam$ is {\em arbitrary}.  Such solutions have been noticed in
the context of (quasi) periodic chains \cite{Baxter:1972wg,
Fabricius:2000yx, Fabricius:2001yy, Baxter:2001sx, Tarasov:2003xz},
and were called in \cite{Fabricius:2000yx} ``exact complete
$p$-strings.''  Such solutions do not lead to new eigenvalues of the
transfer matrix, and therefore, we do not regard such solutions as
admissible.  Nevertheless, Bethe states corresponding to such
solutions are necessary in order to construct the complete
set of states  when one or more tilting modules are spectrum
degenerate \cite{Gainutdinov:2015vba}.

The Bethe states (\ref{Bethestate}) corresponding to such solutions are naively null, since 
\be
\prod_{r=0}^{p-1}{\cal B}(\lam+ r\eta) = 0 \,,
\label{fusionid}
\ee
as already noticed by Tarasov for the (quasi) periodic chain in~\cite{Tarasov:2003xz, Tarasov:1991mf, Tarasov:1992aw}.\footnote{For 
the closed chain, the corresponding product of $B$ operators is a 
component (top-right corner) of a fused \cite{Kulish:1981gi, Kulish:1981bi} monodromy matrix; 
and, for $\eta=i\pi/p$, 
this fused monodromy matrix becomes block diagonal, and therefore the 
top-right corner becomes zero. (See Proposition 5 parts (i) and (ii) 
in \cite{Tarasov:1991mf}, and Lemmas 1.4 and 1.5 in 
\cite{Tarasov:1992aw}.) The same logic applies to the open 
chain, in view of 
the open-chain generalization \cite{Mezincescu:1991ke} of the fusion procedure.}
We proceed, following \cite{Tarasov:2003xz} (see also \cite{Fabricius:2001yy}), by regularizing the solution and  
taking a suitable limit. Therefore, we now define
\be
\eta_{0} \equiv \frac{i\pi}{p}\,, \qquad \mu \equiv \eta-\eta_{0}\,,
\label{eta0}
\ee
and we consider the limit $\mu \rightarrow 0$. 
Given a usual Bethe state $|\lambda_{1} \ldots \lambda_{M} \rangle$ (\ref{Bethestate}),
we define the operators\footnote{For simplicity, we assume here that the $\lambda_i$'s are fixed and 
do not depend on $\mu$. 
In principle, the analysis presented here could be 
generalized by not making any assumptions about the $\lambda_i$'s at the outset, 
which in fact is the approach taken in \cite{Tarasov:2003xz} for the 
closed chain. However, the result of such an analysis is that, in 
order to obtain an eigenvector of the transfer matrix, the 
$\lambda_i$'s must indeed be solutions of the Bethe equations with $\mu\rightarrow 0$.}
\be\label{Bmu-def}
\BTmunew{\lam} =  
\frac{1}{\mu} \prod_{r=0}^{p-1}{\cal B}(\lam+ r\eta + \mu 
x_{r+1}) 
\ee 
and \footnote{The operator (\ref{BTnew-def}) is well 
defined, since $\prod_{r=0}^{p-1}{\cal B}(\lam+ r\eta + \mu x_{r+1}) 
= O(\mu)$ for $\mu\rightarrow 0$, as follows from (\ref{fusionid}) and the 
fact ${\cal B}(u) = {\cal B}(u)\Big\vert_{\mu=0} + O(\mu)$,
and non-zero in general, as follows from examples we studied.
}
\be\label{BTnew-def}
\BTnew{\lam}=  
\lim_{\mu \rightarrow 0} 
\BTmunew{\lam} \,,
\ee
as well as the corresponding new states
\be
\vecTmu{\lam}{\lambda_{1} \ldots \lambda_{M}} = 
\BTmunew{\lam}\, 
|\lambda_{1} \ldots \lambda_{M} \rangle 
\ee
and
\be
\vecT{\lam}{\lambda_{1} \ldots \lambda_{M}} &=&
\BTnew{\lam}\, 
|\lambda_{1} \ldots \lambda_{M} \rangle  
\non \\
&=& \lim_{\mu \rightarrow 0}
\frac{1}{\mu} \prod_{r=0}^{p-1}{\cal B}(\lam+ r\eta + \mu 
x_{r+1})\, |\lambda_{1} \ldots \lambda_{M} \rangle \,,
\label{Tarasovstate}
\ee
where the transfer matrix $t(u)$ and the ${\cal B}$ operators 
(including those used in the construction of the Bethe state $|\lambda_{1} \ldots \lambda_{M} \rangle$ of course)
should be understood to be constructed with generic anisotropy $\eta$ instead of $\eta_{0}$, and
$x_{1}, \ldots, x_{p}$ are still to be determined.
To this end, we obtain the 
off-shell relation for this state (c.f. (\ref{offshell})) 
\be
t(u) \vecTmu{\lam}{\lambda_{1} \ldots \lambda_{M}}  &=& X(u) 
\vecTmu{\lam}{\lambda_{1} \ldots \lambda_{M}}  \non\\
&+& \frac{1}{\mu}\sum_{m=1}^{M} Y_{m}\, B(u) 
\prod_{r=0}^{p-1}{\cal B}(\lam+ r\eta + \mu x_{r+1})\,
\prod_{\scriptstyle{k \ne m}\atop \scriptstyle{k=1}}^M {\cal 
B}(\lambda_{k}) \vac \non\\
&+& \frac{1}{\mu}\sum_{r=0}^{p-1} Z_{r}\, B(u) 
\prod_{\scriptstyle{s \ne r}\atop \scriptstyle{s=0}}^{p-1}{\cal 
B}(\lam+ s\eta + \mu x_{s+1})\,
\prod_{k=1}^M {\cal B}(\lambda_{k}) \vac\,,
\label{offshell2}
\ee
and  the limit $\mu \rightarrow 0$ remains to be performed. 
Evidently, there are now two kinds of ``unwanted'' terms.

It is easy to see from (\ref{Lambda}) that  $X(u)$, which appears 
in the first line of (\ref{offshell2}), is given by
\be
X(u) =
\frac{\sh(2u+2\eta)}{\sh(2u+\eta)}\sh^{2N}(u+\eta)\frac{Q(u-\eta)}{Q(u)}{\cal E}^{-}(u)  +
\frac{\sh(2u)}{\sh(2u+\eta)}\sh^{2N}(u)\frac{Q(u+\eta)}{Q(u)}{\cal E}^{+}(u)  \,,
\ee
where $Q(u)$ is given by (\ref{Q}), 
and the ${\cal E}^{\pm}(u)$ are defined by
\be
{\cal E}^{\pm}(u)  =\frac{{\cal Q}(u\pm\eta)}{{\cal Q}(u)} \,,
\ee 
where
\be
{\cal Q}(u) &=&  \prod_{r=0}^{p-1} 
\sh(u-\lam-(r-\tfrac{1}{2})\eta-\mu 
x_{r+1})
\sh(u+\lam+(r+\tfrac{1}{2})\eta+\mu 
x_{r+1}) \non \\
&=& \prod_{r=0}^{p-1} 
\sh(u-\lam-(r-\tfrac{1}{2})\eta)
\sh(u+\lam+(r+\tfrac{1}{2})\eta) \; +\;  O(\mu)\,.
\label{calQ}
\ee
In the second line of (\ref{calQ}), we keep explicitly only the first term in the expansion around $\mu=0$ and neglect contributions that  
vanish when $\mu$ vanishes.
We see that ${\cal E}^{\pm}(u)  \rightarrow 1$
in the limit $\mu \rightarrow 0$, and therefore $X(u) \rightarrow 
\Lambda(u)$. 

Similarly, from  (\ref{Lambdam}) we find that $Y_{m}$, which appears 
in the second line of (\ref{offshell2}), is given~by 
\be
Y_{m} &=& \ef(u,\lambda_{m}-\tfrac{\eta}{2})
\Bigg[\sh^{2N}(\lambda_{m}+\tfrac{\eta}{2})\, {\cal 
E}^{-}(\lambda_{m}-\tfrac{\eta}{2})
\prod_{\scriptstyle{k \ne m}\atop \scriptstyle{k=1}}^M
\frac{\sh(\lambda_{m}-\lambda_{k}-\eta) 
\sh(\lambda_{m}+\lambda_{k}-\eta)}
{\sh(\lambda_{m}-\lambda_{k}) \sh(\lambda_{m}+\lambda_{k})} 
\non \\
&& -\sh^{2N}(\lambda_{m}-\tfrac{\eta}{2})\, {\cal 
E}^{+}(\lambda_{m}-\tfrac{\eta}{2})
\prod_{\scriptstyle{k \ne m}\atop \scriptstyle{k=1}}^M
\frac{\sh(\lambda_{m}-\lambda_{k}+\eta) 
\sh(\lambda_{m}+\lambda_{k}+\eta)}
{\sh(\lambda_{m}-\lambda_{k}) 
\sh(\lambda_{m}+\lambda_{k})}  \Bigg] \,,
\label{Lambdam2}
\ee
and therefore $Y_{m} \rightarrow \Lambda^{\lambda_{m}}$ as $\mu
\rightarrow 0$.  Hence, the ``unwanted'' terms of the first kind in
(\ref{offshell2}) vanish provided that $\lambda_{1}\,, \ldots \,,
\lambda_{M}$ satisfy the usual Bethe equations (\ref{BAE}) at 
$\eta=\eta_0$.  (The factor $1/\mu$ in the second line 
of~\eqref{offshell2} is canceled by the contribution 
from $\prod_{r=0}^{p-1}{\cal B}(\lam+ r\eta + \mu x_{r+1})$ which 
vanishes as fast as $O(\mu)$ for $\mu\rightarrow 0$, as we noticed 
above.)

Finally, again from  (\ref{Lambdam}) we find that $Z_{r}$, which appears 
in the third line of (\ref{offshell2}), is given by
\be
Z_{r} &=& \ef(u,\lam+(r-\tfrac{1}{2})\eta)
\Bigg[\sh^{2N}(\lam+(r+\tfrac{1}{2})\eta) 
\frac{Q(\lam+(r-\tfrac{3}{2}) 
\eta)}{Q(\lam+(r-\tfrac{1}{2})\eta)} {\cal Z}^{-}_{r} \non \\
&& -\sh^{2N}(\lam+(r-\tfrac{1}{2})\eta)
\frac{Q(\lam+(r+\tfrac{1}{2}) \eta)}{Q(\lam+(r-\tfrac{1}{2})\eta)} {\cal Z}^{+}_{r}
\Bigg] \,,
\label{Lambdam3}
\ee
where 
\be
{\cal Z}^{-}_{r} &=& \prod_{\scriptstyle{s \ne r}\atop \scriptstyle{s=0}}^{p-1}
\frac{\sh((r-s-1)\eta + \mu(x_{r+1}-x_{s+1}))\, 
\sh(2\lam+(r+s-1)\eta)}
{\sh((r-s)\eta)\, 
\sh(2\lam+(r+s)\eta)}\,, \non \\
{\cal Z}^{+}_{r} &=& \prod_{\scriptstyle{s \ne r}\atop \scriptstyle{s=0}}^{p-1}
\frac{\sh((r-s+1)\eta + \mu(x_{r+1}-x_{s+1}))\, 
\sh(2\lam+(r+s+1)\eta)}
{\sh((r-s)\eta)\, 
\sh(2\lam+(r+s)\eta)}\,,
\ee
and we have again neglected contributions that vanish when $\mu$ 
vanishes.
We find
\be
\lim_{\mu\rightarrow 0} \frac{{\cal Z}^{-}_{r}}{\mu} = -(x_{r+1}-x_{r}) 
\frac{\sh(2\lam+2r\eta_{0})}{\sh \eta_{0}\, 
\sh(2\lam+(2r-1)\eta_{0})} \,, \qquad r \ne 0 \,,
\ee 
while for $r=0$ the above result continues to hold except with $x_{0}=x_{p} +p$.
Similarly,
\be
\lim_{\mu\rightarrow 0} \frac{{\cal Z}^{+}_{r}}{\mu} = -(x_{r+2}-x_{r+1}) 
\frac{\sh(2\lam+2r\eta_{0})}{\sh \eta_{0}\, 
\sh(2\lam+(2r+1)\eta_{0})} \,, \qquad r \ne p-1 \,,
\ee 
while for $r=p-1$ the above result continues to hold except with 
$x_{p+1}=x_{1} -p$. We conclude that the ``unwanted'' terms of the 
second kind in (\ref{offshell2}) vanish provided that $x_{1}, 
\ldots, x_{p}$ satisfy
\be
\frac{x_{r+1}-x_{r}}{x_{r+2}-x_{r+1}} =
\left(\frac{\sh(\lam+(r-\frac{1}{2})\eta_{0})}{\sh(\lam+(r+\frac{1}{2})\eta_{0})}\right)^{2N}
\frac{\sh(2\lam+(2r-1)\eta_{0})}{\sh(2\lam+(2r+1)\eta_{0})}
\frac{Q(\lam+(r+\frac{1}{2}) 
\eta_{0})}{Q(\lam+(r-\frac{3}{2}) \eta_{0})}
\label{ratio1}
\ee 
for $r=0, 1, \ldots, p-1$, where 
\be
x_{0}=x_{p} +p\,, \qquad x_{p+1}=x_{1} -p\,,
\label{BC}
\ee
 and $Q(u)$ in (\ref{Q}) is to be evaluated with 
$\eta=\eta_{0}$.

In order to solve (\ref{ratio1}) for $x_{1}, \ldots, x_{p}$, we
now make (along the lines of \cite{Tarasov:2003xz}) the following ansatz
\be
x_{r} = 1 - r - \frac{G(\lam+r\eta_{0})}{F(\lam)} \,, 
\qquad r = 0, \ldots, p+1\,,
\label{xr}
\ee
where $F(u)$ and $G(u)$ are functions with periodicities $\eta_{0}$ and $i \pi$, 
respectively,
\be
F(u+ \eta_{0}) = F(u) \,, \qquad G(u+ i\pi) = G(u)\,.
\label{periodicity}
\ee
Then the boundary conditions (\ref{BC}) are satisfied, and 
\be
\frac{x_{r+1}-x_{r}}{x_{r+2}-x_{r+1}} = 
\frac{H(\lam+r\eta_{0})}{H(\lam+(r+1)\eta_{0})} \,,
\label{ratio2}
\ee
where
\be
H(u) = G(u+r\eta_{0}) - G(u) + F(u) \,.
\label{Gcond}
\ee
The conditions (\ref{periodicity}) and (\ref{Gcond}) can be satisfied by 
setting
\be
F(u) = \frac{1}{p} \sum_{k=0}^{p-1}H(u+ k\eta_{0})\,, \qquad
G(u) = \frac{1}{p} \sum_{k=1}^{p-1}k H(u+ k\eta_{0}) \,.
\label{FuGu}
\ee
Comparing (\ref{ratio1}) and  (\ref{ratio2}), we see that $H(u)$ must 
obey the functional relation
\be
\frac{H(u)}{H(u+\eta_{0})} = \left(\frac{\sh(u-\frac{\eta_{0}}{2})}{\sh(u+\frac{\eta_{0}}{2})}\right)^{2N}
\frac{\sh(2u-\eta_{0})}{\sh(2u+\eta_{0})}\frac{Q(u+\frac{\eta_{0}}{2})}{Q(u-\frac{3\eta_{0}}{2})} \,,
\label{Heqn}
\ee
which is satisfied by\footnote{One can multiply this solution by
any function with periodicity $\eta_{0}$, and it will still 
be a solution of (\ref{Heqn}), though it will not change 
the values of $x_r$'s. We are not aware of any other solutions of the 
functional equation, and expect that this one will be enough to construct the complete basis of eigenstates.}
\be
H(u) = \frac{\sh^{2N}(u-\frac{\eta_{0}}{2}) \sh(2u-\eta_{0})}{Q(u-\frac{\eta_{0}}{2})\,  Q(u-\frac{3\eta_{0}}{2})} \,,
\label{Hu}
\ee

We have therefore proved the following proposition.
\begin{Prop}\label{main-prop-1}
If $|\lambda_{1} \ldots \lambda_{M} \rangle$ is an 
eigenstate of the transfer matrix $t(u)$ with eigenvalue $\Lambda(u)$,
then for any $v\in\mathbb{C}$ the corresponding state $\|\lam; \lambda_{1} \ldots \lambda_{M} 
\rangle\!\rangle$ constructed in  (\ref{Tarasovstate}) using an exact 
complete $p$-string, where $x_{r}$ are given by (\ref{xr}), 
(\ref{FuGu}) and (\ref{Hu}) using~(\ref{Q}), is also an 
eigenstate of the transfer matrix with the same eigenvalue $\Lambda(u)$.
\end{Prop}

By this proposition we see that the operator $\BTnew{\lam}$
in~(\ref{BTnew-def}) maps the specific eigenstate $|\lambda_{1} \ldots
\lambda_{M} \rangle$ defined in (\ref{Bethestate}) to another
eigenstate of $t(u)$. But acting with $\BTnew{\lam}$ on 
other Bethe states does not give in general eigenstates, or saying
differently the operator $\BTnew{\lam}$ does not in general commute
with $t(u)$, as its definition involves Bethe roots $ \lambda_{i}$ via
the function $Q(u)$.

\begin{Rem}\label{rem:Hp2}
For the particular case $p=2$, the $Q(u)$ function obeys 
$Q(u+2\eta_{0}) = Q(u)$, and therefore the ratio of $Q(u)$ functions 
in (\ref{Heqn}) equals $1$, which implies that $H(u)$ can be chosen  
independently of $\{\lambda_{i} \}$, e.g. $H(u) = 
\sh^{2N}(u-\frac{\eta_{0}}{2}) \sh(2u-\eta_{0})$; and therefore 
$\{x_{r} \}$ and thus $\BTnew{v}$ are independent of $\{\lambda_{i} \}$. 
This suggests that $\BTnew{v}$ might be a symmetry of $t(u)$ as it maps any Bethe state to another eigenstate of the {\sl same} eigenvalue. 
We have verified numerically for $p=2$ and up to $N=6$ that
$\BTnew{\lam}$ indeed commutes with $t(u)$ for any complex numbers $u$ and $v$.
\end{Rem}

Several examples of the construction in Proposition 3.1
with $p=2$ can be found in Sec.~\ref{sec:p2}, see e.g. Secs.~\ref{sec:p2N5}, \ref{sec:p2N6}, and~\ref{sec:p2N7}.
For $p>2$, the first appearance of an exact complete 
$p$-string is for the case $p=3, N=8, M=0$, see Section D.6 in 
\cite{Gainutdinov:2015vba}.
We have constructed the vector 
$\vecT{\lam}{-}$ (\ref{Tarasovstate}) numerically for this case, with a
generic value for $v$, and we have verified that it is an eigenvector of
the Hamiltonian with the same eigenvalue as the reference state
(namely, $E=3.5$), yet it is linearly independent from the reference
state.  Moreover, it is a highest-weight vector with spin $j=1$, exactly as
required for the right node of the tilting module $T_1$ 
(recall the structure in~\eqref{Tj-diag} and its description above), which is
spectrum-degenerate with the tilting module $T_4$ 
containing the reference state.

\begin{Rem} The generalization to the case of more than one exact complete
$p$-string is straightforward: a vector with $m$ such $p$-strings is given by
\be
\vecT{\lam_{1}, \ldots, \lam_{m}}{\lambda_{1} \ldots \lambda_{M}} = 
\prod_{i=1}^{m} \BTnew{\lam_{i}}|\lambda_{1} \ldots \lambda_{M} \rangle \,,
\label{Tarasovstategen}
\ee
where $\BTnew{\lam_{i}}$ is constructed as in~\eqref{BTnew-def} and with $\{ x_{i,r} \}$ given by
\be
x_{i,r} = 1 - r - \frac{G(\lam_{i}+r\eta_{0})}{F(\lam)} \,, 
\qquad r = 0, \ldots, p+1\,, \qquad i = 1, \dots\,, m \,,
\label{xrgen}
\ee
 with the same boundary conditions on $x_{i,r}$.
We note that the $S^z$-eigenvalue of~\eqref{Tarasovstategen} is
$\frac{N}{2}-M + mp$ and thus the operators $\prod_{i=1}^{m}
\BTnew{\lam_{i}}$ describe $t(u)$ degeneracies between
$S^z$-eigenspaces that differ by a multiple of $p$.  We stress that
these degeneracies are extra to the degeneracies corresponding to the
action by the divided powers of $U_q sl(2)$ that also change $S^z$ by
$\pm p$.  We discuss below this new type of degeneracies.  An example
with two exact complete $p$-strings (i.e., $m=2$, with $p=2$) is given
in Sec.~\ref{sec:p2N9}.
\end{Rem}

\section{Generalized Bethe states}\label{sec:generalized}

The usual Bethe states (\ref{Bethestate}) are, by construction,
ordinary eigenvectors of the transfer matrix $t(u)$.  In order to
construct generalized eigenvectors (which, as noted in the
Introduction, appear at roots of unity), something different must be
done.  We recall that {\em generalized} eigenvectors $|v\rangle$ are
defined as\footnote{The power in (\ref{geneig}) is $2$ because there
are Jordan cells of maximum rank $2$, and here $|v\rangle$ and
$|v'\rangle$ belong to a Jordan cell of rank $2$.}
\be
\bigl(t(u)-\Lambda(u)\mathbf{1} \bigr)^2 |v\rangle=0 \,,
\label{geneig}
\ee
or equivalently 
\be
t(u)\, |v\rangle=\Lambda(u)\, |v\rangle + |v'\rangle  \quad \mbox{    and   
} \quad t(u)\, |v'\rangle=\Lambda(u)\, |v'\rangle \,.
\label{geneig-2}
\ee
We note that a generalized eigenvector, as $|v\rangle$ in~\eqref{geneig-2}, is defined only 
up to the transformation 
\begin{equation}\label{gen-transf}
|v\rangle \to \alpha  |v\rangle  + \beta  |v'\rangle,\qquad \text{for}\quad \alpha,\beta\in\mathbb{C}.
\end{equation}

Generalized eigenvectors appear only in (direct sums of) the tilting $\qg$-modules $T_j$ with $s(j)$
non-zero, i.e. in the cases where
$T_j$ are indecomposable but reducible, and thus are described by the
diagram in~\eqref{Tj-diag}.  This fact is borne out by the
explicit examples in our previous paper~\cite{Gainutdinov:2015vba},  
see also~\cite{Dubail:2010zz,Vasseur:2011fi} and the proof for $p=2$ in~\cite{Gainutdinov:2012qy}.
As we will see further from an explicit
construction in this section, it is only the states in the head of
$T_j$ -- the
top sub-quotient $\Irr{j-s(j)}$ in~\eqref{Tj-diag} -- on which the Hamiltonian~\eqref{Hamiltonian} 
is non-diagonalizable.
For the case $p=2$, it was already shown in~\cite{Gainutdinov:2012qy} using certain free fermion operators.

\subsection{Introduction and overview}

An important
clue to a Bethe ansatz construction of the generalized eigenvectors can already be learned by considering the simplest case, namely 
a chain with two sites ($N=2$). Indeed, for this case and for generic values of $q$, the 
eigenvectors of the Hamiltonian (\ref{Hamiltonian}) are given 
by 
\be
|{\bf v}_{1}\rangle &=& \vac = |\uparrow\uparrow \rangle = (1,0,0,0)^{T} \,, \non \\
|{\bf v}_{2}\rangle &=& F \vac = q^{-1} |\uparrow\downarrow 
\rangle + |\downarrow\uparrow\rangle = 
(0,q^{-1},1,0)^{T} \,, \non \\
|{\bf v}_{3}\rangle &=& \frac{1}{[2]_{q}}F^{2} \vac = |\downarrow\downarrow \rangle= 
(0,0,0,1)^{T} \,, \non \\
|{\bf v}_{4}\rangle &=& -q |\uparrow\downarrow 
\rangle + |\downarrow\uparrow\rangle = 
(0,-q,1,0)^{T} \,.
\label{eigenvecsbrute}
\ee
The first three vectors, which form a spin-1 representation of $\qg$, 
have the same energy eigenvalue $E_{1}=\tfrac{1}{2}[2]_{q}$, while 
the fourth vector 
(a spin-0 representation) has the energy eigenvalue 
$E_{0}=-\tfrac{3}{2}[2]_{q}$. 
For $p=2$ (i.e., $q=e^{i \pi/2} = i$), the vectors $|{\bf 
v}_{2}\rangle$ and $|{\bf v}_{4}\rangle$ evidently coincide (and 
$E_{1}=E_{0}=0$), 
signaling that the 
Hamiltonian is no longer diagonalizable. A generalized eigenvector of 
the Hamiltonian with generalized eigenvalue $0$
can be constructed from the $q \rightarrow i$ limit of  
an appropriate linear combination of these two 
vectors, e.g.,
\be
|{\bf w}\rangle = \lim_{q \rightarrow i} \frac{1}{[2]_{q}}\bigl( |{\bf v}_{4}\rangle - |{\bf 
v}_{2}\rangle\bigr) =-(0,1,0,0)^{T} \,.
\label{genvecp2ex}
\ee 

Let us now consider the corresponding Bethe ansatz description. For 
generic $q$, the vector $|{\bf v}_{4}\rangle$ is given by
\be
|{\bf v}_{4}\rangle = a(\eta)\, {\cal B}(\nu) \vac \,, \qquad 
\nu = \tfrac{1}{2} \log\Big[- \frac{\sh(\frac{\eta}{2}+\frac{i 
\pi}{4})}{\sh(\frac{\eta}{2}-\frac{i\pi}{4})} \Big] \,,
\label{BAp2}
\ee
where $a$ depends on $\eta$ such that $a(i\pi/2)=0$.
As $q$ approaches $i$ (i.e.,  
$\eta$ approaches $\tfrac{i \pi}{2}$), the Bethe root 
$\nu$ in (\ref{BAp2}) 
goes to infinity. Indeed, setting
$\eta = \tfrac{i \pi}{2} - i \omega^{2}$, we find that 
\be
\nu = -\log \omega + \tfrac{1}{2} \log 2 + O(\omega^{4})
\label{lambdaforsmallomega}
\ee
for $\omega$ near 0.  Expanding the Bethe vector in a series 
about $\omega=0$, we observe that
\be\label{Bnu-exp}
{\cal B}(\nu) \vac = \tfrac{i}{2}\, \omega^{-4} \left( F \vac\right)\big\vert_{\omega=0} +  
O(\omega^{-2}) \,.
\ee
We therefore can subtract $\frac{i}{2}\omega^{-2} F \vac$ from $ \omega^{2}{\cal B}(\nu)\vac$ to get the final result
\be\label{gen-v}
|{\bf v}\rangle &\equiv& \lim_{\omega\rightarrow 0+} \left[ \omega^{2}{\cal 
B}(\nu) \vac
-  \tfrac{i}{2}\, \omega^{-2} F \vac \right] \non\\
&=& (0,0,-1,0)^{T} = 
i|{\bf w}\rangle - |{\bf v}_{2}\rangle\Big\vert_{q=i} \,,
\ee 
which is a {\sl generalized} eigenvector of the Hamiltonian.
Note the similarity of the constructions in (\ref{genvecp2ex}) and 
(\ref{gen-v}): both involve subtracting from a (generically) highest-weight state a contribution proportional 
to $|{\bf v}_2\rangle = F \vac$ and taking the $q\rightarrow i$ 
limit. 
The generalized eigenvector $|{\bf v}\rangle$ is evidently a linear 
combination of the generalized eigenvector $|{\bf w}\rangle$ in (\ref{genvecp2ex}) and the eigenvector $|{\bf 
v}_{2}\rangle$ in (\ref{eigenvecsbrute}), recall that the generalized eigenvector is defined  up to the  transformation~\eqref{gen-transf}.

A construction of generalized Bethe states similar to~\eqref{gen-v} is possible for general values of $N$ and $p$.
We observe from numerical studies 
given in App.~\ref{sec:numerics} that,  as the anisotropy parameter $\eta$ approaches
$\eta_{0}= i \pi/p$ with integer $p \ge 2$, the Bethe
roots corresponding to a generalized eigenvalue contain a string of
length $p' \in \{ 1, 2, \ldots, p-1\}$, whose center (real part) approaches
infinity. 
In more detail, such a string is a set of $p'$ roots differing by $i 
\pi/p'$, e.g. 
\be
\nu_{k}^{\infty} = \nu_{0} + \frac{i \pi}{2p'}(p'-(2k-1)) \,,  \qquad
k=1, \ldots, p' \,,
\label{pprimestring1}
\ee
with $\nu_{0} \rightarrow \infty$.  As we shall 
see below, the value of $p'$ is related to the spin $j$ of the tilting 
module $T_{j}$ (the one containing the corresponding generalized eigenvector) by the simple formula
\be
p' = s(j) \,,
\label{p'formula}
\ee
where $s(j) \in \{ 1, 2, \ldots, p-1\}$ is defined in (\ref{sj}).
For $p=2$, the only 
possibility is $p'=1$, i.e. an infinite real root, as already 
discussed. For $p=3$, the only
possibilities are $p'=1$ and $p'=2$, where the latter consists of the 
pair of roots $\nu_{0} \pm i \pi/4$ with  $\nu_{0} \rightarrow \infty$.
For $p=4$, we can have $p'=1,2,3$;  the $p'=3$ 
case consists of a triplet of roots $\nu_{0}\,, \nu_{0} \pm i \pi/3$ 
with $\nu_{0} \rightarrow \infty$, etc. 
The corresponding Bethe state 
has Bethe roots $\{\nu_{k}^{\infty}\}$ 
tending to infinity in the limit,  and requires a certain subtraction to get a finite vector.
In a nutshell, our construction of  generalized eigenvectors in a tilting module~$T_j$ starts with 
the spin-$j$ highest-weight state that lives in
the right node denoted by
$\Irr{j}$ in  the diagram~\eqref{Tj-diag}.  This state can be constructed using the 
ordinary ABA approach as in~(\ref{Bethestate}). Then, a generalized eigenstate
living in the top node $\Irr{j-s(j)}$ is constructed by applying a certain 
$p'$-string of ${\cal B}(\nu_k)$ operators (with $\nu_k$ as in~\eqref{pprimestring1} but finite $\nu_0$)
on the usual Bethe state in $\Irr{j}$ at generic value of $\eta$, subtracting the image of $F^{p'}$ on the  
 spin-$j$ highest-weight state and taking the limit $\eta\to \eta_0$. 
We give below  details
of the construction with our final claim in Prop.~\ref{prop:gen-vect}, 
while our representation-theoretic interpretation is given in Sec.~\ref{sec:rep-th-descr}.

\subsection{General ABA construction of generalized 
eigenstates}\label{genABAconst}
 
With these observations in mind, let 
\be
|\vec\lambda\rangle \equiv |\lambda_{1} \ldots \lambda_{M} 
\rangle = \prod_{k=1}^{M}{\cal 
B}(\lambda_{k}) \vac
\label{v0onshell}
\ee
denote an on-shell Bethe vector, i.e., an 
ordinary eigenvector of the transfer matrix
\be
t(u)  |\vec\lambda\rangle = \Lambda(u) |\vec\lambda\rangle \,,
\label{ordinary}
\ee 
where the eigenvalue $\Lambda(u)$ is given by (\ref{Lambda}). 
This state is an $\qg$ highest-weight state with spin 
$j=N/2-M$, see (\ref{spinj}). 
Under the already-mentioned assumption that the top node 
$\Irr{j-s(j)}$ of $T_j$ contains generalized eigenstates,
let us construct a generalized eigenvector $\vecG{p'}{\vec\lambda} \equiv \vecG{p'}{ \lambda_{1} \ldots \lambda_{M} }$
whose generalized eigenvalue is also 
$\Lambda(u)$, where $p'=s(j)$. To this end, we now set
\be
\eta=\eta_{0} - i \omega^{2p'}\,, \qquad \eta_{0} = \frac{i\pi}{p}\,,
\label{eta02}
\ee
and look for  a generalized eigenvector as the limit
\be
\vecG{p'}{\vec\lambda} = \lim_{\omega\rightarrow 0+} \vecGomega{p'}{\vec\lambda} \,,
\label{genvec}
\ee
where
\be
\vecGomega{p'}{\vec\lambda}  = \alpha |\vec\nu, \vec\lambda_{\alpha}\rangle + \beta 
F^{p'} |\vec\lambda_{\beta}\rangle \,,
\label{vecGommega}
\ee
with 
\be
|\vec\nu, \vec\lambda_{\alpha}\rangle = \prod_{j=1}^{p'}{\cal B}(\nu_{j})\, 
\prod_{k=1}^{M}{\cal B}(\lambda_{\alpha, k}) \vac 
\,, \qquad 
|\vec\lambda_{\beta}\rangle   =  \prod_{k=1}^{M}{\cal 
B}(\lambda_{\beta, k}) \vac  \,.
\label{genvecmore}
\ee  
Note that the subscripts $\alpha$ and $\beta$ on $\lambda_{\alpha, 
k}$ and $\lambda_{\beta, k}$ are simply labels (i.e., not indices) 
that serve to distinguish $\lambda_{\alpha, k}$ from $\lambda_{\beta, 
k}$ and from $\lambda_{k}$. Note that  $\lambda_k$ is the Bethe solution precisely 
at the root of unity, when $\omega=0$, while $\lambda_{\alpha, 
k}$ and $\lambda_{\beta, k}$ are a priori different functions of $\omega$.
And we assume that,
as $\omega\rightarrow 0+$, 
\be
\nu_{j} &\rightarrow& \nu_{j}^{\infty} \,,\non \\ 
\lambda_{\alpha, k} &\rightarrow& \lambda_{k}\,,\non \\ 
\lambda_{\beta, k} &\rightarrow& \lambda_{k}\,,
\label{limit-ass}
\ee 
where $\nu_{j}^{\infty}$ is given in (\ref{pprimestring1}) with $\nu_0$ diverging as
$\nu_{0} = - \log \omega$. 
However, the $\{ \nu_{j} \}$, $\{ \lambda_{\alpha, k} \}$, $\{ \lambda_{\beta, k} \}$ 
as well as the coefficients $\alpha$ and $\beta$ (actually certain powers of $\omega$) are still to be determined. 
The ${\cal B}$ operators and the transfer matrix $t(u)$ should again 
(as in Section \ref{sec:pstrings})
be understood to be constructed with anisotropy $\eta$ instead of
$\eta_{0}$.
Moreover, $F$ is the $\qg$ generator (see section \ref{sec:qg}) and 
as an operator it also depends 
on $q=e^{\eta}$.

We shall see that the 
state $\vecG{p'}{\vec\lambda}$  
or the limit~\eqref{genvec} is well defined and has the same transfer-matrix (generalized) eigenvalue as 
$|\vec\lambda\rangle$ in (\ref{v0onshell}), and both states
belong to the same tilting module $T_j$, see  Rem.~\ref{rem-Jordan-cell} below.
As  in the usual ABA construction, the state $\vecG{p'}{\vec\lambda}$ 
in our construction also has the maximum value of 
$S^{z}$ in the irreducible subquotient to which it belongs, namely, the top node $\Irr{j-s(j)}$.
We know from (\ref{Szeig}) and (\ref{genvecmore}) that this state has 
$S^{z} = N/2 - M - p' = j - p'$. 
On the other hand, we know from the general structure of tilting modules (\ref{Tj-diag}) that $\vecG{p'}{\vec\lambda}$
has $S^{z}= j - s(j)$. It follows that $p'=s(j)$, as already noted in (\ref{p'formula}).

Next, we observe that for $\omega\rightarrow 0$, the vector $|\vec\nu, \vec\lambda_{\alpha}\rangle$ has the power series expansion:
\be
|\vec\nu, \vec\lambda_{\alpha}\rangle  = c\, \omega^{-2p' N} \left( F^{p'} 
|\vec\lambda\rangle \right)\Big\vert_{\omega=0} + O(\omega^{-2p' 
(N-1)}) \,,
\label{cformula}
\ee
where $c$ is some numerical factor. For $p'=1$, this 
follows from the fact that $B(u) \vac \sim e^{2 N u} F \vac + O(e^{2 
(N-1) u})$ for $u \rightarrow \infty$;
hence, for $u \sim -\log \omega$,  $B(u) \vac \sim 
\omega^{-2N} F \vac +  O(\omega^{-2(N-1)})$. For $p'>1$, the result 
(\ref{cformula}) is a conjecture, which we have checked in many 
examples, see e.g. Secs. \ref{sec:genp3}, \ref{sec:genp4}.
It follows that
\be\label{gen-vec-fin}
\omega^{2p'(N-1)}|\vec\nu, \vec\lambda_{\alpha}\rangle 
- c\, \omega^{-2p'}  F^{p'} |\vec\lambda_{\beta}\rangle = O(\omega^{0})
\ee
for $\omega\rightarrow 0$. We therefore set
\be 
\alpha = \omega^{2 p' (N-1)}\,, \qquad \beta = -c\,  \omega^{-2 p'} \,,
\label{alphabeta}
\ee 
which makes $\vecGomega{p'}{\vec\lambda}$ (\ref{vecGommega}) finite 
for $\omega\rightarrow 0$.

According to the off-shell relation (\ref{offshell}), the transfer
matrix $t(u)$ has the following action on the off-shell Bethe vector $|\vec\nu,
\vec\lambda_{\alpha}\rangle$: 
\be
t(u) |\vec\nu, \vec\lambda_{\alpha}\rangle =
\Lambda_{\alpha}(u) |\vec\nu, \vec\lambda_{\alpha}\rangle +
\sum_{i}\Lambda^{\nu_i}(u)\, B(u) | \hat\nu_{i},  \vec\lambda_{\alpha}\rangle
+ \sum_{i}\Lambda^{\lambda_{\alpha, i}}(u)\, B(u) | \vec\nu,  
\hat\lambda_{\alpha,i}\rangle \,,
\label{offshellalpha}
\ee
where  
a hat over a symbol means that it should be omitted,
i.e.
\be
| \hat\nu_{i},  \vec\lambda_{\alpha}\rangle = 
\prod_{\scriptstyle{j \ne i}\atop \scriptstyle{j=1}}^{p'} {\cal 
B}(\nu_{j}) \prod_{k=1}^{M} {\cal B}(\lambda_{\alpha, 
k}) \vac \,, \qquad
| \vec\nu,  \hat\lambda_{\alpha,i}\rangle =
\prod_{j=1}^{p'} {\cal B}(\nu_{j})
\prod_{\scriptstyle{k \ne i}\atop \scriptstyle{k=1}}^{M}  {\cal B}(\lambda_{\alpha, 
k}) \vac 
\,,
\ee 
and
\be
\Lambda_{\alpha}(u) &=& 
\frac{\sh(2u+2\eta)}{\sh(2u+\eta)}\sh^{2N}(u+\eta)\frac{Q_{\alpha}(u-\eta) 
Q_{\nu}(u-\eta)}{Q_{\alpha}(u) Q_{\nu}(u)} \non \\
&&+ \frac{\sh(2u)}{\sh(2u+\eta)}\sh^{2N}(u)\frac{Q_{\alpha}(u+\eta) Q_{\nu}(u+\eta)}{Q_{\alpha}(u) Q_{\nu}(u)} \,,
\label{Lambdapp}
\ee
and $Q_{\nu}(u)$ and $Q_{\alpha}(u)$ are defined as
\be
Q_{\nu}(u) &=& 
\prod_{j=1}^{p'}\sh\left(u-\nu_{j}+\tfrac{\eta}{2}\right)\sh\left(u+\nu_{j}+\tfrac{\eta}{2}\right) \,, \non \\
Q_{\alpha}(u) &=& 
\prod_{k=1}^{M}\sh\left(u-\lambda_{\alpha, 
k}+\tfrac{\eta}{2}\right)\sh\left(u+\lambda_{\alpha, 
k}+\tfrac{\eta}{2}\right) \,.
\ee
Moreover,  according to~\eqref{Lambdam}, we have
\be
\Lambda^{\nu_i}(u) &=& \ef(u,\nu_{i}-\tfrac{\eta}{2}) 
\Bigg[\sh^{2N}(\nu_{i}+\tfrac{\eta}{2})
\frac{Q_{\alpha}(\nu_{i}-\tfrac{3\eta}{2})}{Q_{\alpha}(\nu_{i}-\tfrac{\eta}{2})}
\prod_{\scriptstyle{j \ne i}\atop \scriptstyle{j=1}}^{p'}
\frac{\sh(\nu_{i}-\nu_{j}-\eta) 
\sh(\nu_{i}+\nu_{j}-\eta)}
{\sh(\nu_{i}-\nu_{j}) \sh(\nu_{i}+\nu_{j})} 
\non \\
&& -\sh^{2N}(\nu_{i}-\tfrac{\eta}{2})
\frac{Q_{\alpha}(\nu_{i}+\tfrac{\eta}{2})}{Q_{\alpha}(\nu_{i}-\tfrac{\eta}{2})}
\prod_{\scriptstyle{j \ne i}\atop \scriptstyle{j=1}}^{p'}
\frac{\sh(\nu_{i}-\nu_{j}+\eta) 
\sh(\nu_{i}+\nu_{j}+\eta)}
{\sh(\nu_{i}-\nu_{j}) 
\sh(\nu_{i}+\nu_{j})}\Bigg] \,,
\label{Lambdaip}
\ee
and
\begin{align}
\Lambda^{\lambda_{\alpha, i}}(u) &= \ef(u,\lambda_{\alpha, i}-\tfrac{\eta}{2})
\Bigg[\sh^{2N}(\lambda_{\alpha, i}+\tfrac{\eta}{2})
\frac{Q_{\nu}(\lambda_{\alpha, 
i}-\tfrac{3\eta}{2})}{Q_{\nu}(\lambda_{\alpha, i}-\tfrac{\eta}{2})}
\prod_{\scriptstyle{j \ne i}\atop \scriptstyle{j=1}}^{M}
\frac{\sh(\lambda_{\alpha, i}-\lambda_{\alpha, j}-\eta) 
\sh(\lambda_{\alpha, i}+\lambda_{\alpha, j}-\eta)}
{\sh(\lambda_{\alpha, i}-\lambda_{\alpha, j}) \sh(\lambda_{\alpha, 
i}+\lambda_{\alpha, j})} 
\non \\
& -\sh^{2N}(\lambda_{\alpha, i}-\tfrac{\eta}{2})
\frac{Q_{\nu}(\lambda_{\alpha,i}+\tfrac{\eta}{2})}{Q_{\nu}(\lambda_{\alpha, i}-\tfrac{\eta}{2})}
\prod_{\scriptstyle{j \ne i}\atop \scriptstyle{j=1}}^{M}
\frac{\sh(\lambda_{\alpha, i}-\lambda_{\alpha, j}+\eta) 
\sh(\lambda_{\alpha, i}+\lambda_{\alpha, j}+\eta)}
{\sh(\lambda_{\alpha, i}-\lambda_{\alpha, j}) 
\sh(\lambda_{\alpha, i}+\lambda_{\alpha, j})}\Bigg] \,.
\label{Lambdaip2}
\end{align}

\medskip

Similarly, the action of the transfer matrix on the off-shell Bethe vector $|\vec\lambda_{\beta}\rangle$
is given by
\be
t(u) |\vec\lambda_{\beta}\rangle  = 
\Lambda_{\beta}(u) |\vec\lambda_{\beta}\rangle +
\sum_{i}\Lambda^{\lambda_{\beta, i}}(u)\,  B(u) |\hat\lambda_{\beta, i}\rangle \,,
\label{offshellbeta}
\ee
where
\be
\Lambda_{\beta}(u) = 
\frac{\sh(2u+2\eta)}{\sh(2u+\eta)}\sh^{2N}(u+\eta)\frac{Q_{\beta}(u-\eta)}{Q_{\beta}(u)} 
+ \frac{\sh(2u)}{\sh(2u+\eta)}\sh^{2N}(u)\frac{Q_{\beta}(u+\eta)}{Q_{\beta}(u)} \,,
\label{Lambdapp2}
\ee
with
\be
Q_{\beta}(u) 
=\prod_{k=1}^{M}\sh\left(u-\lambda_{\beta, 
k}+\tfrac{\eta}{2}\right)\sh\left(u+\lambda_{\beta, 
k}+\tfrac{\eta}{2}\right) \,,
\ee
and
\be
\Lambda^{\lambda_{\beta, i}}(u) &=& \ef(u,\lambda_{\beta, i}-\tfrac{\eta}{2}) 
\Bigg[\sh^{2N}(\lambda_{\beta, i}+\tfrac{\eta}{2})
\prod_{\scriptstyle{j \ne i}\atop \scriptstyle{j=1}}^{M}
\frac{\sh(\lambda_{\beta, i}-\lambda_{\beta, j}-\eta) 
\sh(\lambda_{\beta, i}+\lambda_{\beta, j}-\eta)}
{\sh(\lambda_{\beta, i}-\lambda_{\beta, j}) \sh(\lambda_{\beta, 
i}+\lambda_{\beta, j})} 
\non \\
&& -\sh^{2N}(\lambda_{\beta, i}-\tfrac{\eta}{2})
\prod_{\scriptstyle{j \ne i}\atop \scriptstyle{j=1}}^{M}
\frac{\sh(\lambda_{\beta, i}-\lambda_{\beta, j}+\eta) 
\sh(\lambda_{\beta, i}+\lambda_{\beta, j}+\eta)}
{\sh(\lambda_{\beta, i}-\lambda_{\beta, j}) 
\sh(\lambda_{\beta, i}+\lambda_{\beta, j})}\Bigg] \,.
\label{Lambdaip3}
\ee

We argue in Appendix \ref{sec:proof} that, in order for 
$\vecG{p'}{\vec\lambda}$ (\ref{genvec}) to be a generalized eigenvector of the 
transfer matrix, i.e., it obeys~\eqref{geneig}, it suffices to satisfy the following conditions: 
\be
\lim_{\omega\rightarrow 0+}  \beta \left( \Lambda_{\beta}(u)-\Lambda_{\alpha}(u) \right) &\ne& 0 \,, 
\label{suffcond1I} \\
\lim_{\omega\rightarrow 0+}  \beta \left(  \Lambda_{\beta}(u)-\Lambda_{\alpha}(u) \right)^{2} &=& 0 \,, 
\label{suffcond2I} \\
\lim_{\omega\rightarrow 0+}  \omega^{2N} \beta \Lambda^{\nu_i}(u)  &=& 0 \,, 
\qquad i = 1,\ldots, p'\,,
\label{suffcond3I} \\
\lim_{\omega\rightarrow 0+}  \beta \Lambda^{\lambda_{\alpha, i}}(u)  &=& 0 \,, 
\qquad i = 1,\ldots, M\,,
\label{suffcond4I} \\
\lim_{\omega\rightarrow 0+}  \beta \Lambda^{\lambda_{\beta, i}}(u)  &=& 0 \,, 
\qquad i = 1,\ldots, M\,,
\label{suffcond5I}
\ee 
where $\alpha$ and $\beta$ are given by~\eqref{alphabeta}.
 
Recalling the expressions (\ref{Lambdaip}) and (\ref{Lambdaip2}) for 
$\Lambda^{\nu_i}(u)$ and $\Lambda^{\lambda_{\alpha, i}}(u)$, we see that the conditions 
(\ref{suffcond3I})-(\ref{suffcond4I}) require that $\{\vec\nu, 
\vec\lambda_{\alpha} \}$ be \textit{approximate} solutions (as $\omega\rightarrow 0$) of the Bethe equations\footnote{These are the usual Bethe equations~(\ref{BAE}) but with more Bethe 
roots, since we now have both $\lambda_{\alpha}$'s and~$\nu$'s. The $\nu$'s 
appear in the Bethe equations for the $\lambda_{\alpha}$'s through 
$Q_{\nu}$ functions and vice-versa.}
\begin{multline}
\sh^{2N}(\nu_{i}+\tfrac{\eta}{2})
Q_{\alpha}(\nu_{i}-\tfrac{3\eta}{2})
\prod_{\scriptstyle{j \ne i}\atop \scriptstyle{j=1}}^{p'}
\sh(\nu_{i}-\nu_{j}-\eta) 
\sh(\nu_{i}+\nu_{j}-\eta)
 \\
 =\sh^{2N}(\nu_{i}-\tfrac{\eta}{2})
Q_{\alpha}(\nu_{i}+\tfrac{\eta}{2})
\prod_{\scriptstyle{j \ne i}\atop \scriptstyle{j=1}}^{p'}
\sh(\nu_{i}-\nu_{j}+\eta) 
\sh(\nu_{i}+\nu_{j}+\eta)
\label{BAEnu}
\end{multline}
and
\begin{multline}
\sh^{2N}(\lambda_{\alpha, i}+\tfrac{\eta}{2})
Q_{\nu}(\lambda_{\alpha, 
i}-\tfrac{3\eta}{2})
\prod_{\scriptstyle{j \ne i}\atop \scriptstyle{j=1}}^{M}
\sh(\lambda_{\alpha, i}-\lambda_{\alpha, j}-\eta) 
\sh(\lambda_{\alpha, i}+\lambda_{\alpha, j}-\eta) 
 \\
= \sh^{2N}(\lambda_{\alpha, i}-\tfrac{\eta}{2})
Q_{\nu}(\lambda_{\alpha,i}+\tfrac{\eta}{2})
\prod_{\scriptstyle{j \ne i}\atop \scriptstyle{j=1}}^{M}
\sh(\lambda_{\alpha, i}-\lambda_{\alpha, j}+\eta) 
\sh(\lambda_{\alpha, i}+\lambda_{\alpha, j}+\eta) \,.
\label{BAElama}
\end{multline}
By `approximate solutions' we mean that the equations are satisfied up to a certain order in $\omega$,
not necessarily in all orders, i.e., we solve equations~\eqref{BAEnu} 
and~\eqref{BAElama} in the sense of perturbation theory in the small 
parameter $\omega$, until (\ref{suffcond3I})-(\ref{suffcond4I}) are satisfied.
Similarly for the condition~\eqref{suffcond5I}, it requires that
$\vec\lambda_{\beta}$ be an approximate solution of the Bethe
equations corresponding to $\Lambda^{\lambda_{\beta, i}}(u)$
in~(\ref{Lambdaip3}), 
\be
\lefteqn{\sh^{2N}(\lambda_{\beta, i}+\tfrac{\eta}{2})
\prod_{\scriptstyle{j \ne i}\atop \scriptstyle{j=1}}^{M}\sh(\lambda_{\beta, i}-\lambda_{\beta, j}-\eta) 
\sh(\lambda_{\beta, i}+\lambda_{\beta, j}-\eta)}\non\\
&&=\sh^{2N}(\lambda_{\beta, i}-\tfrac{\eta}{2})
\prod_{\scriptstyle{j \ne i}\atop \scriptstyle{j=1}}^{M}
\sh(\lambda_{\beta, i}-\lambda_{\beta, j}+\eta) 
\sh(\lambda_{\beta, i}+\lambda_{\beta, j}+\eta) \,.
\label{BAElambeta}
\ee

Let us therefore look for a solution $\{\vec\nu, 
\vec\lambda_{\alpha} \}$ of the Bethe  
equations \eqref{BAEnu}-\eqref{BAElama} with $M+p'$ Bethe 
roots that approaches $\{\vec{\nu}^{\,\infty}, 
\vec\lambda \}$ as $\omega \rightarrow 0$, recall our assumption on the limit~\eqref{limit-ass}. We assume that for small
$\omega$ this solution is given by
\be
\nu_{j} &=& -\log \omega + \sum_{k\ge 1}a_{j k} \omega^{2(k-1)}+ 
\frac{i \pi}{2p'}(p'-(2j-1))  \,,  \qquad
j=1, \ldots, p' \,, \non \\
\lambda_{\alpha, j} &=& \lambda_{j} + \sum_{k\ge 1}b_{j k} \omega^{2p'k}
\,,  \qquad j=1, \ldots, M \,,
\label{pprimestring2}
\ee
where the coefficients $\{ a_{j k}\,, b_{j k} \}$ are independent of $\omega$. 
To determine these coefficients, we rewrite the Bethe  
equations \eqref{BAEnu}-\eqref{BAElama} in the form
\be
{\rm BAE}_{k} = 0\,, \qquad k = 1, \ldots, M+p'\,,
\label{BEgk}
\ee
where ${\rm BAE}_{k}$ is defined as the difference of the left-hand and right-hand 
sides. 
We insert (\ref{eta02}) and (\ref{pprimestring2}) into 
(\ref{BEgk}), perform series expansions about $\omega=0$,
and solve the resulting equations for $\{ a_{j k}\,, b_{j k} \}$, starting from the most singular terms 
in the series expansions (the most singular term has obviously a finite
order in $\omega$). 
In practice,  the conditions (\ref{suffcond3I})-(\ref{suffcond4I}) 
are satisfied by keeping sufficiently many terms in the expansion (\ref{pprimestring2}).

Similarly, we can find a solution $\vec\lambda_{\beta}$ of the Bethe 
equations  
(\ref{BAElambeta}) with $M$ Bethe 
roots that approaches $\vec\lambda$ as $\omega \rightarrow 0$. We assume that for small
$\omega$ this solution is given by
\be
\lambda_{\beta, j} &=& \lambda_{j} + \sum_{k\ge 1}c_{j k} \omega^{2p'k}
\,,  \qquad j=1, \ldots, M \,,
\label{pprimestring3}
\ee
and we solve for the coefficients $\{ c_{j k} \}$ in a similar way.
We find in practice that, by keeping sufficiently many terms in the expansion 
(\ref{pprimestring3}), the condition (\ref{suffcond5I}) is also satisfied.
In general, $\vec\lambda_{\beta} \ne \vec\lambda_{\alpha}$.

We then find by doing explicit expansion using
(\ref{pprimestring2}) and (\ref{pprimestring3}), with the same number of terms in the sums as in the previous step,
that $\Lambda_{\beta}(u)-\Lambda_{\alpha}(u)$ (recall the definitions
(\ref{Lambdapp}), (\ref{Lambdapp2})) is of order $\omega^{2 p'}$ 
\be\label{b-a}
\Lambda_{\beta}(u)-\Lambda_{\alpha}(u)  = O(\omega^{2 p'}) \,.
\ee
For the choice of $\beta$ in (\ref{alphabeta}), it follows that both
conditions (\ref{suffcond1I}) and (\ref{suffcond2I}) are also
satisfied.

We have therefore demonstrated the following proposition, assuming that our conjecture~(\ref{cformula}) is true.

\begin{Prop} \label{prop:gen-vect} 
For anisotropy $\eta=i \pi/p$ with integer $p \ge 2$,
given a Bethe eigenvector~$|\vec\lambda\rangle$ in~\eqref{v0onshell} 
of the transfer matrix $t(u)$ with eigenvalue $\Lambda(u)$ 
(\ref{ordinary}),
a generalized eigenvector of rank 2 with the same generalized eigenvalue 
is given by 
\be
\vecG{p'}{\vec\lambda} = 
\lim_{\omega\rightarrow0+}\left[ \omega^{2 p' (N-1)} |\vec\nu, \vec\lambda_{\alpha}\rangle 
-c\,  \omega^{-2 p'} F^{p'} |\vec\lambda_{\beta}\rangle \right]\,,
\label{genvec2}
\ee
where $p'$ equals $N-2M +1$ modulo $p$,
the vectors $|\vec\nu, \vec\lambda_{\alpha}\rangle$ and $|\vec\lambda_{\beta}\rangle$ are given by 
(\ref{genvecmore}), $c$~is given by (\ref{cformula}), and 
$\vec\nu$, $\vec\lambda_{\alpha}$, and $\vec\lambda_{\beta}$ are given by 
the series expansions (\ref{pprimestring2}) and (\ref{pprimestring3}), 
whose coefficients are determined by the Bethe equations 
(\ref{BAEnu}), (\ref{BAElama}), and~(\ref{BAElambeta}) up to a certain order in $\omega$ such that
(\ref{suffcond3I})-(\ref{suffcond5I}) are satisfied.
\end{Prop}

\begin{Rem}\label{rem-Jordan-cell}
In this remark, we address the problem of constructing the whole
Jordan cell for the transfer matrix -- the states $|v'\rangle$ and
$|v\rangle$ in~\eqref{geneig-2} --
or what is the corresponding eigenvector $|v'\rangle$ for
the generalized eigenvector $|v\rangle=\vecG{p'}{\vec\lambda}$ constructed in~\eqref{genvec2}?
We give two arguments, one is computational and uses the results of
App.~\ref{sec:proof} where we stated Cor.~\ref{cor:vv} in the end.  It
states that under the assumptions made in Prop.~\ref{prop:gen-vect}
$|v'\rangle$ is non-zero and equals $\kappa F^{p'}|\vec\lambda\rangle$
where $\kappa$ is the limit in~\eqref{suffcond1I}.
The other argument is less technical and counts only degeneracies.
First, the state $|v'\rangle$ should have  the same $S^z=N/2-M-p'$ as
$|v\rangle=\vecG{p'}{\vec\lambda}$ has.  Note further that $|v\rangle$
is in the same tilting module $T_{j=N/2-M}$ as the initial Bethe state
$|\vec\lambda\rangle$ because the two states have the same eigenvalue
$\Lambda(u)$ of the transfer matrix $t(u)$, and the ordinary Bethe
states of the same $M$ value are non-degenerate (with respect to
$t(u)$) at roots of unity~\cite{Gainutdinov:2015vba}. 
Indeed, if the
generalized eigenstate $|v\rangle$ would belong to another copy of
$T_{j=N/2-M}$ not containing $|\vec\lambda\rangle$, we could obtain by
acting on $|v\rangle$
with ($p'$ power of) the raising $\qg$ generator $E$ a
highest-weight state, see the action in App.~\ref{app:proj-mod-base},
which is another Bethe state\footnote{or complete $p$-string operators
on a Bethe state with $M'$ lower than $M$ by a multiple of $p$.}, say
$|\vec\lambda'\rangle$, with the same $M$ and by construction the 
same eigenvalue
$\Lambda(u)$ as $|v\rangle$, which 
contradicts  the
non-degeneracy result in~\cite{Gainutdinov:2015vba}, and thus
$|\vec\lambda'\rangle\sim|\vec\lambda\rangle$.
Further, the weight $S^z=N/2-M-p'$ is only doubly degenerate in
$T_{j=N/2-M}$: $|v\rangle=\vecG{p'}{\vec\lambda}$ and the vector
$F^{p'}|\vec\lambda\rangle$ in the bottom of $T_j$ have this weight.
We thus have~\eqref{geneig-2} with
\be
|v'\rangle \sim F^{p'}|\vec\lambda\rangle.
\ee
We have also checked this result explicitly for the 
examples in Secs.~\ref{sec:genp2}\,-\,\ref{sec:genp4}.
\end{Rem}

We note that Prop.~\ref{prop:gen-vect}  gives only sufficient 
conditions on existence of the generalized
eigenvectors, and the construction if the conditions are satisfied.  
Their actual existence is  clear in the examples we consider below. 
We give in 
Secs.~\ref{sec:genp2}\,-\,\ref{sec:genp4} explicit examples of constructing 
the Jordan cells and generalized eigenvectors for $p=2,3,4$ using the construction in Prop.~\ref{prop:gen-vect}. 
Readers who are interested more in these 
examples can skip the next subsection where we go back to representation theory and tilting modules.

\subsection{Representation-theoretic description}\label{sec:rep-th-descr}
 We give here a representation-theoretic interpretation of our
construction in Prop.~\ref{prop:gen-vect} by analyzing the
contribution of $V_j$'s to different tilting modules in the
root-of-unity limit.  Then, we also discuss the problem of counting
the (generalized) eigenvectors using this analysis.

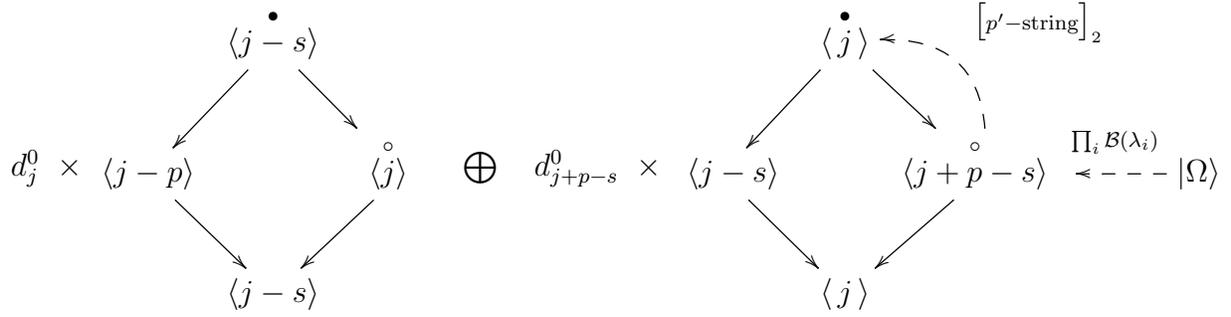
\begin{figure}
\begin{equation*}
\xymatrix@R=39pt@C=0pt@M=0pt@W=0pt{
\mbox{}&\\
&d^0_j \;\; \times\;\\
}
\xymatrix@R=22pt@C=4pt@W=3pt@M=4pt{
&\stackrel{\bullet}{\Irr{j-s}}\ar[dl]\ar[dr]&\\
\Irr{j-p}\ar[dr]
&&\mbox{}\;\stackrel{\circ}{\Irr{j}}\;\;\;\;\;\ar[dl]\\
&\Irr{j-s}&
} 
\xymatrix@R=39pt@C=0pt@M=0pt@W=0pt{
\mbox{}&\\
&\bigoplus\quad d^0_{j+p-s} \;\; \times\\
}\!\!\!
\xymatrix@R=22pt@C=4pt@W=3pt@M=4pt{
&\stackrel{\bullet}{\Irr{\,j\,}}\ar[dl]\ar[dr]&&&&\\
{\mbox{}\quad\Irr{j-s}\;}\ar[dr]
&&\stackrel{\circ}{\Irr{j+p-s}}\;\;\ar[dl]\ar@/_2pc/@{-->}[ul]_{\bigl[p'-\text{string}\bigr]_2}&&&
\vac\ar@{-->}[lll]_{\mbox{}\quad\;\;\prod_i {\cal B}(\lambda_i)}\\
&\Irr{\,j\,}&&&&
} 
\label{Vj-Tj}
\end{equation*}
\caption{The subquotient structure of the tilting module $d_j^0T_j
\oplus d_{j+p-s}^0T_{j+p-s}$ with the solid arrows corresponding to
the action of $\qg$;  here,  $s\equiv s(j)$ for brevity and $s(j)$ is
defined in~\eqref{sj}. The spin-$j$
highest-weight states (ordinary
eigenstates) live in the node $\stackrel{\circ}{\Irr{j}}$, on the left,
while the spin-$j$ generalized eigenstates are in 
$\stackrel{\bullet}{\Irr{j}}$, on the right part of the diagram,
and are constructed from spin-$(j+p-s)$
highest-weight states -- the curvy arrow  for the 
$p'$-string of ${\cal B}(\nu_k)$ operators and $[\ldots]_2$ denotes the 
subtraction of the action of $F^{p'}$ (the 
construction of Prop.~\ref{prop:gen-vect}), with
$p'=p-s$.  
The horizontal dashed arrow corresponds to the action of a product of ${\cal B}(\lambda_i)$
operators (the ordinary ABA construction).
}
\label{fig:pstring}
\end{figure}

We begin with  the decomposition of the spin chain at 
{\em generic} $q$
\be
\bigl(V_{\frac{1}{2}}\bigr)^{\otimes N} = \bigoplus_{j=0(1/2)}^{N/2} d_j V_j,
\label{decomposition-gen}
\ee
where  the multiplicity $d_j$ of the spin-$j$ representation 
$V_{j}$ is defined in~\eqref{dj}.  It is instructive to compare this
decomposition with the one~\eqref{decomposition} at roots of unity in 
terms of tilting modules $T_j$ with multiplicities $d^0_j\leq d_j$, see the
expression in~\eqref{dj0}. We will consider further only 
those values of $j$ for which $2j+1$ modulo $p$ 
is nonzero (that is, $s(j)$ defined in (\ref{sj}) is nonzero),
i.e., when $T_j$ are indecomposable but reducible and thus contain 
generalized eigenvectors, 
recall the discussion after~\eqref{gen-transf}. The multiplicity $d^0_j$ is then strictly less than~$d_j$.
Each such $T_j$ contains $V_j$ as a proper
submodule. The corresponding spin-$j$ highest-weight state lives in
the node denoted by
$\stackrel{\circ}{\Irr{j}}$ in the left half of 
Fig.~\ref{fig:pstring}. This state can be constructed using the 
ordinary ABA approach as in (\ref{Bethestate}).
The rest $d_j-d^0_j=d^0_{j+p-s(j)}$
of the initial number of $V_j$'s are not submodules but
{\sl sub-quotients} in another tilting module -- in $T_{j+p-s(j)}$ (recall
the discussion in Sec.~\ref{sec:Tj-str}.) 
Being `sub-quotient' here means that the spin-$j$ states lose the property ``highest-weight''  in the root-of-unity limit. 
These states are generalized eigenstates of $t(u)$.  They live in the node  
$\stackrel{\bullet}{\Irr{j}}$ in the right half of
Fig.~\ref{fig:pstring}.

We therefore expect that $d^0_j$ of the spin-$j$ Bethe states have a
well-defined  limit as $q$ approaches a root 
of unity and give ordinary  
$t(u)$-eigenstates
  living in $\stackrel{\circ}{\Irr{j}}$; and on the
other side, we expect irregular behavior of the $d_j-d^0_j$ Bethe
states -- the corresponding Bethe roots~$\nu_k$ go to infinity as~\eqref{pprimestring1} -- such that an appropriate limit
gives the generalized eigenstates
living in $\stackrel{\bullet}{\Irr{j}}$. 
By  Prop.~\ref{prop:gen-vect}, we
construct the latter states by applying the 
$p'$-string of ${\cal B}(\nu_k)$ operators
on the usual Bethe states living in the node
$\stackrel{\circ}{\Irr{j+p-s}}$ in $T_{j+p-s}$ 
and subtracting the image of $F^{p'}$ (as in~\eqref{gen-vec-fin}) on the  
spin-$(j+p-s)$ highest-weight state that guarantees absence of diverging terms in the limit.
We sketched this  in the right half of
Fig.~\ref{fig:pstring} where the subtraction is schematically denoted by $[\ldots]_2$.  Note that the difference in the highest
$S^z$-eigenvalues
in $\stackrel{\bullet}{\Irr{j}}$ and 
in  $\stackrel{\circ}{\Irr{j+p-s}}$ is $p-s(j)$. 
(Recall that the number $j$ in $\langle j\rangle$
corresponds to the spin $S^z=j$ value of the highest-weight vector.) 
Hence, the number $p'$ in the $p'$-string equals $p-s(j)$.
Similarly, we should use a string of length
$p'=s(j)$ to construct generalized eigenstates in
$\stackrel{\bullet}{\Irr{j-s}}$ out of Bethe eigenstates from
$\stackrel{\circ}{\Irr{j}}$, in the left part of
Fig.~\ref{fig:pstring}, as anticipated in (\ref{p'formula}) and in  Prop.~\ref{prop:gen-vect}.

We give finally a comment about counting the (generalized) eigenstates.
The
limit of ordinary Bethe states gives  as many linearly independent 
states as the number of admissible
solutions of the Bethe equations at the root of unity, and we know~\cite{Gainutdinov:2015vba}
that there can be deviations of this number from $d^0_j$ (it is less than $d^0_j$ in general).   Taking into account the 
deviations $n_j$ studied in~\cite{Gainutdinov:2015vba} we should thus have $d^0_j-n_j$ linearly independent eigenstates
and the number $d^0_{j+p-s}-n_{j+p-s}$ of linearly independent
generalized eigenstates of spin-$j$. To construct the missing
eigenstates of spin-$j$ or highest-weight states in $\stackrel{\circ}{\Irr{j}}$, we should use the exact complete
$p$-strings from Sec.~\ref{sec:pstrings}.
We believe that the same complete
$p$-strings construction can be applied to generalized eigenvectors 
and it recovers the total number $d^0_{j+p-s}$ of generalized eigenvectors of spin-$j$.

\subsection*{Examples}

We now illustrate the general construction (\ref{genvec2}) with 
several explicit examples.

\subsection{$p=2$}\label{sec:genp2} 

As already noted, for $p=2$, the only possibility is $p'=1$, i.e. an
infinite real root. For even~$N$ and irrespectively of the value of $M$,
the small-$\omega$ behavior of this root is given by  
\be
\nu = -\log \omega +  O(\omega^{0}) \,,
\label{nusmallomega}
\ee
as in (\ref{lambdaforsmallomega}) and (\ref{pprimestring2}).
We find that the construction (\ref{genvec2}) produces a generalized 
eigenvector irrespectively of the values of the $O(\omega^{0})$ and higher-order terms.
Hence, for $p=2$ and even~$N$,
the generalized eigenvector $\vecG{1}{\vec\lambda}$ corresponding to the on-shell Bethe vector $|\vec\lambda\rangle$
with any value of $M$ is given by
\be
\vecG{1}{\vec\lambda} = \lim_{\omega\rightarrow0+} \Big[\omega^{2(N-1)}  {\cal B}(\nu)  -c\,  
\omega^{-2} F \Big]|\vec\lambda\rangle\,,
\label{vecG1}
\ee
for some ``non-universal'' constant $c$ and $\nu$ is given by (\ref{nusmallomega}). 
 We denote by $\vecG{1}{-}$ the result for 
the reference state (no Bethe roots) $|\vec\lambda\rangle = \vac$.
For odd $N$, there is no solution of the form (\ref{nusmallomega}), 
which is in correspondence with the fact 
that the Hamiltonian is diagonalizable at odd $N$.

For example, we have explicitly computed (\ref{vecG1})
with $|\vec\lambda\rangle = \vac$
for 
$N=4,6,8$ using {\tt Mathematica}, and we have verified that the 
result $\vecG{1}{-}$
is a generalized eigenvector of the Hamiltonian (\ref{Hamiltonian}), with 
generalized eigenvalue 0: 
\be
H^{2}\vecG{1}{-} =0 \,, \qquad 
H \vecG{1}{-} \sim F \vac \,,
\ee
where we use $\sim$ to denote equality up to some 
nonzero numerical factor.

\subsection{$p=3$}\label{sec:genp3} 

For $p=3$, both $p'=1$ and $p'=2$ are possible. 

\subsubsection{$p'=1$}\label{sec:genp3pp1}

Let us first consider the case $p'=1$, $N=6$ and $M=0$. 
Following the procedure explained in (\ref{pprimestring2}) and 
immediately below, we find that the corresponding $\nu$ is given by 
\be
\nu= -\log \omega +\tfrac{1}{4}\log 3 - \tfrac{\sqrt{3}}{12}\omega^{2}
+ O(\omega^{4}) \,,
\ee
and the corresponding vector $\vecG{1}{-}$ 
is a generalized eigenvector of the Hamiltonian (\ref{Hamiltonian}) 
with generalized eigenvalue 5/2:
\be
(H- \tfrac{5}{2})^{2} \vecG{1}{-} =0 \,, \qquad 
(H- \tfrac{5}{2}) \vecG{1}{-} \sim F \vac \,.
\ee

\subsubsection{$p'=2$}\label{sec:genp3pp2}

Let us now consider $p'=2$. An example is the case $N=4$ and $M=0$, for which $\vec\nu$
(\ref{pprimestring2}) is given by
\be
\nu_{1}\,, \nu_{2} = \pm \tfrac{i\pi}{4} -\log \omega + \tfrac{1}{8}\log(\tfrac{243}{4}) \mp 
\tfrac{2i\sqrt{2}}{3^{5/4}}\omega^{2} - 
\tfrac{13\sqrt{3}}{36}\omega^{4}
+ O(\omega^{6}) \,.
\ee 
We have explicitly verified that 
$\vecG{2}{-}$ is a generalized eigenvector of the Hamiltonian, with 
generalized eigenvalue 3/2:
\be
(H- \tfrac{3}{2})^{2} \vecG{2}{-} =0 \,, \qquad 
(H- \tfrac{3}{2}) \vecG{2}{-} \sim F^2 \vac \,.
\ee

Another example is the case $N=6$ and  $M=1$. This is our first 
example with $M>0$ (and $p>2$), which makes this case particularly interesting. 
There are 4 solutions of 
the Bethe equations (\ref{BAE}) with $p=3, N=6, M=1$, and let us 
focus here on the simplest $\lambda=\tfrac{1}{2}\log 2 \approx 0.346574$.
By following the procedure described around 
(\ref{pprimestring2})-(\ref{pprimestring3}), we obtain
\be
\nu_{1}\,, \nu_{2} &=& \pm \tfrac{i\pi}{4} -\log \omega + 
\tfrac{1}{8}\log(108) \mp 
\tfrac{19i\sqrt{2}}{16*3^{3/4}}\omega^{2} - \tfrac{34493}{21888\sqrt{3}}\omega^{4}
+ O(\omega^{6}) \,, \non \\
\lambda_{\alpha}  &=& \tfrac{1}{2}\log 2  - 
\tfrac{3}{16}\sqrt{3}\omega^{4} + O(\omega^{8})\,, \non \\
\lambda_{\beta}  &=& \tfrac{1}{2}\log 2  - 
\tfrac{1}{4}\sqrt{3}\omega^{4} + O(\omega^{8})\,.
\label{p3N6M1pp2}
\ee
Note that $\lambda_{\alpha} \ne \lambda_{\beta}$.
We have explicitly verified that the corresponding vector
$\vecG{2}{\lambda}$ (\ref{genvec2}) is a generalized eigenvector of
the Hamiltonian with generalized eigenvalue $-3/2$,
\be
(H+ \tfrac{3}{2})^{2} \vecG{2}{\lambda} =0 \,, \qquad 
(H+ \tfrac{3}{2}) \vecG{2}{\lambda} \sim F^2|\lambda\rangle \,.
\ee

\subsection{$p=4$}\label{sec:genp4}

For $p=4$, we can have $p'=1, 2, 3$, but we illustrate here only two 
of these three possibilities.  

\subsubsection{$p'=1$}\label{sec:genp4pp1}

Let us first consider $p'=1$. An example is the case $p=4, N=4, M=0, p'=1$, 
for which~$\nu$ from~(\ref{pprimestring2}) is given by 
\be
\nu= -\log \omega +\tfrac{1}{4}\log 2 - \tfrac{1}{4}\omega^{2}
+ O(\omega^{4}) \,,
\ee
and the corresponding vector $\vecG{1}{-}$ 
is a generalized eigenvector of the Hamiltonian with generalized eigenvalue 
$3\sqrt{2}/2$, 
\be
(H- \tfrac{3}{2}\sqrt{2})^{2} \vecG{1}{-} =0 \,, \qquad 
(H- \tfrac{3}{2}\sqrt{2}) \vecG{1}{-} \sim F \vac \,.
\ee

Another example is the case $p=4, N=6, M=1, p'=1$, which (as the 
example in Eq. (\ref{p3N6M1pp2})) has $M>0$. There are 5 solutions of 
the Bethe equations (\ref{BAE}) with $p=4, N=6, M=1$, and let us 
focus here on the simplest $\lambda=\tfrac{1}{2}{\rm arcsinh}(1) \approx 0.440687$.
We find 
\be
\nu &=&  -\log \omega + 
\tfrac{1}{4}\log 2 -
\tfrac{1}{6}\omega^{2} + O(\omega^{4}) \,, \non \\
\lambda_{\alpha}  &=& \tfrac{1}{2}{\rm arcsinh}(1)  - 
\tfrac{5}{12}\sqrt{2}\omega^{2} + O(\omega^{4}) \,, \non \\
\lambda_{\beta}  &=& \tfrac{1}{2}{\rm arcsinh}(1)  - 
\tfrac{1}{2}\sqrt{2}\omega^{2} + O(\omega^{4}) \,.
\ee
We have explicitly verified that the corresponding vector
$\vecG{1}{\lambda}$ (\ref{genvec2}) is a generalized eigenvector of
the Hamiltonian with generalized eigenvalue $\sqrt{2}/2$, 
\be
(H- \tfrac{1}{2}\sqrt{2})^{2} \vecG{1}{\lambda} =0 \,, \qquad 
(H- \tfrac{1}{2}\sqrt{2}) \vecG{1}{\lambda} \sim F |\lambda\rangle \,.
\ee

\subsubsection{$p'=3$}\label{sec:genp4pp3}

Let us now consider $p'=3$. An example is the case $p=4, N=6, M=0, 
p'=3$, for which $\vec\nu$ (\ref{pprimestring2}) is given by
\be
\nu_{1} = \tfrac{i\pi}{3} -\log \omega + \tfrac{1}{12}\log(1352) -
(-\tfrac{1}{13})^{1/3}\omega^{2}-\tfrac{53}{12}(-\tfrac{1}{13})^{2/3}\omega^{4}
-\tfrac{3847}{3744}\omega^{6} + O(\omega^{8})\,, \non \\
\nu_{2} = -\log \omega + \tfrac{1}{12}\log(1352) +
(\tfrac{1}{13})^{1/3}\omega^{2}-\tfrac{53}{12}(\tfrac{1}{13})^{2/3}\omega^{4} 
-\tfrac{3847}{3744}\omega^{6} + O(\omega^{8})
\,, \quad \nu_{3} = \nu_{1}^{*} \,.
\ee
We have explicitly verified that the corresponding vector $\vecG{3}{-}$ (\ref{genvec2})
is a generalized eigenvector of the Hamiltonian with generalized eigenvalue 
$5\sqrt{2}/2$, 
\be
(H- \tfrac{5}{2}\sqrt{2})^{2} \vecG{3}{-} =0 \,, \qquad 
(H- \tfrac{5}{2}\sqrt{2}) \vecG{3}{-} \sim F^{3} \vac \,.
\ee

\section{Complete sets of eigenstates for 
$p=2$}\label{sec:p2}

For the case $p=2$, the decomposition of the space of states into tilting modules depends 
fundamentally on the parity of $N$:

\subsection*{Even $N$}

For $p=2$ and even $N$, the decomposition (\ref{decomposition}) consists 
of tilting modules $T_{j}$ of dimension $4j$, where $j$ is an 
integer. Recall the diagram in~\eqref{Tj-diag}: each such module has a 
right node (or simple subquotient)  ${\bf R}_{j}$ of dimension $j+1$, a bottom 
node   ${\bf 
B}_{j}$ of dimension $j$, a top 
node   ${\bf T}_{j}$ of dimension $j$, and a left node 
  ${\bf L}_{j}$ of dimension $j-1$ (provided 
that $j>1$).
We use the basis and $\qg$-action in $T_j$ in
App.~\ref{app:proj-mod-base} to make the following statements.  The
right node consists of the vectors \footnote{The basis' construction
\eqref{basis-Rp2}-\eqref{basis-Lp2}  is just an example of 
 the general one~\eqref{basis-R}-\eqref{basis-L} for any $p$ in the beginning of the next section. We found it is more convenient to describe the basis here as well.}
\be\label{basis-Rp2}
{\bf R}_{j}:\quad |v\rangle\,, f |v\rangle\,, f^{2} |v\rangle\,, \ldots \,,  f^{j} |v\rangle\,,
\ee
where $|v\rangle$ can be either a usual Bethe state or a state 
constructed from an exact complete 2-string; and $f$ is the $s\ell(2)$ 
lowering generator from $\qg$. 
The bottom node consists of the vectors obtained by acting on the 
right node with the $\qg$ lowering generator $F$ 
\be
{\bf B}_{j}:\quad F |v\rangle\,, F f |v\rangle\,, F f^{2} |v\rangle\,, \ldots \,, F f^{j-1} |v\rangle\,.
\ee
The top  node consists of the \textit{generalized} 
eigenvectors
\be
{\bf T}_{j}:\quad  \vecG{1}{v}\,, f  \vecG{1}{v}\,, f^{2} \vecG{1}{v}\,, \ldots \,, f^{j-1}  \vecG{1}{v}\,,
\ee
where $\vecG{1}{v}$ is given by (\ref{vecG1}) with $|\vec\lambda \rangle 
= |v\rangle $. 
Finally, the left  node ${\bf L}_j$  consists of 
(ordinary) eigenvectors.
We first introduce states  obtained by acting on the 
top node with~$F$ 
\be\label{basis-Lp2}
\tilde{{\bf L}}_{j}:\quad F \vecG{1}{v}\,, F f \vecG{1}{v}\,, F f^{2} \vecG{1}{v}\,, \ldots \,, F f^{j-2} \vecG{1}{v}\,.
\ee
Together with~\eqref{basis-Rp2}, they form a basis in the direct sum ${\bf L}_j \oplus {\bf R}_j$, the states in ${\bf L}_j$ are linear combinations of those in $\tilde{{\bf L}}_{j}$ and ${\bf R}_{j}$. For later convenience, we will refer to $\tilde{{\bf L}}_{j}$ instead of ${\bf L}_{j}$, see more details in Sec.~\ref{sec:pgt2} for the general case.

We note that the generalized eigenvectors appear only in the top
node.

\subsection*{Odd $N$}

For $p=2$ and odd $N$, the decomposition (\ref{decomposition}) consists 
of {\em irreducible} tilting modules $T_{j} = V_{j}$ of dimension $2j+1$, where $j$ is half-odd 
integer -- indeed, the number $s(j)$ is zero for all these $j$, and all
$T_j$ are then irreducible following the discussion in
Sec.~\ref{sec:Tj-str}.  Starting from a
highest-weight vector $|v\rangle$, the remaining vectors of the 
multiplet are obtained by applying $F$ and powers of $f$.
For odd $N$ there are only ordinary eigenvectors (i.e., no 
generalized eigenvectors), which is in agreement with~\cite{Gainutdinov:2012qy}.

\subsection*{Examples}
We now illustrate the above general framework by exhibiting ABA
constructions of complete sets of $2^{N}$ (generalized) eigenvectors
for the cases $N=4,5,6$.  For each of these cases, we have explicitly
verified that the vectors are indeed (generalized)
eigenvectors of the Hamiltonian (\ref{Hamiltonian}) and are linearly
independent. The needed admissible solutions of 
the Bethe equations for $p=2$ are given in Appendix C of 
\cite{Gainutdinov:2015vba}.

We also consider 
selected eigenvectors for the cases $N=7, 9$ in order to further illustrate 
the construction in Sec.~\ref{sec:pstrings}.
We emphasize that when one or more modules in the decomposition (\ref{decomposition}) 
are spectrum-degenerate (which can occur for either odd or even $N$),
it is necessary to use 
this construction  (\ref{Tarasovstate}), (\ref{Tarasovstategen})  based on exact 
complete 2-strings.

\subsection{$N=4$}

For  $p=2, N=4$, the decomposition (\ref{decomposition}) is given by
\be
2 T_1 \oplus  T_2 \,. \non 
\ee
The $T_{2}$ consists of the following 8 vectors:  
\be
 & {\bf R}_{2}:& \quad |v\rangle\,, f |v\rangle\,, f^{2} |v\rangle \,, \non \\
T_{2}:\qquad & {\bf B}_{2}:& \quad F |v\rangle\,, F f |v\rangle\,, \non \\
 & {\bf T}_{2}:& \quad \vecG{1}{v}\,, f \vecG{1}{v}\,, \non \\
 & \tilde{{\bf L}}_{2}:& \quad F \vecG{1}{v}\,,
\label{N4T2} 
\ee
where $|v\rangle = \vac$ is the reference state (\ref{reference}), 
and $ \vecG{1}{v}= \vecG{1}{-}$. Each of the two 
copies of $T_{1}$ consists of the following 4 vectors: 
\be
& {\bf R}_{1}:& \quad |v\rangle\,, f |v\rangle\,,  \non \\
T_{1}:\qquad& {\bf B}_{1}:& \quad F |v\rangle\,,  \non \\
& {\bf T}_{1}:& \quad \vecG{1}{v}\,,
\label{N4T1} 
\ee
where $|v\rangle = {\cal B}(\lambda) \vac$, 
$\vecG{1}{v}= \vecG{1}{\lambda}$, 
and $\lambda$ is an admissible solution 
of the Bethe equations with $N=4$ and $M=1$, of which there are two: 
$\lambda=0.440687$ and $\lambda=0.440687 + \tfrac{i\pi}{2}$.
All together we thus find $2^{4}=16$ vectors.

\subsection{$N=5$}\label{sec:p2N5} 

For  $p=2, N=5$, the space of states decomposes into a direct sum 
of irreducible representations
\be
5 V_{\frac{1}{2}} \oplus 4 V_{\frac{3}{2}} \oplus V_{\frac{5}{2}} \,. \non 
\ee

The $V_{\frac{5}{2}}$, with dimension 6, has the reference state  
$\vac$ as its highest weight state.
As noted in Appendix D of \cite{Gainutdinov:2015vba}, this module is 
spectrum-degenerate with one copy of $V_{\frac{1}{2}}$; 
the latter has dimension 2 and highest weight $\vecT{v_{1}}{-}$ i.e. an eigenvector constructed 
from an exact perfect 2-string and no other Bethe roots (\ref{Tarasovstate}), 
where $v_{1}$ is an arbitrary number 
(for arbitrary  $v_1$ and $v_1'$ we get two vectors different only by a scalar).
The other four copies of $V_{\frac{1}{2}}$ 
also have dimension 2, with highest-weight vectors
${\cal B}(\lambda_{1})\, {\cal 
B}(\lambda_{2}) \vac $, where $\{ \lambda_{1}, 
\lambda_{2}\}$ is an admissible solution of the Bethe equations with 
$N=5$ and $M=2$, of which there are four:
\be
&&\{0.337138, 0.921365\}\,, \qquad\qquad \{0.337138+ \tfrac{i\pi}{2}, 
0.921365\}\,, 
\non \\
&&\{0.337138, 0.921365+ \tfrac{i\pi}{2}\}\,, \quad 
\{0.337138+ \tfrac{i\pi}{2}, 0.921365+ \tfrac{i\pi}{2}\}\,. \non 
\ee
Finally, each of the four copies of $V_{\frac{3}{2}}$ has dimension 4
and the highest-weight vector ${\cal B}(\lambda) \vac$, where $\lambda$ is an admissible solution 
of the Bethe equations with $N=5$ and $M=1$, of which there are four: 
$\lambda=0.337138$, $\lambda=0.337138 + \tfrac{i\pi}{2}$, 
$\lambda=0.921365$, $\lambda=0.921365 + \tfrac{i\pi}{2}$.
All together we find $2^{5}=32$ vectors.

\subsection{$N=6$}\label{sec:p2N6} 

For  $p=2, N=6$, the decomposition (\ref{decomposition}) is given by
\be
5 T_1 \oplus 4 T_2 \oplus T_3 \,. \non 
\ee
The $T_{3}$ consists of the following 12 vectors: 
\be
& {\bf R}_{3}:& \quad |v\rangle\,, f |v\rangle\,, f^{2} |v\rangle \,, f^{3} 
|v\rangle \,, \non \\
T_{3}:\qquad & {\bf B}_{3}:& \quad F |v\rangle\,, F f |v\rangle\,, F f^{2} |v\rangle\,, \non \\
& {\bf T}_{3}:& \quad \vecG{1}{v}\,, f \vecG{1}{v}\,,  f^{2} \vecG{1}{v}\,, \non \\
& \tilde{{\bf L}}_{3}:& \quad F \vecG{1}{v}\,, F f \vecG{1}{v}\,,
\label{N6T3} 
\ee
where $|v\rangle = \vac$ is the reference state.  As noted in 
Appendix D of \cite{Gainutdinov:2015vba}, this module is 
spectrum-degenerate with one copy of $T_{1}$; 
the latter has the 
basis~(\ref{N4T1}) 
where $|v\rangle=\vecT{v_{1}}{-}$ is a generalized 
eigenvector constructed from an exact perfect 2-string and no other 
Bethe roots, and $v_{1}$ is an arbitrary number. The 
remaining four copies of $T_{1}$ also have the 
basis~(\ref{N4T1}), 
where 
$|v\rangle = {\cal B}(\lambda_{1})\, {\cal 
B}(\lambda_{2}) \vac $, and $\{\lambda_{1}, 
\lambda_{2}\}$ is an admissible solution of the Bethe equations with 
$N=6$ and $M=2$, of which there are four: 
\be
&&\{0.274653, 0.658479\}\,, \qquad\qquad \{0.274653+ \tfrac{i\pi}{2},  
0.658479\}\,, 
\non \\
&&\{0.274653,  0.658479+ \tfrac{i\pi}{2}\}\,, \quad 
\{0.274653+ \tfrac{i\pi}{2},  0.658479+ \tfrac{i\pi}{2}\}\,. \non 
\ee
Finally, each of the four copies of 
$T_{2}$ has the 
basis~(\ref{N4T2}) 
where $|v\rangle = {\cal B}(\lambda) \vac$, and $\lambda$ is an admissible solution 
of the Bethe equations with $N=6$ and $M=1$, of which there are four: 
$\lambda=0.274653$, $\lambda=0.274653 + \tfrac{i\pi}{2}$, 
$\lambda=0.658479$, $\lambda=0.658479 + \tfrac{i\pi}{2}$.
All together we find $2^{6}=64$ vectors.

\subsection{$N=7$}\label{sec:p2N7} 

For  $p=2, N=7$, the decomposition (\ref{decomposition}) is given by
\be
14 V_{\frac{1}{2}} \oplus 14 V_{\frac{3}{2}} \oplus 6 V_{\frac{5}{2}} 
\oplus V_{\frac{7}{2}} \,. \non 
\ee
For this case we do not enumerate all the eigenvectors, focusing 
instead on those constructed with exact complete 2-strings.  

As noted in Appendix D of \cite{Gainutdinov:2015vba}, $V_{\frac{7}{2}}$ is 
spectrum-degenerate with {\em two} copies of $V_{\frac{3}{2}}$. The former, 
with dimension 8, has the reference state $\vac$ as its highest weight state.
The latter have dimension 4 and have highest weights $\vecT{\lam_{i}}{-}$ where $i = 
1,2$,  i.e. two eigenvectors constructed 
from exact perfect 2-strings and no other Bethe roots. We have 
explicitly verified that, provided
$\lam_{1} \ne \lam_{2}$ (but otherwise arbitrary), the eigenvectors $\vecT{\lam_{1}}{-}$ and 
$\vecT{\lam_{2}}{-}$ are indeed linearly independent.

Moreover, the 6 $V_{\frac{5}{2}}$ are spectrum-degenerate with 6 
$V_{\frac{1}{2}}$. The former, with dimension 6, have highest-weight vectors ${\cal B}(\lambda) \vac$, where $\lambda$ is an admissible solution 
of the Bethe equations with $N=7$ and $M=1$, of which there are six: 
\be
&& 0.232336\,, \qquad\qquad 0.525032\,, \qquad\qquad 1.09163\,,
\non \\
&& 0.232336+ \tfrac{i\pi}{2} \,, \qquad 0.525032+ \tfrac{i\pi}{2}\,, \qquad 1.09163+ \tfrac{i\pi}{2}\,. \non 
\ee
Each of the corresponding $V_{\frac{1}{2}}$, with dimension 2, has the highest-weight vector $\vecT{v_{1}}{\lambda}$ i.e. an eigenvector constructed 
from an exact perfect 2-string ($v_{1}$ is arbitrary) and the Bethe 
root $\lambda$. These are the first examples of the construction (\ref{Tarasovstate}) 
that we meet involving a Bethe state other than the reference state.
However, since here $p=2$, then (as noted 
in Rem.~\ref{rem:Hp2}) the $\{ 
x_{r} \}$ used in this construction do not depend on 
$\lambda$.

\subsection{$N=9$}\label{sec:p2N9}

For  $p=2, N=9$, the decomposition (\ref{decomposition}) is given by
\be
42 V_{\frac{1}{2}} \oplus 48 V_{\frac{3}{2}} \oplus 27 V_{\frac{5}{2}} 
\oplus 8V_{\frac{7}{2}} \oplus V_{\frac{9}{2}} \,. \non 
\ee
Again for this case we do not enumerate all the eigenvectors, focusing 
instead on those constructed with exact complete 2-strings.  

As noted in Appendix D of \cite{Gainutdinov:2015vba}, $V_{\frac{9}{2}}$ is 
spectrum-degenerate with {\em three} copies of $V_{\frac{5}{2}}$ as well as 
with {\em two} copies of $V_{\frac{1}{2}}$. The module $V_{\frac{9}{2}}$, 
with dimension 10, has the reference state $\vac$ as its highest weight state.
The $V_{\frac{5}{2}}$ have dimension 6 and have highest-weight vectors $\vecT{\lam_{i}}{-}$ where $i = 
1,2,3$,  i.e. three eigenvectors constructed 
from exact perfect 2-strings and no other Bethe roots. We have 
explicitly verified that, provided
$\lam_{1} \ne \lam_{2} \ne \lam_{3}$ (but otherwise arbitrary), the 
three eigenvectors $\vecT{\lam_{i}}{-}$ are indeed linearly 
independent. The two $V_{\frac{1}{2}}$, each with dimension 2, are particularly interesting, 
since they have highest-weight vectors $\vecT{\lam_{i}, \lam_{j}}{-}$, i.e. 
with two exact perfect 2-strings (\ref{Tarasovstategen}). (This is 
the first, and in fact only, such example that we meet in this work.)
We have 
explicitly verified that there are precisely two such linearly 
independent vectors. The modules $V_{\frac{9}{2}} \oplus 3 
V_{\frac{5}{2}} \oplus 2 V_{\frac{1}{2}}$ account altogether for the 
32 eigenvectors with eigenvalue 0, 
as we observed in~\cite{Gainutdinov:2015vba}.

Moreover, each of the 8 $V_{\frac{7}{2}}$ is spectrum-degenerate with 
2 copies of $V_{\frac{3}{2}}$. The former, with dimension 8, have highest-weight vectors
${\cal B}(\lambda) \vac$, where $\lambda$ is an admissible solution 
of the Bethe equations with $N=9$ and $M=1$, of which there are eight: 
\be
&& 0.178189\,, \qquad\qquad 0.381455\,, \qquad\qquad 0.658479\,, \qquad\qquad 1.21812\,,
\non \\
&& 0.178189+ \tfrac{i\pi}{2} \,, \qquad 0.381455+ \tfrac{i\pi}{2}\,, \qquad 0.658479+ \tfrac{i\pi}{2}\,, \qquad 1.21812+ \tfrac{i\pi}{2}\,. \non 
\ee
The corresponding $V_{\frac{3}{2}}$, with dimension 4, have highest-weight vectors $\vecT{\lam_{i}}{\lambda}$ where $i = 
1,2$,  i.e. two eigenvectors constructed 
from exact perfect 2-strings and the Bethe root $\lambda$. Similarly 
to the case $N=7$ (section \ref{sec:p2N7}), we have 
explicitly verified that, provided
$\lam_{1} \ne \lam_{2}$ (but otherwise arbitrary), the eigenvectors 
$\vecT{\lam_{1}}{\lambda}$ and 
$\vecT{\lam_{2}}{\lambda}$ are indeed linearly independent;
and the $\{ x_{r} \}$ do not depend on $\lambda$.

The remaining 24 $V_{\frac{5}{2}}$ (i.e., those that are not 
spectrum-degenerate with $V_{\frac{9}{2}}$, as discussed above) are spectrum-degenerate with 24 
$V_{\frac{1}{2}}$. The former, with dimension 6, have highest-weight vectors
${\cal B}(\lambda_{1}){\cal B}(\lambda_{2}) \vac$, where 
$\{\lambda_{1}, \lambda_{2}\}$ is an admissible solution 
of the Bethe equations with $N=9$ and $M=2$, of which there are 24. 
The corresponding $V_{\frac{1}{2}}$, with dimension 2, have highest 
weights $\vecT{v_{1}}{\lambda_{1}, \lambda_{2}}$.

\section{Complete sets of eigenstates for $p>2$}\label{sec:pgt2}

We now exhibit ABA constructions of complete sets of $2^{N}$ 
(generalized) eigenvectors for various values of $p>2$ and $N$.
The decomposition (\ref{decomposition}) consists 
of tilting modules $T_{j}$ of dimension $2j+1$ if $s(j)=0$, see~\eqref{sj}, and of dimension  $4j+2- 2s(j) = 2pr$, where $j$ is an 
integer or half-odd integer, and we set 
\be
2j+1 \equiv rp + s\qquad \text{and} \qquad s\equiv s(j)
\ee
 for brevity. Recall the diagram in~\eqref{Tj-diag}: each $T_j$ with non-zero $s(j)$ has a 
right node (or simple subquotient)  ${\bf R}_{j}$ of dimension $s(r+1)$, a bottom 
node   ${\bf 
B}_{j}$ of dimension $(p-s)r$, a top 
node   ${\bf T}_{j}$ of dimension $(p-s)r$, and a left node 
  ${\bf L}_{j}$ of dimension $s(r-1)$ (provided 
that $r>1$).
We use the basis~\eqref{left-proj-basis-plus} and $\qg$-action in
$T_j$ in App.~\ref{app:proj-mod-base} to make the following
statements.  The right node consists of the vectors
\be\label{basis-R}
{\bf R}_{j}:\quad \rightpr_{k,l} = F^{k}f^l |v\rangle\,,\quad 0\le k\le s-1,\quad 0\le l\le  r \,,
\ee
where $|v\rangle$ can be either a usual Bethe state or a state 
constructed from an exact complete $p$-string 
-- it is a highest-weight vector; and $f$ is the $s\ell(2)$ 
lowering 
``divided power'' generator from $\qg$. 
The bottom node consists of the vectors obtained by acting on the 
right node with the $\qg$ lowering generator $F$ 
\be
{\bf B}_{j}:\quad \botpr_{n,m} = F^{s+n}f^m |v\rangle\,,\quad 0\le 
n\le p-s-1,\quad 0\le m\le  r-1 \,.
\ee
The top  node consists of the \textit{generalized} 
eigenvectors 
\be
{\bf T}_{j}:\quad \toppr_{n,m} = F^{n}f^m  \vecG{s}{v}\,,\quad 0\le n\le p-s-1,\quad 0\le m\le  r-1 \,,
\ee
where $\vecG{s}{v}$ is given by (\ref{genvec2}). 
Finally, the left  node ${\bf L}_j$ consists of the (ordinary) eigenvectors  $\leftpr_{n,m}$. To construct the basis $\{\leftpr_{n,m}\}$ in the left node ${\bf L}_j$, we first introduce states obtained by acting on the 
top node with $F^{p-s}$:
\be\label{basis-L}
\tilde{{\bf L}}_{j}:\quad \tilde{\leftpr}_{n,m} = F^{p-s+n}f^m  \vecG{s}{v}\,,\quad 0\le n\le s-1,\quad 0\le m\le  r-2 \,.
\ee
Together with~\eqref{basis-R}, they form a basis in the direct sum ${\bf L}_j \oplus {\bf R}_j$.
The vectors $\tilde{\leftpr}_{n,m}$  do not belong to ${\bf L}_j$, they are a linear combination of  $\leftpr_{n,m}$ and  $\rightpr_{n,m+1}$: $\tilde{\leftpr}_{n,m}  =\frac{1}{r}(\rightpr_{n,m+1}-\leftpr_{n,m})$, compare with the $F$ action in App.~\ref{app:proj-mod-base}. We will use below the basis elements $\tilde{\leftpr}_{n,m}$ instead  $\leftpr_{n,m}$.

In all the examples below,  we have explicitly
checked that the vectors in~\eqref{basis-R}-\eqref{basis-L} 
are indeed (generalized)
eigenvectors of the Hamiltonian (\ref{Hamiltonian}) and are linearly
 independent, and thus give a basis in $T_j$ as they should. We have also verified
by the explicit 
construction of the states that the dimensions of 
the nodes in $T_j$ coincide with the values given by \eqref{Tj-diag} 
and \eqref{dimj} 
and reviewed just above.
We remind the reader that all the needed admissible solutions of the Bethe equations 
(\ref{BAE}) are given in  
Appendix E in \cite{Gainutdinov:2015vba}.

\subsection{$p=3, N=4$}\label{sec:p3N4}

For  $p=3, N=4$, the decomposition (\ref{decomposition}) is given by
\be
T_0 \oplus  3 T_1 \oplus T_{2} \,. \non 
\ee

The $T_{2}$ (dimension 6) has the following basis, 
see~\eqref{basis-R}-\eqref{basis-L} for $r=1$ (i.e. ${\bf L}_2$ is absent) and $s=2$:
\begin{itemize}
    \item right node ${\bf R}_{2}$ consisting of 4 ordinary 
     eigenvectors (namely, $\vac\,, f \vac\,,  
    F \vac\,, F f \vac$); 
    \item bottom node ${\bf B}_{2}$ consisting of 1 ordinary vector 
    ($F^{2} \vac$) ; and
   \item top node ${\bf T}_{2}$ consisting of 1 generalized eigenvector
$\vecG{2}{-}$, which is described in 
section \ref{sec:genp3pp2}.
\end{itemize}

Each of the three $T_{1}$ are irreducible representations of 
dimension 3 consisting of a highest-weight vector 
${\cal B}(\lambda) \vac$ plus two more states obtained by lowering 
with $F$.
The three admissible solutions of (\ref{BAE}) with $p=3$, $N=4$, and $M=1$ 
are $\lambda=0.243868\,, \lambda=0.658479 \,, 
\lambda=0.902347 + \tfrac{i\pi}{2}$.
 
The $T_{0}$ (dimension 1) consists of the vector 
${\cal B}(\lambda_{1})\, {\cal B}(\lambda_{2}) \vac$, where 
$\{\lambda_{1}\,, \lambda_{2} \} = \{ 0.256013,\\ 0.857073 \}$ is the 
admissible solution of (\ref{BAE}) with $p=3$, $N=4$, $M=2$.

All together we find $2^{4}=16$ vectors.

\subsection{$p=3, N=5$}\label{sec:p3N5}

For  $p=3, N=5$, the decomposition (\ref{decomposition}) is given by
\be
T_{\frac{1}{2}} \oplus  4 T_{\frac{3}{2}} \oplus T_{\frac{5}{2}} \,. \non 
\ee

Each of the four $T_{\frac{3}{2}}$ (dimension 6) has the following basis, 
see~\eqref{basis-R}-\eqref{basis-L} for $r=1$ (i.e. ${\bf 
L}_{\frac{3}{2}}$ is absent) and $s=1$:
\begin{itemize}
    \item right node ${\bf R}_{\frac{3}{2}}$ consisting of the two ordinary 
     eigenvectors $|v\rangle$ and $f |v\rangle$, where $|v\rangle = {\cal 
    B}(\lambda)\, \vac$; 
    \item bottom node ${\bf B}_{\frac{3}{2}}$ consisting of the two 
    ordinary vectors $F |v\rangle$ and $F^{2} |v\rangle$; and
   \item top node ${\bf T}_{\frac{3}{2}}$ consisting of the two generalized 
   eigenvectors $\vecG{1}{\lambda}$ and $F \vecG{1}{\lambda}$.
\end{itemize}
The four admissible solutions of (\ref{BAE}) with $p=3$, $N=5$, and $M=1$ 
are $\lambda= 0.189841\,, \lambda= 0.447048\,, \lambda=1.08394\,, 
\lambda= 0.636889  + \tfrac{i\pi}{2}$.

The $T_{\frac{5}{2}}$ is an irreducible representation of 
dimension 6 consisting of a highest-weight vector 
$\vac$ plus five more vectors obtained by lowering 
with $F$ and/or $f$.

The $T_{\frac{1}{2}}$ is an irreducible representation of dimension 2 
consisting of the highest-weight vector 
${\cal B}(\lambda_{1})\, {\cal B}(\lambda_{2}) \vac$, plus the vector 
obtained by lowering with $F$, where 
$\{\lambda_{1}\,, \lambda_{2} \} = \{ 0.201117\,, 0.504773  \}$ is the 
admissible solution of (\ref{BAE}) with $p=3$, $N=5$, $M=2$.

All together we find $2^{5}=32$ vectors.

\subsection{$p=3, N=6$}\label{sec:p3N6}

For  $p=3, N=6$, the decomposition (\ref{decomposition}) is given by
\be
T_0 \oplus  9 T_1 \oplus 4 T_{2} \oplus T_{3} \,. \non 
\ee

The $T_{3}$ (dimension 12) has the following basis, 
see~\eqref{basis-R}-\eqref{basis-L} for $r=2$ and $s=1$:
\begin{itemize}
    \item right node ${\bf R}_{3}$ consisting of 3 ordinary 
     eigenvectors (namely, $\vac\,, f \vac\,,  
    f^{2} \vac$); 
    \item bottom node ${\bf B}_{3}$ consisting of 4 ordinary 
     eigenvectors (namely, $F \vac$, $Ff \vac$,  
    $F^{2} \vac$, $F^{2}f \vac$); 
    \item top node ${\bf T}_{3}$ consisting of 4 generalized eigenvectors
($\vecG{1}{-}$, which is described in section \ref{sec:genp3pp1}, 
plus 3 more obtained by lowering with $f$ and/or $F$, 
namely $f \vecG{1}{-}$, $F \vecG{1}{-}$, $F f \vecG{1}{-}$); and
     \item left node, 
     or rather $\tilde{{\bf L}}_{3}$,  consisting of 1 ordinary 
eigenvector obtained by lowering the generalized eigenvector 
($F^{2}\vecG{1}{-}$).
\end{itemize}

Each of the four $T_{2}$ (dimension 6) has the following basis:
\begin{itemize}
    \item right node ${\bf R}_{2}$ consisting of 
    4 ordinary eigenvectors ($|\lambda\rangle = {\cal B}(\lambda) 
    \vac$ plus 3 more obtained 
by lowering, namely, $f |\lambda\rangle\,, F |\lambda\rangle\,, F f 
|\lambda\rangle$); 
     \item bottom node ${\bf B}_{2}$ consisting of 1 ordinary 
     eigenvector ($F^{2} |\lambda\rangle$); and
     \item top node ${\bf T}_{2}$ consisting of the corresponding generalized eigenvector 
     $\vecG{2}{\lambda}$, an example 
     of which is described in section \ref{sec:genp3pp2}.
\end{itemize}
\nobreak
The four admissible solutions of (\ref{BAE}) with $p=3, N=6, M=1$ are 
$\lambda=0.155953\,, \lambda=0.346574 \,, \lambda=0.658479 \,, 
\lambda=0.502526 +\tfrac{i\pi}{2}$.

Each of the nine $T_{1}$ are irreducible representations of 
dimension 3 consisting of a highest-weight vector 
${\cal B}(\lambda_{1})\, {\cal B}(\lambda_{2}) \vac$ plus 2 more 
obtained by lowering with $F$. The nine admissible solutions $\{\lambda_{1}\,, 
\lambda_{2} \}$ of (\ref{BAE}) with $p=3, N=6, M=2$ are
\be
 && \{ 0.36275, 0.765051 \}\,,  \{ 0.16097, 0.774681 \}\,,  \{ 
 0.706816 \pm 0.526679 i \}\,, \non \\
 && \{ 0.151629, 1.00054  + \tfrac{i \pi}{2} \}\,,  \{ 0.331821, 
 0.969804+ \tfrac{i \pi}{2} \}\,, 
 \{ 0.47492 + \tfrac{i \pi}{2}, 1.23081 + \tfrac{i \pi}{2}\}\,, \non \\
 && \{ 0.164318, 0.376118\}\,,  \{ 0.583386, 0.853782 + \tfrac{i 
  \pi}{2} \}\,,  \{ 0.977905, 0.595372 + \tfrac{i \pi}{2} \}\,. \non
\ee

The $T_{0}$ (which has dimension 1) consists of the vector
${\cal B}(\lambda_{1})\, {\cal B}(\lambda_{2})\, {\cal 
B}(\lambda_{3}) \vac$, where
$\{\lambda_{1}\,, \lambda_{2}\,, \lambda_{3} \} = \{ 0.168223, 
0.39058, 0.980264\}$ is the admissible solutions of (\ref{BAE}) with 
$p=3, N=6, M=3$.

All together we thus find $2^{6}=64$ vectors.

\subsection{$p=4, N=4$}\label{sec:p4N4}
For  $p=4, N=4$, the decomposition (\ref{decomposition}) is given by
\be
2 T_0 \oplus  2 T_1 \oplus T_{2} \,. \non 
\ee

The $T_{2}$ (dimension 8) has the following basis, 
see~\eqref{basis-R}-\eqref{basis-L} for $r=1$ (i.e. ${\bf L}_2$ is absent) and $s=1$:
\begin{itemize}
    \item right node ${\bf R}_{2}$ consisting of 2 
ordinary eigenvectors (the reference state $\vac$ and $f \vac$); 
     \item bottom node ${\bf B}_{2}$ consisting of 3 ordinary 
     eigenvectors (namely, $F \vac\,, F^{2} \vac\,,  
    F^{3} \vac$); and
     \item top node ${\bf T}_{2}$ consisting of 3 generalized eigenvectors
($\vecG{1}{-}$, which is described in section \ref{sec:genp4pp1}, plus 2 more obtained by 
lowering with $F$). 
\end{itemize}

Each of the two $T_{1}$ are irreducible representations of 
dimension 3 consisting of a highest-weight vector 
${\cal B}(\lambda) \vac$ plus 2 more obtained by lowering with $F$.
The two admissible solutions of (\ref{BAE}) with $p=4, N=4, M=1$ 
are $\lambda=0.173287\,, \lambda=0.440687$.
 
Each of the two $T_{0}$ (dimension 1) consists of the vector 
${\cal B}(\lambda_{1})\, {\cal B}(\lambda_{2}) \vac$, where 
$\{\lambda_{1}\,, \lambda_{2} \} = \{ 0.186864, 0.582103 \}\,,  \{ 
0.703959 \pm 0.429694 i \} $ are the two admissible solutions of 
(\ref{BAE}) with $p=4, N=4, M=2$.

All together we find $2^{4}=16$ vectors.

\subsection{$p=4, N=6$}\label{sec:p4N6}

For  $p=4, N=6$, the decomposition (\ref{decomposition}) is given by
\be
4 T_0 \oplus  4 T_1 \oplus 5 T_{2} \oplus T_{3} \,. \non 
\ee

The $T_{3}$ (dimension 8) has the following basis,   
see~\eqref{basis-R}-\eqref{basis-L} for $r=1$ (i.e. ${\bf L}_3$ is absent) and $s=3$:
\begin{itemize}
    \item right node ${\bf R}_{3}$ consisting of 6 
ordinary eigenvectors (namely, $\vac\,, f \vac$,\break
$F \vac\,, F^{2} \vac\,,  F f \vac\,, 
F^{2} f \vac$); 
    \item bottom node ${\bf B}_{3}$ consisting of 1 ordinary 
    eigenvector ($F^{3} \vac$ ); and
    \item top node ${\bf T}_{3}$ consisting of 1 generalized eigenvector
$\vecG{3}{-}$, which is described in 
section \ref{sec:genp4pp3}.
\end{itemize}

Each of the five $T_{2}$ (dimension 8) has the following basis:
\begin{itemize}
    \item right node ${\bf R}_{2}$ consisting of 2 ordinary eigenvectors 
($|\lambda\rangle = {\cal B}(\lambda) \vac$ and 
$f|\lambda\rangle$); 
     \item bottom node ${\bf B}_{2}$ consisting of 3 ordinary 
     eigenvectors ($F |\lambda\rangle\,, F^{2} |\lambda\rangle\,, 
     F^{3} |\lambda\rangle$); and
     \item top node ${\bf T}_{2}$ consisting of 3 generalized 
     eigenvectors ($\vecG{1}{\lambda}$, an example of which is described in section 
     \ref{sec:genp4pp1}, plus 2 more obtained by lowering with $F$).
\end{itemize}
The 
corresponding five admissible solutions of (\ref{BAE}) with $p=4, N=6, M=1$
are $\lambda=0.111447$, $\lambda=0.243868$, 
$\lambda=0.440687$, $\lambda=0.902347$, and
$\lambda=0.769926 +\tfrac{i\pi}{2}$.

Each of the four $T_{1}$ are irreducible representations of 
dimension 3 consisting of the highest-weight vector 
${\cal B}(\lambda_{1})\, {\cal B}(\lambda_{2}) \vac$ plus 2 more 
obtained by lowering with $F$. The four admissible solutions $\{\lambda_{1}\,, 
\lambda_{2} \}$ of (\ref{BAE}) with $p=4, N=6, M=2$ are
\be
 && \{ 0.260368, 0.516935 \}\,,  \{ 0.11923, 0.269157 \}\,,   \non \\
 &&  \{ 0.116959, 0.523048 \} \,, \{ 0.393822 \pm  0.39281 i  \} \,.
\ee

Each of the four  $T_{0}$ (dimension 1) consists of the vector
${\cal B}(\lambda_{1})\, {\cal B}(\lambda_{2})\, {\cal 
B}(\lambda_{3}) \vac$. The four admissible solutions 
$\{\lambda_{1}\,, \lambda_{2}\,, \lambda_{3} \}$ of (\ref{BAE}) with 
$p=4, N=6, M=3$ are
\be
 && \{ 0.124053, 0.285872, 0.670931 \}\,,  \{ 0.116697, 0.77288 \pm  
 0.427941 i \}\,,   \non \\
 && \{ 0.261262, 0.749721 \pm  0.425077 i \}\,,  \{ 0.583433, 0.593097 
 \pm  0.402559 i \}\,.
\ee

All together we thus find $2^{6}=64$ vectors.

\section{Discussion}\label{sec:discuss}

We have seen that, when $q$ is a root of unity 
($q=e^{i\pi/p}$ with integer $p\ge 2$), the
$\qg$-invariant open spin-1/2 XXZ chain has two new types of
eigenvectors: eigenvectors corresponding to continuous solutions of
the Bethe equations (exact complete $p$-strings), and generalized
eigenvectors.  We have proposed here general ABA constructions for
these two new types of eigenvectors.  The construction for exact
complete $p$-strings (\ref{Tarasovstate}), (\ref{Tarasovstategen}) is
a generalization of the one proposed by Tarasov \cite{Tarasov:2003xz}
for the closed chain, while the construction of generalized
eigenvectors (\ref{genvec2}) is new.  We have demonstrated in examples
with various values of $p$ and $N$ that these constructions are indeed
sufficient for obtaining the complete set of (generalized)
eigenvectors of the model.  

The model (\ref{Hamiltonian})
at primitive roots of unity is 
related to 
the unitary $(p-1,p)$  conformal Minimal Models,
by restricting to the first $p-1$ irreducible tilting modules (see e.g. \cite{Pasquier:1989kd}),
as well as to logarithmic conformal field 
theories if one keeps all the tilting modules~\cite{Read:2007qq, Gainutdinov:2012mr}. 
We expect that our results 
can be easily generalized to the case of
rational (non-integer) values of $p$, which is related to 
non-unitary
Minimal Models. 
Indeed, for rational $p=a/b$, with $a$, $b$ coprime and $a>b$, there
are two different cases $q^a=\pm1$, {\sl i.e.,} $b$ even or odd.  For
odd $b$ (or $q^a=-1$ and $a$ can be odd or even), we have obviously the
same structure of the tilting $\qg$ modules, as the structure depends
only on the conditions on $q$ and it is the same as for $b=1$.  The
repeated tensor products of the fundamental $\qg$ representations (or
the spin-chains) are decomposed in the same way as well (replacing $p$
by $a$, of course) and thus with the same multiplicities $d_j^0$, and
therefore our construction of the generalized eigenstates should be
the same but using $a$ instead of $p$, i.e., the $p'$ in the
$p'$-string takes values from 1 to $a-1$, etc.  For even $b$ (or
$q^a=1$ and odd $a$), a more careful analysis is required.  
According to~\cite{Chari:1994pz}, for the case of $q^a=1$, the tilting modules have the same structure as in Sec.~\ref{sec:Tj-str}, where
 one should
again replace $p$ by $a$, and the multiplicities in the tensor
products are also identical to what we had here.  The only real
difference will be in the values of the Bethe roots, as the spectrum
of the Hamiltonian is different for different choices of $a$ and $b$,
and thus the continuum limit too. 
We also expect that similar
constructions can be used for quantum-group invariant spin chains at
roots of unity with higher spin and/or rank of the quantum-group
symmetry. It would be interesting to consider 
similar constructions for supersymmetric ($\mathbb{Z}_2$-graded) spin chains, 
such as the $U_q sl(2|1)$-invariant chain \cite{Foerster:1993fp}.
Of course, the algebraic Bethe ansatz would require nesting for rank greater than one,
which would render the corresponding constructions more complicated.

We are currently investigating the symmetry operators -- generators of a non-abelian symmetry of the transfer-matrix $t(u)$ -- responsible for
the higher degeneracies of the model, which are signaled by the
appearance of continuous solutions of the Bethe equations, whose
corresponding eigenvectors are obtained by the construction of section
\ref{sec:pstrings}. It would also be interesting to find a
group-theoretic understanding of the construction in section 
\ref{sec:generalized} of generalized eigenvectors,
e.g. within the context
of the quantum affine algebra $U_{q} \widehat{sl}(2)$ or rather its
coideal $q$-Onsager subalgebra at roots of unity \cite{Baseilhac:2016}.

\section*{Acknowledgments}
AMG  thanks Hubert Saleur for valuable discussions, and the IPhT in 
Saclay for its hospitality.
The work of AMG was supported by DESY and  Humboldt fellowships, C.N.R.S. and RFBR-grant 13-01-00386.
RN thanks Rodrigo Pimenta and Vitaly Tarasov
for valuable discussions, and the DESY 
theory group for its hospitality.
The work of RN was supported in part by the National Science
Foundation under Grant PHY-1212337, and by a Cooper fellowship.

\appendix

\section{Tilting $\qg$-modules at roots of unity}\label{app:proj-mod-base}
We explicitly describe here the $\qg$ action in the tilting modules $T_j$ for $q=e^{i\pi/p}$ and integer $p\geq 2$. For $2j+1\leq p$, these modules are irreducible  
of dimension $2j+1=s(j)\equiv s$, recall our convention~\eqref{sj}, 
and have the basis 
$\{\stprp_{n},\, 0\leq n\leq s{-}1\}$
where  $\stprp_{0}$ is the highest-weight
vector and 
the action  is 
\begin{align}
  K \stprp_{n} &=
   \q^{s - 1 - 2n} \stprp_{n},\qquad
  &h\, \stprp_{n} &= 0,\label{basis-lusz-irrep-r1-1}\\
  E \stprp_{n} &=
   [n]_q[s - n]_q\, \stprp_{n - 1},\qquad
  &e\, \stprp_{n} &=  0,\label{basis-lusz-irrep-r1-2}\\
  F \stprp_{n} &= \stprp_{n + 1},\qquad
  &f\, \stprp_{n} &= 0,\label{basis-lusz-irrep-r1-3}
\end{align}
where we set $\stprp_{-1}=\stprp_{s}=0$.

For $2j+1>p$, the $T_j$'s are identified\footnote{The identification is easy to see using the diagram~\eqref{Tj-diag} with the formula for dimensions~\eqref{dimj} and the general decomposition of the spin-chain over $\qg$ in terms of projective covers in~\cite[Sec. 3.2]{Gainutdinov:2012mr}.} with projective $\qg$-modules from~\cite{Bushlanov:2009cv} denoted there by
 $\PP^{\alpha}_{p-s(j),r}$ with  $s\equiv s(j)$ and $r$ defined from the equation  $2j+1 = rp + s(j)$, i.e., $s$ an integer $1\leq s\leq p-1$ and
$r\geq1$, and $\alpha\equiv \alpha(r) = (-1)^{r-1}$.  
Using the identification and the known basis and action~\cite{Bushlanov:2009cv}  in~ $\PP^{\alpha}_{p-s(j),r}$, we give below the $\qg$-action in $T_j$'s.

For $r > 1$ and $2j+1$ is not zero modulo $p$,  
$T_{j}$ has the basis
\begin{equation}\label{left-proj-basis-plus}
  \{\toppr_{n,m},\botpr_{n,m}\}_{\substack{0\le n\le p-s-1\\0\le m\le r-1}}
  \cup\{\leftpr_{k,l}\}_{\substack{0\le k\le s-1\\0\le l\le
  r-2}}
  \cup\{\rightpr_{k,l}\}_{\substack{0\le k\le s-1\\0\le l\le
  r}},
\end{equation}
where $\{\toppr_{n,m}\}_{\substack{0\le n\le p-s-1\\0\le m\le
    r-1}}$ is the basis
in to the top module ${\bf T}_{j}$ in~\eqref{Tj-diag},
$\{\botpr_{n,m}\}_{\substack{0\le n\le p-s-1\\0\le m\le r-1}}$
is the basis in the bottom ${\bf B}_{j}$, $\{\leftpr_{k,l}\}_{\substack{0\le k\le
    s-1\\0\le l\le r-2}}$ is the basis in the left ${\bf L}_{j}$, and
$\{\rightpr_k\}_{\substack{0\le k\le s-1\\0\le l\le r}}$ is the basis in
the right module ${\bf R}_{j}$. In thus introduced basis, the $s\ell(2)$-generators $e$, $f$ and $h$ of $\qg$
act in ${\bf T}_{j}$ as in the $r$-dimensional $s\ell(2)$-module:
\begin{align}
&h\, \toppr_{n,m} =  \half(r-1-2m)\toppr_{n,m},
&e\, \toppr_{n,m} =  m(r-m)\toppr_{n,m-1},\qquad
&f\, \toppr_{n,m} =  \toppr_{n,m+1}
\end{align}
where we set $\toppr_{n,-1}=\toppr_{n,r}=0$, and identically in ${\bf B}_{j}$, while for ${\bf R}_{j}$ the action is
\begin{align}\label{sl2-act-Rj}
&h\, \rightpr_{k,l} =  \half(r-2l)\rightpr_{k,l},
&e\, \rightpr_{k,l} =  l(r+1-l)\rightpr_{k,l-1},\qquad
&f\, \rightpr_{k,l} =  \rightpr_{k,l+1}
\end{align}
where we set $\rightpr_{n,-1}=\rightpr_{n,r+1}=0$, and identically in ${\bf L}_{j}$ but with the replacement of $r$ by $r-2$ in~\eqref{sl2-act-Rj}.
 The $\qg$-action of the three other generators $E$, $F$, and $K$ in the basis~\eqref{left-proj-basis-plus} is given by
{\allowdisplaybreaks 
\begin{align*}
  K\toppr_{n,m}&=\alpha\q^{p-s-1-2n}\, \toppr_{n,m}, \quad 0\le n\le p-s-1,\quad 0\le m\le r-1,\\
  K\leftpr_{k,m}&=-\alpha\q^{s-1-2k}\, \leftpr_{k,m}, \quad 0\le k\le s-1,\quad 0\le m\le r-2,\\
  K\rightpr_{k,m}&=-\alpha\q^{s-1-2k}\, \rightpr_{k,m}, \quad 0\le k\le s-1,\quad 0\le m\le r,\\
  K\botpr_{n,m}&=\alpha\q^{p-s-1-2n}\, \botpr_{n,m}, \quad 0\le n\le p-s-1,\quad 0\le m\le r-1,\\
    F\toppr_{n,m}&=
  \begin{cases}
    \toppr_{n+1,m}, &0\le n\le p-s-2,\\
    \frac{1}{r}\rightpr_{0,m+1}-\frac{1}{r}\leftpr_{0,m}, & n=p-s-1 
    \quad(\leftpr_{0,r-1}\equiv0),\\
  \end{cases}
   \quad 0\le m\le r-1,\\
  F\leftpr_{k,m}&=
  \begin{cases}
    \leftpr_{k+1,m}, &0\le k\le s-2,\\
    \botpr_{0,m+1}, & k=s-1,\\
  \end{cases}
  \quad 0\le m\le r-2,\\
  F\rightpr_{k,m}&=
  \begin{cases}
    \rightpr_{k+1,m}, & 0\le k\le s-2,\\
    \botpr_{0,m}, & k=s-1\quad (\botpr_{0,r}\equiv0),\\
  \end{cases}
  \quad 0\le m\le r,\\
  F\botpr_{n,m}&= \botpr_{n+1,m}, \quad 1\le n\le p-s-1,
  \quad 0\le m\le r-1 \quad (\botpr_{p-s,m}\equiv 0),\\
  E\toppr_{n,m}&=
  \begin{cases}
    \alpha[n]_q[p-s-n]_q\, \toppr_{n-1,m}+ \alpha g\, \botpr_{n-1,m}, &1\le n\le p-s-1,\\
    \alpha g\bigl(\frac{r-m}{r}\rightpr_{s-1,m}+\frac{m}{r}\leftpr_{s-1,m-1}\bigr), & n=0,\\
  \end{cases}
  \quad 0\le m\le r-1,\\
  E\leftpr_{k,m}&=
  \begin{cases}
	-\alpha[k]_q[s-k]_q\, \leftpr_{k-1,m}, &1\le k\le s-1, \\
    \alpha g(m-r+1)\botpr_{p-s-1,m}, & k=0,\\
  \end{cases}
  \quad 0\le m\le r-2,\\
  E\rightpr_{k,m}&=
  \begin{cases}
    -\alpha[k]_q[s-k]_q\, \rightpr_{k-1,m}, &1\le k\le s-1,\\
    \alpha g m\, \botpr_{p-s-1,m-1}, & k=0,\\
  \end{cases}
  \quad 0\le m\le r,\\
  E\botpr_{n,m}&= \alpha[n]_q[p-s-n]_q\, \botpr_{n-1,m}, \quad 1\le n\le
  p-s-1, \quad 0\le m\le r-1 \quad (\botpr_{-1,m}\equiv 0),
\end{align*}}
where $g=\frac{(-1)^{p}[s]_q}{[p-1]_q!}$.
For $r=1$, the basis~\eqref{left-proj-basis-plus} does not contain
$\{\leftpr_{k,l}\}_{\substack{0\le k\le p-s-1\\0\le l\le
r-2}}$ terms and we imply $\leftpr_{k,l}~\equiv~0$ in the
action. Then, the formulas for the action are the same as above.

\section{Large $p'$-strings}\label{sec:numerics}

We provide here some numerical evidence that the Bethe equations 
(\ref{BAE}) have solutions of the form (\ref{pprimestring1}) i.e. 
\be
\nu_{k}^{\infty} = \nu_{0} + \frac{i \pi}{2p'}(p'-(2k-1)) \,,  \qquad
k=1, \ldots, p'  \,, \qquad p' = s(j)  \,,
\label{pprimestring1again}
\ee
with $\nu_{0} \rightarrow \infty$ as $\eta \rightarrow \eta_{0} = i \pi/p$ 
with integer $p \ge 2$; and that the corresponding transfer-matrix eigenvalues become 
degenerate in this limit. Such ``large $p'$-string'' solutions play a key role in 
the construction described in section \ref{sec:generalized} of generalized 
eigenvectors. For convenience, in this section we set 
$\eta = i \pi/p$ with $p$ {\em real}, and we study the limit that $p$ 
approaches an integer.

\subsection{$p=3\,, \quad p'=1$}

Let us consider the case $N=6$. For $p=3$, we know 
\cite[Table 4(b)]{Gainutdinov:2015vba} that the transfer-matrix eigenvalue corresponding to the 
reference state ($M=0, j=3$ ) has degeneracy $12$; while away from $p=3$, we 
find that this degeneracy splits into $7+5$. In view of 
(\ref{dimVj}) describing the transfer-matrix degeneracy, the 
corresponding two solutions of the Bethe equations must have $M=0$ and 
$M=1$, respectively. The latter solution is our $p'$-string with 
$p'=s(3)=1$. As $p$ approaches 3, this real Bethe root becomes 
 ``large'' i.e. tends to infinity. 
This solution corresponds to the generalized eigenvector $\vecG{1}{-}$ in the 
tilting module $T_{3}$ discussed in sections \ref{sec:genp3pp1} and \ref{sec:p3N6}.

\subsection{$p=3\,, \quad p'=2$}

\begin{figure}
\centering
\subfloat[]{\includegraphics[width=5.5cm]{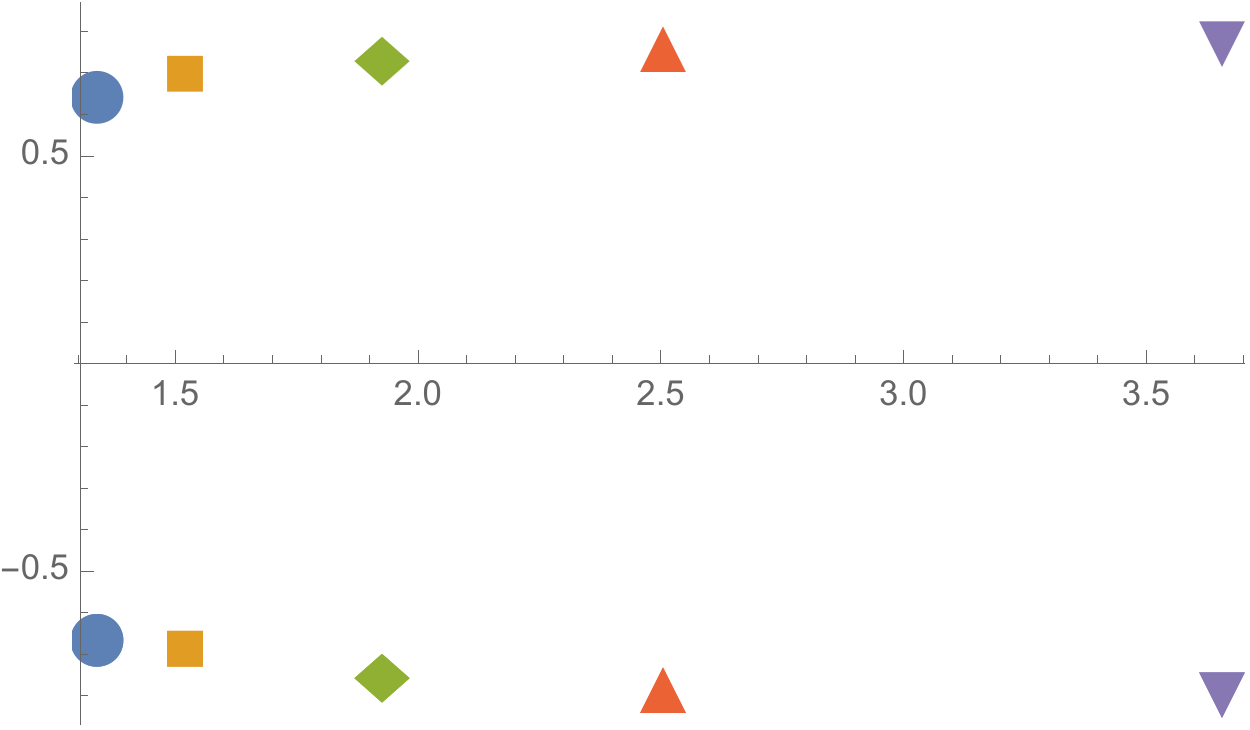}}
\hspace{2cm}
\subfloat[]{\includegraphics[width=5.5cm]{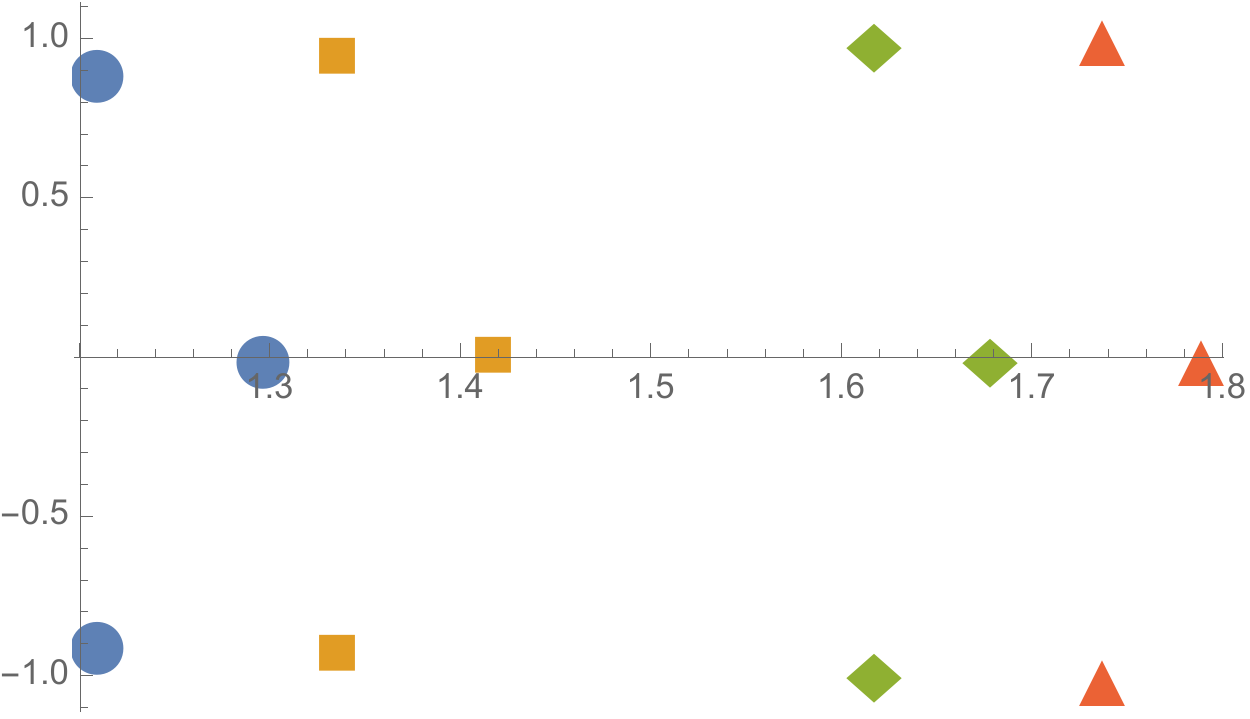}}
\caption{(a) 2-strings for $N=4$ and $p=3.1 (\bullet), 3.05 
({\scriptscriptstyle\blacksquare}), 3.01 ({\scriptstyle\blacklozenge}), 
3.001 ({\scriptstyle\blacktriangle}), 3.00001 ({\scriptstyle\blacktriangledown})$ 
(b) 3-strings for $N=6$ and $p=4.1 (\bullet), 4.05 ({\scriptscriptstyle\blacksquare}), 
4.01 ({\scriptstyle\blacklozenge}), 4.005 ({\scriptstyle\blacktriangle})$} 
\label{fig:strings}
\end{figure}

Let us first consider the case $N=4$. For $p=3$, we know 
\cite[Table 4(b)]{Gainutdinov:2015vba} that the transfer-matrix eigenvalue corresponding to the 
reference state ($M=0, j=2$) has degeneracy~$6$; while away from $p=3$, we 
find that this degeneracy splits into $5+1$. In view of 
(\ref{dimVj}), the corresponding solutions of the Bethe equations must have $M=0$ and 
$M=2$, respectively. The latter solution is our $p'$-string with 
$p'= s(2) = 2$. Figure \ref{fig:strings}(a) shows a plot in the complex plane of the latter solution for values of 
$p$ near 3. We observe that, as $p$ approaches 3, the real part 
increases, and the imaginary parts approach $\pm \pi/4$. 
This solution
corresponds to the generalized eigenvector $\vecG{2}{-}$
in the tilting module $T_{2}$  discussed in sections~\ref{sec:genp3pp2} and~\ref{sec:p3N4}.

Let us next consider the case $N=6$. For $p=3$, we know 
\cite[Table 3(b)]{Gainutdinov:2015vba} that there 
are 4 transfer-matrix eigenvalues corresponding to solutions of the 
Bethe equations with  $M=1$, $j=2$, each of which has degeneracy 6.
Away from $p=3$, we find that this degeneracy splits into $5+1$. 
In view of (\ref{dimVj}), the corresponding solutions of the Bethe 
equations must have $M=1$ and 
$M=3$, respectively. We indeed find 4 solutions with $M=3$ 
that consist of a real root and a 2-string, such that, as $p$ approaches 3,
the real root remains small, the center of the 2-string becomes 
large, and the imaginary parts of the 2-string approach $\pm \pi/4$. 
These solutions correspond to the generalized eigenvector $\vecG{2}{\lambda}$
with $p'= s(2) = 2$
in the tilting module $T_{2}$ discussed in sections 
\ref{sec:genp3pp2} and \ref{sec:p3N6}.

\subsection{$p=4\,, \quad p'=1$}

Let us consider the case $N=4$. For $p=4$, we know 
\cite[Table 4(c)]{Gainutdinov:2015vba} that the transfer-matrix eigenvalue corresponding to the 
reference state ($M=0, j=2$) has degeneracy 8; while away from $p=4$, we 
find that this degeneracy splits into $5+3$. In view of 
(\ref{dimVj}), the corresponding solutions of the Bethe equations must have $M=0$ and 
$M=1$, respectively. The latter solution is our $p'$-string with 
$p'=s(2)=1$. As $p$ approaches 4, this real Bethe root becomes large. 
This solution
corresponds to the generalized eigenvector $\vecG{1}{-}$ in the 
tilting module $T_{2}$ discussed in sections \ref{sec:genp4pp1} and \ref{sec:p4N4}.

\subsection{$p=4\,, \quad p'=3$}

Let us consider the case $N=6$. For $p=4$, we know  
\cite[Table 4(c)]{Gainutdinov:2015vba} that the transfer-matrix eigenvalue corresponding to the 
reference state  ($M=0$, $j=3$) has degeneracy 8; while away from $p=4$, we 
find that this degeneracy splits into $7+1$. In view of 
(\ref{dimVj}), the corresponding solutions of the Bethe equations must have $M=0$ and 
$M=3$, respectively. The latter solution is our $p'$-string with 
$p'=s(3)=3$. Figure \ref{fig:strings}(b) shows a plot in the complex plane of the latter solution for values of 
$p$ near 4. We observe that, as $p$ approaches 4, the real part becomes 
large, and the nonzero imaginary parts approach $\pm \pi/3$. 
This solution
corresponds to the generalized eigenvector $\vecG{3}{-}$ in the 
tilting module $T_{3}$  discussed in sections~\ref{sec:genp4pp3} and~\ref{sec:p4N6}.

\section{Special off-shell relation}\label{sec:special}

We derive here an off-shell relation for Bethe vectors of the special
form $B(u) \prod_{j} B(v_{j}) \vac$ (i.e., with an ``extra''
factor $B(u)$, whose argument is the same as that of the transfer
matrix $t(u)$), 
which we need in Appendix \ref{sec:proof} to
derive an off-shell relation for generalized eigenvectors.
The proof is a generalization of the one developed by Izergin and Korepin 
\cite{Izergin:1982hy} for repeated Bethe roots.

We begin by recalling the basic exchange relations 
\cite{Sklyanin:1988yz}  that are needed to derive the 
usual off-shell relation (\ref{offshell}) 
\be
A(u)\, B(v) &=& f(u,v)\, B(v)\, A(u) + g(u,v)\, B(u)\, A(v) + w(u,v)\, 
B(u)\, D(v)\,, 
\label{ABexchange}\\
D(u)\, B(v) &=& h(u,v)\, B(v)\, D(u) + k(u,v)\, B(u)\, D(v) + n(u,v)\, 
B(u)\, A(v)\,, 
\label{DBexchange}
\ee
where
\be
f(u,v) &=& \frac{\sh(u-v-\eta) \sh(u+v)}{\sh(u-v) \sh(u+v+\eta)} 
\equiv \frac{\tilde f(u,v)}{\sh(u-v)} \,, \non \\
g(u,v) &=& \frac{\sh\eta \sh(2v)}{\sh(u-v) \sh(2v+\eta)} 
\equiv \frac{\tilde g(u,v)}{\sh(u-v)} \,, \non \\
w(u,v) &=& -\frac{\sh\eta}{\sh(u+v+\eta)}\,, 
\label{fgw}
\ee
and 
\be
h(u,v) &=& \frac{\sh(u-v+\eta) \sh(u+v+2\eta) }{\sh(u-v) \sh(u+v+\eta)}
\equiv \frac{\tilde h(u,v)}{\sh(u-v)}\,, \non \\
k(u,v) &=& -\frac{\sh\eta \sh(2u+2\eta)}{\sh(u-v) \sh(2u+\eta)} 
\equiv \frac{\tilde k(u,v)}{\sh(u-v)} \,, \non \\
n(u,v)  &=& \frac{\sh\eta \sh(2u+2\eta) \sh(2v)}{\sh(2u+\eta) 
\sh(2v+\eta) \sh(u+v+\eta)}\,.
\label{hkn}
\ee

The entire difficulty stems from the fact that the exchange relations
(\ref{ABexchange}), (\ref{DBexchange}) become singular when the two
spectral parameters coincide.  (See (\ref{fgw}) and (\ref{hkn}).)
We can nevertheless derive regular
exchange relations at $u=v$ by multiplying both sides of the exchange
relations by $\sh(u-v)$, differentiating with respect to $v$, and then
letting $u \rightarrow v$. In this way, we arrive at
\be
A(v)\, B(v) &=& \varphi(v) B(v)\, A(v) + \frac{\sh \eta 
\sh(2v)}{\sh(2v+\eta)}\left(B'(v)\, A(v) - B(v)\, A'(v) \right) \non 
\\
&&-\frac{\sh \eta}{\sh(2v+\eta)} B(v)\, D(v) \,,
\label{ABspecial}
\ee
where
\be
\varphi(v) = - \frac{\partial}{\partial v}\left[  \tilde f(u,v) +  
\tilde g(u,v) \right]\Big\vert_{u=v}  \,,
\ee
and 
\be
D(v)\, B(v) &=& \psi(v) B(v)\, D(v) + \frac{\sh \eta 
\sh(2v+2\eta)}{\sh(2v+\eta)}\left(B(v)\, D'(v) - B'(v)\, D(v) \right) \non 
\\
&&+\frac{\sh \eta \sh(2v) \sh(2v+2\eta)}{\sh^{3}(2v+\eta)} B(v)\, A(v) \,,
\label{DBspecial}
\ee
where
\be
\psi(v) = - \frac{\partial}{\partial v}\left[  \tilde h(u,v) +  
\tilde k(u,v) \right]\Big\vert_{u=v}  \,.
\ee

The transfer matrix (\ref{transfer}) can be reexpressed as 
\be
t(u) = a(u)\, A(u) + d(u)\, D(u) \,,
\label{transfer2}
\ee
where 
\be
a(u) = e^{-u}\frac{\sh(2u+2\eta)}{\sh(2u+\eta)}\,, \qquad
d(u) = e^{u+\eta} \,.
\ee
The reference state (\ref{reference}) is an eigenstate of $A(u)$ and $D(u)$,
\be
A(u) \vac = \alpha(u) \vac \,, \qquad
D(u) \vac = \delta(u) \vac \,,
\ee
where the corresponding eigenvalues are given by
\be
\alpha(u) = e^{u} \sh^{2N}(u+\eta)  \,, \qquad
\delta(u) = e^{-u-\eta} \frac{\sh(2u)}{\sh(2u+\eta)}\sh^{2N}u \,.
\ee

The action of $a(u) A(u)$ on the vector $B(u)  
\prod_{j} B(v_{j}) \vac$ produces three types of terms (instead 
of the usual two)
\be
a(u) A(u) \Big[B(u)  
\prod_{j} B(v_{j}) \vac \Big] &=& \Gamma^{(0)}(u)\, B(u)  
\prod_{j} B(v_{j}) \vac + B^{2}(u) \sum_{l} \Gamma^{(1)} 
_{l}(u)\, 
\prod_{j \ne l} B(v_{j}) \vac \non \\
& &+ \Gamma^{(2)}(u)\, B'(u)  
\prod_{j} B(v_{j}) \vac \,.
\label{Aspecial}
\ee
The coefficient $\Gamma^{(0)}(u)$ is obtained using the first, third
and fourth terms in (\ref{ABspecial}) and then the first term in
(\ref{ABexchange}) or (\ref{DBexchange}), yielding
\be
\Gamma^{(0)}(u) &=& a(u)\Bigg\{ \varphi(u) \alpha(u) \prod_{j}f(u,v_{j}) 
-\frac{\sh \eta}{\sh(2u+\eta)}\delta(u) \prod_{j}h(u,v_{j}) \non \\
&&- \frac{\sh \eta \sh(2u)}{\sh(2u+\eta)} 
\frac{\partial}{\partial u}\Big[\alpha(u) \prod_{j}f(u,v_{j}) 
\Big] \Bigg\}
\,.
\ee
For $\Gamma^{(1)}_{l}(u)$, we rewrite the vector as 
$B(v_{l}) \left[\prod_{j\ne l} B(v_{j}) \right] B(u) \vac$; 
using the second and third terms in (\ref{ABexchange}) for 
$A(u) B(v_{l})$, and then using exclusively the first terms in the exchange 
relations, we obtain
\be
\Gamma^{(1)}_{l}(u) &=& a(u) \Bigg\{g(u,v_{l}) 
\Big[\prod_{j\ne l} f(v_{l}, v_{j}) \Big] f(v_{l},u) \alpha(v_{l}) 
\non \\
&& + w(u,v_{l}) 
\Big[\prod_{j\ne l} h(v_{l}, v_{j}) \Big] h(v_{l},u) \delta(v_{l}) 
\Bigg\}
\,.
\ee
Finally, with the help of (\ref{ABspecial}), we readily obtain
\be
\Gamma^{(2)}(u) = a(u) \frac{\sh \eta \sh(2u)}{\sh(2u+\eta)} 
\alpha(u) \prod_{j}f(u, v_{j}) \,.
\ee

Similarly, acting with $d(u) D(u)$ also generates three terms
\be
d(u) D(u) \Big[B(u)  
\prod_{j} B(v_{j}) \vac \Big] &=& \Upsilon^{(0)}(u)\, B(u)  
\prod_{j} B(v_{j}) \vac + B^{2}(u) \sum_{l} \Upsilon^{(1)} 
_{l}(u)\, 
\prod_{j \ne l} B(v_{j}) \vac \non \\
& &+ \Upsilon^{(2)}(u)\, B'(u)  
\prod_{j} B(v_{j}) \vac \,,
\label{Dspecial}
\ee
with
\be
\Upsilon^{(0)}(u) &=& d(u) \Bigg\{ \psi(u)\delta(u) \prod_{j}h(u,v_{j}) 
+ \frac{\sh \eta \sh(2u)\sh(2u+2\eta)}{\sh^{3}(2u+\eta)}\alpha(u) 
\prod_{j}f(u,v_{j}) \non\\
&&+\frac{\sh \eta \sh(2u+2\eta)}{\sh(2u+\eta)} 
\frac{\partial}{\partial u}\Big[\delta(u) \prod_{j}h(u,v_{j}) \Big] \Bigg\}
\,, \non \\
\Upsilon^{(1)}_{l}(u) &=& d(u)  \Bigg\{ k(u,v_{l}) 
\Big[\prod_{j\ne l} h(v_{l}, v_{j}) \Big] h(v_{l},u) \delta(v_{l}) 
+ n(u,v_{l}) 
\Big[\prod_{j\ne l} f(v_{l}, v_{j}) \Big] f(v_{l},u) 
\alpha(v_{l})  \Bigg\}\,, \non \\
\Upsilon^{(2)}(u) &=& - d(u) \frac{\sh \eta 
\sh(2u+2\eta)}{\sh(2u+\eta)} \delta(u)  \prod_{j}h(u,v_{j}) \,.
\ee

Combining the results (\ref{transfer2}), (\ref{Aspecial}), (\ref{Dspecial}), 
we arrive at the desired off-shell relation for the transfer matrix
\be
t(u) \Big[B(u)  
\prod_{j} B(v_{j}) \vac \Big] &=& \Lambda^{(0)}(u)\, B(u)  
\prod_{j} B(v_{j}) \vac + B^{2}(u) \sum_{l} \Lambda^{(1)} 
_{l}(u)\, 
\prod_{j \ne l} B(v_{j}) \vac \non \\
& &+ \Lambda^{(2)}(u)\, B'(u)  
\prod_{j} B(v_{j}) \vac \,,
\label{specialoffshell}
\ee
where
\be
\Lambda^{(0)}(u) &=&\Gamma^{(0)}(u) + \Upsilon^{(0)}(u) \,, \non \\
\Lambda^{(1)}_{l}(u) &=&\Gamma^{(1)}_{l}(u) + \Upsilon^{(1)}_{l}(u) \,, \non \\
\Lambda^{(2)}(u) &=&\Gamma^{(2)}(u) + \Upsilon^{(2)}(u) \,.
\ee
After some algebra, we find
\be
\Lambda^{(0)}(u) &=& \sh^{2N}(u+\eta) \Big\{c_{1}(u) \prod_{j}f(u,v_{j})
 - c_{2}(u) 
\frac{\partial}{\partial u}\big[ \prod_{j}f(u,v_{j}) \big] 
\Big\} \non \\
&&+ \sh^{2N}(u) \Big\{c_{3}(u)  \prod_{j}h(u,v_{j}) + c_{2}(u) 
\frac{\partial}{\partial u}\big[\prod_{j}h(u,v_{j}) \big] 
\Big\} \,,
\ee
where
\be
c_{1}(u) &=& \frac{1}{\sh^{3}(2u+\eta)}\big[\sh(2u+2\eta) \sh^{2}(2u+\eta) - 2 \sh^{2}\eta 
\sh(2u+2\eta) \non \\
&&- 4 N \ch^{2}(u+\eta) \sh \eta \sh(2u) \sh(2u+
\eta)\big] \,, \non \\
c_{2}(u) &=& \frac{1}{\sh^{2}(2u+\eta)} \sh \eta \sh(2u) \sh(2u+2\eta) \,, \non \\
c_{3}(u) &=& \frac{1}{\sh^{3}(2u+\eta)}\big[\sh(2u) \sh^{2}(2u+\eta)+ 
2 \sh^{2}\eta \sh(2u+2\eta) \non \\
&&+ 4 N \ch^{2}u \sh \eta  \sh(2u+\eta) 
\sh(2u+2\eta)] \,.
\ee
Moreover,
\be
\Lambda^{(1)}_{l}(u) &=&\ef(u,v_{l}) \Big[\sh^{2N}(v_{l}+\eta) 
f(v_{l},u) \prod_{j\ne l}f(v_{l},v_{j})  \non \\
&&-\sh^{2N}(v_{l})  h(v_{l},u) \prod_{j\ne l}h(v_{l},v_{j}) \Big] \,,
\ee
where $\ef(u,v)$ is defined in (\ref{fuv}); and finally,
\be
\Lambda^{(2)}(u) &=& \frac{\sh \eta \sh(2u) 
\sh(2u+2\eta)}{\sh^{2}(2u+\eta)}\Big[\sh^{2N}(u+\eta) 
\prod_{j}f(u,v_{j}) \non \\
&&- \sh^{2N}(u) \prod_{j}h(u,v_{j}) \Big] \,.
\ee

\section{Off-shell relation for a generalized eigenvector}\label{sec:proof}

We derive here an off-shell relation for the vector
$\vecG{p'}{\vec\lambda}$ (\ref{genvec}), which leads 
to the set (\ref{suffcond1I})-(\ref{suffcond5I})
of sufficient conditions for this vector to be a generalized 
eigenvector of the transfer matrix.

Since $\vecG{p'}{\vec\lambda}$ should be a 
generalized eigenvector of the transfer matrix $t(u)$ of rank 2, we proceed 
to compute the action of $(t(u) - \Lambda_{\alpha}(u))^{2}$ on the 
off-shell vector $\vecGomega{p'}{\vec\lambda}$: 
\be
\left(t(u) - \Lambda_{\alpha}(u)\right)^{2} \vecGomega{p'}{\vec\lambda} =
t(u)^{2} \vecGomega{p'}{\vec\lambda} 
-2  \Lambda_{\alpha}(u) t(u) \vecGomega{p'}{\vec\lambda}
+ \Lambda_{\alpha}(u)^{2}  \vecGomega{p'}{\vec\lambda}
\,. \label{threeterms}
\ee
We now evaluate in turn the three terms on the RHS of (\ref{threeterms}). We begin with the
first term, which is the most difficult, since it requires a
nontrivial step.  Using (\ref{vecGommega}) and the off-shell relations
(\ref{offshellalpha}), (\ref{offshellbeta}), we obtain 
\be
t(u)^{2} \vecGomega{p'}{\vec\lambda} &=&
t(u) \left[ \alpha t(u) |\vec\nu, \vec\lambda_{\alpha}\rangle 
+ \beta t(u) F^{p'} |\vec\lambda_{\beta}\rangle
\right] \non \\
&=& t(u) \Bigg[ \alpha \Lambda_{\alpha}(u) |\vec\nu, \vec\lambda_{\alpha}\rangle
+ \alpha \sum_{i}\Lambda^{\nu_i}(u)\, B(u) | \hat\nu_{i},  \vec\lambda_{\alpha}\rangle
+  \alpha \sum_{i}\Lambda^{\lambda_{\alpha, i}}(u)\, B(u) | \vec\nu,  
\hat\lambda_{\alpha,i}\rangle \non\\
& & \qquad + \beta F^{p'} \Lambda_{\beta}(u) |\vec\lambda_{\beta}\rangle
+ \beta F^{p'} \sum_{i}\Lambda^{\lambda_{\beta, i}}(u)\,  B(u) |\hat\lambda_{\beta, i}\rangle
\Bigg] = \ldots
\label{stage1}
\ee
In passing to the second line, we have made use of the important fact (\ref{qgsymm}) that the 
transfer matrix commutes with $F$. Continuing the calculation, we obtain
\be
\hspace{-0.2in}\ldots
&=& \alpha \Lambda_{\alpha}(u) \left[ \Lambda_{\alpha}(u) |\vec\nu, \vec\lambda_{\alpha}\rangle
+ \sum_{i}\Lambda^{\nu_i}(u)\, B(u) | \hat\nu_{i},  \vec\lambda_{\alpha}\rangle
+ \sum_{i}\Lambda^{\lambda_{\alpha, i}}(u)\, B(u) | \vec\nu,  
\hat\lambda_{\alpha,i}\rangle \right]  \non\\
& & + \alpha \sum_{i} \Lambda^{\nu_i}(u)\, t(u)  B(u) | \hat\nu_{i}, \vec\lambda_{\alpha}\rangle
+  \alpha \sum_{i}\Lambda^{\lambda_{\alpha, i}}(u)\, t(u) B(u) | \vec\nu, \hat\lambda_{\alpha,i}\rangle \non\\
& & + \beta F^{p'} \Lambda_{\beta}(u) \left[ \Lambda_{\beta}(u) |\vec\lambda_{\beta}\rangle
+ \sum_{i}\Lambda^{\lambda_{\beta, i}}(u)\,  B(u) |\hat\lambda_{\beta, i}\rangle 
\right]  \non\\
& & + \beta F^{p'} \sum_{i}\Lambda^{\lambda_{\beta, i}}(u)\,  t(u) B(u) |\hat\lambda_{\beta, i}\rangle
 = \ldots 
\label{stage2}
\ee 
Now comes the nontrivial step, when we evaluate the action of $t(u)$ 
on $B(u) |\cdots \rangle$ in three terms in (\ref{stage2}). Indeed, the off-shell relation  (\ref{offshell}) 
cannot be applied, since its derivation assumes that the argument of 
the transfer matrix (namely, $u$) does not 
coincide with any of the arguments of the $B$ operators used to  
construct the vector, which evidently is not the case for the vectors $B(u) |\cdots \rangle$.
We use instead the special off-shell relation 
(\ref{specialoffshell}), which we rewrite in a more condensed form here as
\be
t(u) B(u) | \vec \mu \rangle = \tilde\Lambda^{\mu}(u) B(u) |\vec \mu  \rangle  + 
\sum_{i} \tilde{\tilde\Lambda}^{\mu, \mu_{i}}(u)\, B^{2}(u) |\hat \mu_{i} \rangle + 
\tilde{\tilde{\tilde\Lambda}}^{\mu}(u) B'(u) | \vec \mu \rangle\,,
\ee
where $\vec\mu$ can be $\vec\nu, \vec\lambda_{\alpha},  
\vec\lambda_{\beta}$, etc.
Fortunately, we shall not need the explicit expressions for the 
coefficients $\tilde\Lambda^{\mu}(u)$, $\tilde{\tilde\Lambda}^{\mu, \mu_{i}}(u)$ 
and $\tilde{\tilde{\tilde\Lambda}}^{\mu}(u)$, which 
can be deduced from the results in Appendix~\ref{sec:special}.

Continuing the calculation from the point (\ref{stage2}), we conclude 
that 
{\allowdisplaybreaks 
\be
\lefteqn{t(u)^{2} \vecGomega{p'}{\vec\lambda}
=  \alpha \Lambda_{\alpha}(u)^{2} |\vec\nu, 
\vec\lambda_{\alpha}\rangle} \non \\
&& + \alpha \Lambda_{\alpha}(u) \sum_{i} \Lambda^{\nu_i}(u)\, 
B(u) | \hat\nu_{i},  \vec\lambda_{\alpha}\rangle 
+\alpha \Lambda_{\alpha}(u)  \sum_{i}\Lambda^{\lambda_{\alpha, i}}(u)\, 
B(u) | \vec\nu,  \hat\lambda_{\alpha,i}\rangle
\non \\
&& +\alpha \sum_{i} \Lambda^{\nu_i}(u) \Bigg[
\tilde\Lambda^{\nu_{i}}(u) B(u) | \hat\nu_{i}, \vec\lambda_{\alpha}\rangle  + 
\sum_{j} \tilde{\tilde\Lambda}^{\nu_{i}, \nu_{j}}(u)\, B^{2}(u) | \hat\nu_{i},  
\hat\nu_{j},\vec\lambda_{\alpha}\rangle \non \\
&&\qquad\qquad + 
\sum_{j} \tilde{\tilde\Lambda}^{\nu_{i}, \lambda_{\alpha, j}}(u)\, B^{2}(u) | \hat\nu_{i},  
\hat\lambda_{\alpha, j}\rangle
+\tilde{\tilde{\tilde\Lambda}}^{\nu_{i}}(u) B'(u) | \hat\nu_{i}, \vec\lambda_{\alpha}\rangle
\Bigg]\non \\
&& +\alpha \sum_{i} \Lambda^{\lambda_{\alpha, i}}(u) \Bigg[
\tilde\Lambda^{\lambda_{\alpha, i}}(u) B(u) | \vec\nu, \hat\lambda_{\alpha,i}\rangle  + 
\sum_{j} \tilde{\tilde\Lambda}^{\lambda_{\alpha, i}, \nu_{j}}(u)\, B^{2}(u) |   
\hat\nu_{j},\hat\lambda_{\alpha,i}\rangle \non \\
&&\qquad\qquad + 
\sum_{j} \tilde{\tilde\Lambda}^{\lambda_{\alpha,i}, \lambda_{\alpha, j}}(u)\, B^{2}(u) | \vec\nu,  
\hat\lambda_{\alpha,i}, \hat\lambda_{\alpha, j}\rangle
+\tilde{\tilde{\tilde\Lambda}}^{\lambda_{\alpha, i}}(u) B'(u) | \vec\nu, \hat\lambda_{\alpha,i}\rangle
\Bigg]\non \\
&& + \beta \Lambda_{\beta}(u)^{2} F^{p'} |\vec\lambda_{\beta}\rangle 
+  \beta \Lambda_{\beta}(u) \sum_{i}\Lambda^{\lambda_{\beta, i}}(u)\,  F^{p'} B(u) |\hat\lambda_{\beta, i}\rangle 
\non \\
&& + \beta F^{p'} \sum_{i}\Lambda^{\lambda_{\beta, i}}(u)\,  \Bigg[
\tilde\Lambda^{\lambda_{\beta, i}}(u) B(u) |\hat\lambda_{\beta, i}\rangle +
\sum_{j} \tilde{\tilde\Lambda}^{\lambda_{\beta, i}, \lambda_{\beta, 
j}}(u)\, B^{2}(u) |\hat\lambda_{\beta, i}, \hat\lambda_{\beta, 
j}\rangle \non\\
&&\qquad\qquad + \tilde{\tilde{\tilde\Lambda}}^{\lambda_{\beta, i}}(u) B'(u) 
|\hat\lambda_{\beta, i}\rangle \Bigg]
\,.
\label{firsttermresult}
\ee
}

The second term in (\ref{threeterms}) is much simpler to evaluate: 
\be
\lefteqn{-2 \Lambda_{\alpha}(u) t(u) \vecGomega{p'}{\vec\lambda}
= -2  \Lambda_{\alpha}(u) \left[ \alpha t(u) |\vec\nu, \vec\lambda_{\alpha}\rangle
+ \beta F^{p'} t(u)  |\vec\lambda_{\beta}\rangle 
\right]} \non \\
&&= -2  \Lambda_{\alpha}(u) \Bigg[ \alpha \Lambda_{\alpha}(u) |\vec\nu, \vec\lambda_{\alpha}\rangle
+ \alpha \sum_{i}\Lambda^{\nu_i}(u)\, B(u) | \hat\nu_{i},  
\vec\lambda_{\alpha}\rangle  \label{secondttermresult}\\
&&+ \alpha \sum_{i}\Lambda^{\lambda_{\alpha, i}}(u)\, B(u) | \vec\nu,  
\hat\lambda_{\alpha,i}\rangle 
+ \beta F^{p'} \Lambda_{\beta}(u) |\vec\lambda_{\beta}\rangle
+ \beta F^{p'} \sum_{i}\Lambda^{\lambda_{\beta, i}}(u)\,  B(u) |\hat\lambda_{\beta, i}\rangle
\Bigg] \,. \non
\ee

Finally, the third term in (\ref{threeterms}) immediately gives 
\be
\Lambda_{\alpha}(u)^{2}  \vecGomega{p'}{\vec\lambda} =
\Lambda_{\alpha}(u)^{2}\left(  \alpha |\vec\nu, \vec\lambda_{\alpha}\rangle + \beta 
F^{p'} |\vec\lambda_{\beta}\rangle \right) \,.
\label{thirdtermresult}
\ee

Collecting all the terms from (\ref{firsttermresult}), (\ref{secondttermresult}), (\ref{thirdtermresult}), 
we finally obtain the desired off-shell relation
{\allowdisplaybreaks 
\be
\lefteqn{\left(t(u) - \Lambda_{\alpha}(u)\right)^{2} 
\vecGomega{p'}{\vec\lambda}  =
\beta \left(\Lambda_{\beta}(u) - \Lambda_{\alpha}(u)\right)^{2} 
F^{p'} |\vec\lambda_{\beta}\rangle} \non\\ 
&&+ \alpha \sum_{i} \Lambda^{\nu_i}(u) \left( 
\tilde\Lambda^{\nu_{i}}(u) -\Lambda_{\alpha}(u) \right) 
B(u) | \hat\nu_{i}, \vec\lambda_{\alpha}\rangle \non \\
&&+ \alpha \sum_{i} \Lambda^{\lambda_{\alpha, i}}(u) \left( 
\tilde\Lambda^{\lambda_{\alpha, i}}(u) -\Lambda_{\alpha}(u) \right) 
B(u)| \vec\nu, \hat\lambda_{\alpha,i}\rangle \non \\
&&+ \beta  \sum_{i}\Lambda^{\lambda_{\beta, i}}(u)\,  
\left[\tilde\Lambda^{\lambda_{\beta, i}}(u) + \Lambda_{\beta}(u) 
-2  \Lambda_{\alpha}(u) \right]
F^{p'} B(u) |\hat\lambda_{\beta, i}\rangle \non\\
&&+ \alpha \sum_{i} \Lambda^{\nu_i}(u) \Bigg[
\sum_{j} \tilde{\tilde\Lambda}^{\nu_{i}, \nu_{j}}(u)\, B^{2}(u) | \hat\nu_{i},  
\hat\nu_{j},\vec\lambda_{\alpha}\rangle 
+ \sum_{j} \tilde{\tilde\Lambda}^{\nu_{i}, \lambda_{\alpha, j}}(u)\, B^{2}(u) | \hat\nu_{i},  
\hat\lambda_{\alpha, j}\rangle\non \\
&&\qquad\qquad 
+\tilde{\tilde{\tilde\Lambda}}^{\nu_{i}}(u) B'(u) | \hat\nu_{i}, \vec\lambda_{\alpha}\rangle
\Bigg]\non \\
&&+ \alpha \sum_{i} \Lambda^{\lambda_{\alpha, i}}(u) \Bigg[
\sum_{j} \tilde{\tilde\Lambda}^{\lambda_{\alpha, i}, \nu_{j}}(u)\, B^{2}(u) |   
\hat\nu_{j},\hat\lambda_{\alpha,i}\rangle 
+ \sum_{j} \tilde{\tilde\Lambda}^{\lambda_{\alpha,i}, \lambda_{\alpha, j}}(u)\, B^{2}(u) | \vec\nu,  
\hat\lambda_{\alpha,i}, \hat\lambda_{\alpha, j}\rangle \non \\
&&\qquad\qquad
+\tilde{\tilde{\tilde\Lambda}}^{\lambda_{\alpha, i}}(u) B'(u) | \vec\nu, \hat\lambda_{\alpha,i}\rangle
\Bigg]\non \\
&&+ \beta F^{p'} \sum_{i}\Lambda^{\lambda_{\beta, i}}(u)\,  \Bigg[
\sum_{j} \tilde{\tilde\Lambda}^{\lambda_{\beta, i}, \lambda_{\beta, 
j}}(u)\, B^{2}(u) |\hat\lambda_{\beta, i}, \hat\lambda_{\beta, 
j}\rangle  + \tilde{\tilde{\tilde\Lambda}}^{\lambda_{\beta, i}}(u) B'(u) 
|\hat\lambda_{\beta, i}\rangle \Bigg] \non\\
&\xrightarrow{\;\omega\to0\;}&  0 \,, 
\label{cond1}
\ee
}
whose RHS we demand to vanish in the limit $\omega \rightarrow 0+$.

Since $\vecG{p'}{\vec\lambda}$ should {\em not} be an 
ordinary eigenvector of the transfer matrix, we also require that
$(t(u) - \Lambda_{\alpha}(u)) \vecGomega{p'}{\vec\lambda} $ should 
{\em not} vanish in the limit $\omega \rightarrow 0+$. This means 
that we also require 
\be
\lefteqn{\left(t(u) - \Lambda_{\alpha}(u)\right) \vecGomega{p'}{\vec\lambda} 
= \left(t(u) - \Lambda_{\alpha}(u)\right) \left( \alpha 
|\vec\nu, \vec\lambda_{\alpha}\rangle 
+ \beta F^{p'} |\vec\lambda_{\beta}\rangle \right)} \non \\
&&= \alpha \left[ \Lambda_{\alpha}(u) |\vec\nu, \vec\lambda_{\alpha}\rangle +
\sum_{i}\Lambda^{\nu_i}(u)\, B(u) | \hat\nu_{i},  \vec\lambda_{\alpha}\rangle
+ \sum_{i}\Lambda^{\lambda_{\alpha, i}}(u)\, B(u) | \vec\nu,  
\hat\lambda_{\alpha,i}\rangle
\right] \non\\
&& \qquad + \beta F^{p'} \left[ \Lambda_{\beta}(u) |\vec\lambda_{\beta}\rangle +
\sum_{i}\Lambda^{\lambda_{\beta, i}}(u)\,  B(u) |\hat\lambda_{\beta, i}\rangle \right]
-  \Lambda_{\alpha}(u) \left( \alpha 
|\vec\nu, \vec\lambda_{\alpha}\rangle 
+ \beta F^{p'} |\vec\lambda_{\beta}\rangle \right) \non \\
&&= \beta \left( 
\Lambda_{\beta}(u)-\Lambda_{\alpha}(u) \right) F^{p'} |\vec\lambda_{\beta}\rangle 
+ \beta \sum_{i}\Lambda^{\lambda_{\beta, i}}(u)\,  F^{p'} B(u) |\hat\lambda_{\beta, i}\rangle
\non \\
&&+\alpha \sum_{i}\Lambda^{\nu_i}(u)\, B(u) | \hat\nu_{i},  \vec\lambda_{\alpha}\rangle
+ \alpha \sum_{i}\Lambda^{\lambda_{\alpha, i}}(u)\, B(u) | \vec\nu,  
\hat\lambda_{\alpha,i}\rangle  \non \\
&\xrightarrow{\;\omega\to0\;}&  |v'\rangle \ne 0 \,,
\label{cond2}
\ee
where $|v'\rangle$ was introduced in (\ref{geneig-2}).

In order to satisfy both conditions (\ref{cond1}) 
and (\ref{cond2}), we conjecture that it suffices to have: 
\be
\lim_{\omega\rightarrow 0+}  \beta \left( \Lambda_{\beta}(u)-\Lambda_{\alpha}(u) \right) &\ne& 0 \,, 
\label{suffcond1} \\
\lim_{\omega\rightarrow 0+}  \beta \left(  \Lambda_{\beta}(u)-\Lambda_{\alpha}(u) \right)^{2} &=& 0 \,, 
\label{suffcond2} \\
\lim_{\omega\rightarrow 0+}  \omega^{2N} \beta \Lambda^{\nu_i}(u)  &=& 0 \,, 
\qquad i = 1,\ldots, p'\,,
\label{suffcond3} \\
\lim_{\omega\rightarrow 0+}  \beta \Lambda^{\lambda_{\alpha, i}}(u)  &=& 0 \,, 
\qquad i = 1,\ldots, M\,,
\label{suffcond4} \\
\lim_{\omega\rightarrow 0+}  \beta \Lambda^{\lambda_{\beta, i}}(u)  &=& 0 \,, 
\qquad i = 1,\ldots, M\,,
\label{suffcond5}
\ee 
where the limit in the first line~\eqref{suffcond1} is supposed to be finite. Indeed, 
the conditions (\ref{suffcond1}), (\ref{suffcond2}) and 
(\ref{suffcond5}) are fairly obvious. The condition (\ref{suffcond4}) 
is less evident, since
it is instead $\alpha \Lambda^{\lambda_{\alpha, i}}(u)$ that appears
in (\ref{cond1}) and (\ref{cond2}).  However, some of the terms with
this factor also
contain the vector $B(u)| \vec\nu, \hat\lambda_{\alpha,i}\rangle$ 
which is of order $\omega^{-2 p' N}$ according to (\ref{cformula}).
Hence, we need $\omega^{-2 p' N} \alpha \Lambda^{\lambda_{\alpha,
i}}(u)$ to vanish as $\omega \rightarrow 0$, which is equivalent to
(\ref{suffcond4}), since $\alpha$ and $\beta$ are given by
(\ref{alphabeta}).\footnote{We have explicitly verified in several
examples that the contributions from the $\tilde\Lambda$ and $\tilde{\tilde{\tilde\Lambda}}$
terms in (\ref{cond1}) are finite in the $\omega \rightarrow 0$ 
limit. Some of the $\tilde{\tilde\Lambda}$ terms in (\ref{cond1}) 
are divergent, as are the vectors that they multiply; nevertheless, the 
total degree of divergence is consistent with (\ref{suffcond3}) and (\ref{suffcond4}).}
The condition (\ref{suffcond3}) has a similar explanation: although 
$\alpha \Lambda^{\nu_i}(u)$ appears in (\ref{cond1}) and (\ref{cond2}), some of the terms with
this factor also contain the vector $| \hat{\nu}_i, \ldots \rangle$, 
which is missing the factor ${\cal B}(\nu_{i})$, and therefore is of order $\omega^{-2 (p'-1) 
N}$. Hence, we require $\omega^{-2 (p'-1) N} \alpha 
\Lambda^{\nu_i}(u)$ to vanish in the limit.

\begin{Cor}\label{cor:vv}
As a corollary of the expression in~\eqref{cond2} and if the
sufficient conditions above are satisfied, the limit of $\left(t(u) -
\Lambda_{\alpha}(u)\right) \vecGomega{p'}{\vec\lambda}$ equal
$\left(t(u) - \Lambda(u)\right) \vecG{p'}{\vec\lambda}$ is non-zero
and proportional to $F^{p'} |\vec\lambda\rangle$.  Indeed, the
proportionality coefficient is~\eqref{suffcond1} and finite non-zero
by the assumption, while the limit of $F^{p'}
|\vec\lambda_{\beta}\rangle$ is $F^{p'} |\vec\lambda\rangle$ and it is
non-zero due to our special choice of $p'=s(j)$ -- it is a state in
the bottom node of the tilting module $T_j$, recall the discussion
just above~\eqref{cformula} and Sec.~\ref{sec:rep-th-descr}.
\end{Cor}


\begin{thebibliography}{10}

\bibitem{Pasquier:1989kd}
V.~Pasquier and H.~Saleur, ``{Common Structures Between Finite Systems and
  Conformal Field Theories Through Quantum Groups},''
\href{http://dx.doi.org/10.1016/0550-3213(90)90122-T}{{\em Nucl.Phys.}
  {\bfseries B330} (1990) 523}.

\bibitem{Sklyanin:1988yz}
E.~K. Sklyanin, ``{Boundary Conditions for Integrable Quantum Systems},''
\href{http://dx.doi.org/10.1088/0305-4470/21/10/015}{{\em J.Phys.} {\bfseries
  A21} (1988) 2375--289}.

\bibitem{Alcaraz:1987uk}
F.~C. Alcaraz, M.~N. Barber, M.~T. Batchelor, R.~J. Baxter, and G.~R.~W.
  Quispel, ``{Surface Exponents of the Quantum XXZ, Ashkin-Teller and Potts
  Models},''
\href{http://dx.doi.org/10.1088/0305-4470/20/18/038}{{\em J.Phys.} {\bfseries
  A20} (1987) 6397}.

\bibitem{Dubail:2010zz}
J.~Dubail, J.~L. Jacobsen, and H.~Saleur, ``{Conformal field theory at central
  charge $c = 0$: a measure of the indecomposability (b) parameters},''
  \href{http://dx.doi.org/10.1016/j.nuclphysb.2010.02.016}{{\em Nucl.Phys.}
  {\bfseries B834} (2010) 399--422},
\href{http://arxiv.org/abs/1001.1151}{{\ttfamily arXiv:1001.1151 [math-ph]}}.

\bibitem{Vasseur:2011fi}
R.~Vasseur, J.~L. Jacobsen, and H.~Saleur, ``{Indecomposability parameters in
  chiral Logarithmic Conformal Field Theory},''
  \href{http://dx.doi.org/10.1016/j.nuclphysb.2011.05.018}{{\em Nucl.Phys.}
  {\bfseries B851} (2011) 314--345},
\href{http://arxiv.org/abs/1103.3134}{{\ttfamily arXiv:1103.3134 [math-ph]}}.

\bibitem{MorinDuchesne:2011kd}
A.~Morin-Duchesne and Y.~Saint-Aubin, ``{The Jordan Structure of Two
  Dimensional Loop Models},''
  \href{http://dx.doi.org/10.1088/1742-5468/2011/04/P04007}{{\em J.Stat.Mech.}
  {\bfseries 1104} (2011) P04007},
\href{http://arxiv.org/abs/1101.2885}{{\ttfamily arXiv:1101.2885 [math-ph]}}.

\bibitem{Gainutdinov:2015vba}
A.~Gainutdinov, W.~Hao, R.~I. Nepomechie, and A.~J. Sommese, ``{Counting
  solutions of the Bethe equations of the quantum group invariant open XXZ
  chain at roots of unity},''
  \href{http://dx.doi.org/10.1088/1751-8113/48/49/494003}{{\em J.Phys.}
  {\bfseries A48} (2015) 494003},
\href{http://arxiv.org/abs/1505.02104}{{\ttfamily arXiv:1505.02104 [math-ph]}}.

\bibitem{Baxter:1972wg}
R.~J. Baxter, ``{Eight vertex model in lattice statistics and one-dimensional
  anisotropic Heisenberg chain. I. Some fundamental eigenvectors},''
\href{http://dx.doi.org/10.1016/0003-4916(73)90439-9}{{\em Annals Phys.}
  {\bfseries 76} (1973) 1--24}.

\bibitem{Fabricius:2000yx}
K.~Fabricius and B.~M. McCoy, ``{Bethe's equation is incomplete for the XXZ
  model at roots of unity},''
  \href{http://dx.doi.org/10.1023/A:1010380116927}{{\em J.Statist.Phys.}
  {\bfseries 103} (2001) 647--678},
\href{http://arxiv.org/abs/cond-mat/0009279}{{\ttfamily arXiv:cond-mat/0009279
  [cond-mat.stat-mech]}}.

\bibitem{Fabricius:2001yy}
K.~Fabricius and B.~M. McCoy,
  \href{http://dx.doi.org/10.1007/978-1-4612-0087-1_6}{``{Evaluation parameters
  and Bethe roots for the six vertex model at roots of unity},''} in {\em
  MathPhys Odyssey 2001}, M.~Kashiwara and T.~Miwa, eds., pp.~119--144.
\newblock Springer, 2002.
\newblock
\href{http://arxiv.org/abs/cond-mat/0108057}{{\ttfamily arXiv:cond-mat/0108057
  [cond-mat]}}.
\newblock

\bibitem{Baxter:2001sx}
R.~J. Baxter, ``{Completeness of the Bethe ansatz for the six and eight vertex
  models},'' \href{http://dx.doi.org/10.1023/A:1015437118218}{{\em
  J.Statist.Phys.} {\bfseries 108} (2002) 1--48},
\href{http://arxiv.org/abs/cond-mat/0111188}{{\ttfamily arXiv:cond-mat/0111188
  [cond-mat]}}.

\bibitem{Tarasov:2003xz}
V.~O. Tarasov, ``{On Bethe vectors for the XXZ model at roots of unity},''
  \href{http://dx.doi.org/10.1023/B:JOTH.0000049576.42200.77}{{\em J. Math.
  Sciences} {\bfseries 125} (2005) 242--248},
  \href{http://arxiv.org/abs/math/0306032}{{\ttfamily arXiv:math/0306032
  [math.QA]}}.

\bibitem{Hou:1991tc}
B.-Y. Hou, K.-J. Shi, Z.-X. Yang, and R.-H. Yue, ``{Integrable quantum chain
  and the representation of quantum group $SU_q(2)$},''
{\em J.Phys.} {\bfseries A24} (1991) 3825--3836.

\bibitem{Mezincescu:1991rb}
L.~Mezincescu and R.~I. Nepomechie, ``{Quantum algebra structure of exactly
  soluble quantum spin chains},''
\href{http://dx.doi.org/10.1142/S0217732391002931}{{\em Mod.Phys.Lett.}
  {\bfseries A6} (1991) 2497--2508}.

\bibitem{Izergin:1982hy}
A.~Izergin and V.~Korepin, ``{Pauli principle for one-dimensional bosons and
  the algebraic Bethe ansatz},''
\href{http://dx.doi.org/10.1007/BF00400323}{{\em Lett.Math.Phys.} {\bfseries 6}
  (1982) 283--288}.

\bibitem{Gainutdinov:2012qy}
A.~Gainutdinov, H.~Saleur, and I.~Y. Tipunin, ``{Lattice W-algebras and
  logarithmic CFTs},''
  \href{http://dx.doi.org/10.1088/1751-8113/47/49/495401}{{\em J.Phys.}
  {\bfseries A47} no.~49, (2014) 495401},
\href{http://arxiv.org/abs/1212.1378}{{\ttfamily arXiv:1212.1378 [hep-th]}}.

\bibitem{Kulish:1991np}
P.~P. Kulish and E.~K. Sklyanin, ``{The general $U_q sl(2)$ invariant XXZ
  integrable quantum spin chain},''
{\em J.Phys.} {\bfseries A24} (1991) L435--L439.

\bibitem{Chari:1994pz}
V.~Chari and A.~Pressley, ``A guide to quantum groups,''
\newblock CUP,
1994.
\newblock

\bibitem{Bushlanov:2009cv}
P.~Bushlanov, B.~Feigin, A.~Gainutdinov, and I.~Y. Tipunin, ``{Lusztig limit of
  quantum $sl(2)$ at root of unity and fusion of $(1,p)$ Virasoro logarithmic
  minimal models},''
  \href{http://dx.doi.org/10.1016/j.nuclphysb.2009.03.016}{{\em Nucl.Phys.}
  {\bfseries B818} (2009) 179--195},
\href{http://arxiv.org/abs/0901.1602}{{\ttfamily arXiv:0901.1602 [hep-th]}}.

\bibitem{Gainutdinov:2012mr}
A.~M. Gainutdinov and R.~Vasseur, ``{Lattice fusion rules and logarithmic
  operator product expansions},''
  \href{http://dx.doi.org/10.1016/j.nuclphysb.2012.11.004}{{\em Nucl.Phys.}
  {\bfseries B868} (2013) 223--270},
\href{http://arxiv.org/abs/1203.6289}{{\ttfamily arXiv:1203.6289 [hep-th]}}.

\bibitem{Martin:1991pk}
P.~P. Martin and D.~S. Mcanally, ``{On commutants, dual pairs and nonsemisimple
  algebras from statistical mechanics},''
\href{http://dx.doi.org/10.1142/S0217751X92003987}{{\em Int.J.Mod.Phys.}
  {\bfseries A7S1B} (1992) 675--705}.

\bibitem{Read:2007qq}
N.~Read and H.~Saleur, ``{Associative-algebraic approach to logarithmic
  conformal field theories},''
  \href{http://dx.doi.org/10.1016/j.nuclphysb.2007.03.033}{{\em Nucl.Phys.}
  {\bfseries B777} (2007) 316--351},
\href{http://arxiv.org/abs/hep-th/0701117}{{\ttfamily arXiv:hep-th/0701117
  [hep-th]}}.

\bibitem{Tarasov:1991mf}
V.~O. Tarasov, ``{Cyclic monodromy matrices for the $R$ matrix of the six vertex
  model and the chiral Potts model with fixed spin boundary conditions},''
\href{http://dx.doi.org/10.1142/S0217751X92004129}{{\em Int. J. Mod. Phys.}
  {\bfseries A7S1B} (1992) 963--975}.

\bibitem{Tarasov:1992aw}
V.~Tarasov, ``{Cyclic monodromy matrices for $sl(n)$ trigonometric $R$ matrices},''
  \href{http://dx.doi.org/10.1007/BF02096799}{{\em Commun. Math. Phys.}
  {\bfseries 158} (1993) 459--484},
\href{http://arxiv.org/abs/hep-th/9211105}{{\ttfamily arXiv:hep-th/9211105
  [hep-th]}}.

\bibitem{Kulish:1981gi}
P.~P. Kulish, N.~{\relax Yu}. Reshetikhin, and E.~K. Sklyanin, ``{Yang-Baxter
  Equation and Representation Theory. 1.},''
\href{http://dx.doi.org/10.1007/BF02285311}{{\em Lett. Math. Phys.} {\bfseries
  5} (1981) 393--403}.

\bibitem{Kulish:1981bi}
P.~P. Kulish and E.~K. Sklyanin, ``{Quantum spectral transform method. Recent
  developments},''
{\em Lect. Notes Phys.} {\bfseries 151} (1982) 61--119.

\bibitem{Mezincescu:1991ke}
L.~Mezincescu and R.~I. Nepomechie, ``{Fusion procedure for open chains},''
{\em J. Phys.} {\bfseries A25} (1992) 2533--2544.

\bibitem{Foerster:1993fp}
A.~Foerster and M.~Karowski, ``{The Supersymmetric $t - J$ model with quantum
  group invariance},''
\href{http://dx.doi.org/10.1016/0550-3213(93)90377-2}{{\em Nucl. Phys.}
  {\bfseries B408} (1993) 512--534}.

\bibitem{Baseilhac:2016}
P.~Baseilhac, A.~M. Gainutdinov, and T.T.~Vu, ``{Cyclic tridiagonal pairs, higher
  order Onsager algebras and orthogonal polynomials},''  in preparation.

\end{thebibliography}

\providecommand{\href}[2]{#2}\begingroup\raggedright\endgroup

\end{document}